\documentclass[fleqn,usenatbib]{mnras}
\usepackage{newtxtext,newtxmath}
\usepackage{lscape}
\usepackage{rotate}
\usepackage{rotating}
\usepackage{hyperref}
\usepackage[T1]{fontenc}
\usepackage[paper=portrait,pagesize]{typearea}
\usepackage{amsmath}
\usepackage{mathtools}

\newbox\grsign \setbox\grsign=\hbox{$>$} \newdimen\grdimen \grdimen=\ht\grsign
\newbox\simlessbox \newbox\simgreatbox
\setbox\simgreatbox=\hbox{\raise.5ex\hbox{$>$}\llap
     {\lower.5ex\hbox{$\sim$}}}\ht1=\grdimen\dp1=0pt
\setbox\simlessbox=\hbox{\raise.5ex\hbox{$<$}\llap
     {\lower.5ex\hbox{$\sim$}}}\ht2=\grdimen\dp2=0pt
\def\simgreater{\mathrel{\copy\simgreatbox}}
\def\simless{\mathrel{\copy\simlessbox}}
\newbox\simppropto
\setbox\simppropto=\hbox{\raise.5ex\hbox{$\sim$}\llap
     {\lower.5ex\hbox{$\propto$}}}\ht2=\grdimen\dp2=0pt

\usepackage{color}

\usepackage{textcomp}
\def\pm{\textpm}

\usepackage{graphicx}	
\UseRawInputEncoding

\title[Substructure in the Galactic stellar halo]{The chemical characterisation of halo substructure in the Milky Way based on APOGEE}

\author[D. Horta et al.]{
Danny Horta$^{1,2}$,\thanks{E-mail: D.HortaDarrington@2018.ljmu.ac.uk}
Ricardo P. Schiavon$^{1}$,
J. Ted Mackereth$^{3,4}$,
David H. Weinberg$^{5}$,
\newauthor
Sten Hasselquist$^{6}$,
Diane Feuillet$^{7}$,
Robert W. O'Connell$^{8}$,
Borja Anguiano$^{8}$,
\newauthor
Carlos Allende-Prieto$^{9,10}$,
Rachael L. Beaton$^{11}$,
Dmitry Bizyaev$^{12}$
Katia Cunha$^{13,14}$,
\newauthor
Doug Geisler$^{15,16,17}$,
D. A. Garc\'ia-Hern\'andez$^{9,10}$,
Jon Holtzman$^{12}$,
Henrik J\"onsson$^{18}$,
\newauthor
Richard R. Lane$^{19}$,
Steve R. Majewski$^{8}$,
Szabolcs M\'esz\'aros$^{20,21}$,
Dante Minniti$^{22,23}$,
\newauthor
Christian Nitschelm$^{24}$,
Matthew Shetrone$^{25}$,
Verne V. Smith$^{26}$,
Gail Zasowski$^{27}$
\\
$^{}$Affiliations are listed at the end of the paper}
\date{Accepted XXX. Received YYY; in original form ZZZ}
\pubyear{2021}
\begin{document}
\label{firstpage}
\pagerange{\pageref{firstpage}----\pageref{lastpage}}
\maketitle
\begin{abstract}
Galactic haloes in a $\Lambda$-CDM universe are predicted to host today a swarm of debris resulting from cannibalised dwarf galaxies. The chemo-dynamical information recorded in their stellar populations helps elucidate their nature, constraining the assembly history of the Galaxy.  Using data from APOGEE and \textit{Gaia}, we examine the chemical properties of various halo substructures, considering elements that sample various nucleosynthetic pathways. The systems studied are Heracles, \textit{Gaia}-Enceladus/Sausage (GES), the Helmi stream, Sequoia, Thamnos, Aleph, LMS-1, Arjuna, I'itoi, Nyx, Icarus, and Pontus.  Abundance patterns of all substructures are cross-compared in a statistically robust fashion.  Our main findings include: {\it i)} the chemical properties of most substructures studied match qualitatively those of dwarf Milky Way satellites, such as the Sagittarius dSph. Exceptions are Nyx and Aleph, which are chemically similar to disc stars, implying that these substructures were likely formed \textit{in situ}; {\it ii)} Heracles differs chemically from {\it in situ} populations such as Aurora and its inner halo counterparts in a statistically significant way. The differences suggest that the star formation rate was lower in Heracles than in the early Milky Way; {\it iii)} the chemistry of Arjuna, LMS-1, and I'itoi is indistinguishable from that of GES, suggesting a possible common origin; {\it iv)} all three Sequoia samples studied are qualitatively similar.  However, only two of those samples present chemistry that is consistent with GES in a statistically significant fashion; {\it v)} the abundance patterns of the Helmi stream and Thamnos are different from all other halo substructures.
\end{abstract}

\section{Introduction}
\label{Introduction}
"How did the Milky Way form?" is likely the most fundamental question facing the Field of Galactic archaeology. When posed in a cosmological context, the $\Lambda$-CDM model predicts that the Galaxy formed in great measure via the process of hierarchical mass assembly. In this scenario, nearby satellite galaxies are consumed by the Milky Way due to them being attracted to its deeper gravitational potential, and as a result merge with the Galaxy. In such cases, these merger events shape the formation and evolution of the Milky Way. Therefore, an understanding of the assembly history of the Milky Way in the context of $\Lambda$-CDM depends critically on the determination of the properties of the systems accreted during the Galaxy's history, including their masses and chemical compositions. Moreover, the merger history of the Galaxy has a direct impact on its resulting stellar populations at present time, and plays a vital role in shaping its components.

 Since the seminal work by \citet{Searle1978}, many studies have aimed at characterising the stellar populations of the Milky Way, linking them to either an "\emph{in situ}" or accreted origin. Albeit detection of substructure in phase space has worked extremely well for the identification of on-going and/or recent accretion events (\citealp[e.g. Sagittarius dSph,][]{Ibata1994}; \citealp[Helmi stream,][]{Helmi1999}), the identification of accretion events early in the life of the Milky Way has proven difficult due to phase-mixing. A possible solution to this conundrum resides in the use of additional information, typically in the form of detailed chemistry and/or ages (\citealp[e.g.,][]{Nissen2010,Hawkins2015,Hayes2018,Haywood2018,Mackereth2019b,Das2020,Montalban2020,Horta2021, Hasselquist2021, Buder2022,Carrillo2022}).

The advent of large spectroscopic surveys such as APOGEE \citep{Majewski2017}, GALAH \citep{DeSilva2015}, SEGUE \citep{Segue2009}, RAVE \citep{Steinmetz2020}, LAMOST \citep{Zhao2012}, H3 \citep{Conroy2019}, amongst others, in combination with the outstanding astrometric data supplied by the \emph{Gaia} satellite \citep{Gaia2018,Gaiaedr3}, revolutionised the field of Galactic archaeology, shedding new light into the mass assembly history of the Galaxy.

The core of the Sagittarius dSph system and its still forming tidal stream \citep[][]{Ibata1994} have long served as an archetype for dwarf galaxy mergers in the Milky Way. Moreover, in the past few years, several phase-space substructures have been identified in the Field of the Galactic stellar halo that are believed to be the debris of satellite accretion events, including the \emph{Gaia}-Enceladus/Sausage (\citealp[GE/S,][]{Helmi2018,Belokurov2018,Haywood2018,Mackereth2019b}), Heracles \citep{Horta2021}, Sequoia (\citealp[][]{barba2019,Matsuno2019,Myeong2019}), Thamnos 1 and 2 \citep{Koppelman2019b}, Nyx \citep[][]{Necib2020}, , LMS-1 \citep[][]{Yuan2020}\footnote{This structure also goes by the name of Wukong \citep[][]{Naidu2020}.}, the substructures identified using the H3 survey: namely Aleph, Arjuna, and I'itoi \citep[][]{Naidu2020}, Icarus \citep[][]{Refiorentin2021}, Cetus \citep[][]{Newberg2009}, and Pontus \citep[][]{Malhan2022}. While the identification of these substructures is helping constrain our understanding of the mass assembly history of the Milky Way, their association with any particular accretion event still needs to be clarified. Along those lines, predictions from numerical simulations suggest that a single accretion event can lead to multiple substructures in phase space \citep[e.g.,][]{Jean2017, Koppelman2020}. Therefore, in order to ascertain the reality and/or distinction of these accretion events, one must combine phase-space information with detailed chemical compositions for large samples.

Previous studies dedicated to characterising the chemical properties of halo substructures have been based on either large samples from spectroscopic surveys, focusing on a relatively small number of elemental abundances (\citealp[e.g.,][]{Hayes2020,Feuillet2021,Hasselquist2021,Buder2022}), or more detailed studies of smaller samples from follow-up programs at an even higher resolution (\citealp[e.g.,][]{Monty2020,Limberg2021,Matsuno2022,Matsuno2022_seq, Naidu2022}). This is the first attempt at mapping the detailed abundance patterns of all halo substructures reported thus far in the literature.

In this work we set out to combine the latest data releases from the APOGEE and \emph{Gaia} surveys in order to dynamically determine and chemically characterise previously identified halo substructures in the Milky Way. We attempt, where possible, to define the halo substructures using kinematic information only, so that the distributions of stellar populations in various chemical planes can be studied in an unbiased fashion. This allows us to understand in more detail the reality and nature of these identified halo substructures, as chemical abundances encode more pristine fossilised records of the formation environment of stellar populations in the Galaxy.

For the convenience of the reader we hope to facilitate navigation of the paper by supplying a detailed list of its contents, as follows:
\begin{itemize}
    \item Section~\ref{data} describes the data and the selection criteria adopted to select the parent sample upon which this work is based. This is followed in Section~\ref{method} by a detailed account of the criteria adopted to define each substructure, building on techniques from previous work. For the keen reader interested solely on the study of the substructures' chemical compositions, we suggest jumping straight into Section~\ref{chemistry}.
    \item Section~\ref{kinematics} very briefly presents the resulting  distributions of the various structures in the orbital energy vs angular momentum plane.
    \item Section~\ref{chemistry} presents a qualitative examination of the  halo substructures in different chemical  planes. Specifically, we show the $\alpha$ elements in Section~\ref{sec_alphas}, the iron-peak elements in Section~\ref{sec_ironpeak}, the odd-Z elements in Section~\ref{sec_oddZ}, the carbon and nitrogen abundances in Section~\ref{sec_cn}, a neutron capture element (namely, Ce) in Section~\ref{sec_neutron_capture}, and the [Mg/Mn]-[Al/Fe] chemical composition plane in Section~\ref{sec_other_chem}.  Readers interested in the quantitative results may skip this section and go straight to Section~\ref{sec_abundances}.
    \item Section~\ref{sec_abundances} describes a statistical technique we developed to calculate a quantitative estimate of the chemical similitude between any two substructures. We run comparisons between all possible pairs of substructures and the results are encapsulated in the form of a confusion matrix in  Fig~\ref{confusion_matrix}.  
    \item We then discuss our results in the context of previous work in Section~\ref{discussion}, and summarise our conclusions in Section~\ref{conclusion}. 
\end{itemize}

\section{Data and sample} 
\label{data} 
This paper combines the latest data release \citep[DR17,][]{sdss2021} of the SDSS-III/IV (\citealp{Eisenstein2011,Blanton2017}) and APOGEE survey (\citealp[][]{Majewski2017}) with distances and astrometry determined from the early third data release from the \textit{Gaia} survey \citep[EDR3,][]{Gaia2020}. The celestial coordinates and radial velocities supplied by APOGEE \citep[][Holtzman et al, in prep]{Nidever2015}, when combined with the proper motions and inferred distances \citep{Leung2019} based on \textit{Gaia} data, provide complete 6-D phase space information for over $\sim$700,000 stars in the Milky Way, for most of which exquisite abundances for up to $\sim$20 different elements have been determined.

All data supplied by APOGEE are based on observations collected by (almost) twin high-resolution multi-fibre spectrographs \citep{Wilson2019} attached to the 2.5m Sloan telescope at Apache Point Observatory \citep{Gunn2006}
and the du Pont 2.5~m telescope at Las Campanas Observatory
\citep{BowenVaughan1973}. Elemental abundances are derived from
automatic analysis of stellar spectra using the ASPCAP
pipeline \citep{Perez2015} based on the FERRE\footnote{github.com/callendeprieto/ferre} code \citep[][]{Prieto2006} and the line lists from \citet{Cunha2017} and \citet[][]{Smith2021}. The spectra themselves were
reduced by a customized pipeline \citep{Nidever2015}. For details on
target selection criteria, see \citet{Zasowski2013} for APOGEE, \citet{Zasowski2017} for APOGEE-2, \citet[][]{Beaton2021} for APOGEE north, and \citet[][]{Santana2021} for APOGEE south.

We make use of the distances for the APOGEE DR17 catalogue generated by \citet{Leung2019b}, using the \texttt{astroNN} python package \citep[for a full description, see][]{Leung2019}. These distances are determined using a re-trained \texttt{astroNN} neural-network software, which predicts stellar luminosity from spectra using a training set comprised of stars with both APOGEE DR17 spectra and \textit{Gaia} EDR3 parallax measurements \citep{Gaia2020}. The
model is able to simultaneously predict distances and account for
the parallax offset present in \textit{Gaia}-EDR3, producing high precision,
accurate distance estimates for APOGEE stars, which match well
with external catalogues \citep[][]{Hogg2019} and standard candles like red clump stars \citep[][]{Bovy2014}. We note that the systematic bias in distance measurements at large distances for APOGEE DR16 as described in \citet[][]{Bovy2019} have been reduced drastically in APOGEE DR17. Therefore, we are confident that this bias will not lead to unforeseen issues during the calculation of the orbital parameters. Our samples are contained within a distance range of $\sim$20 kpc and have a mean $d_{\mathrm{err}}$/$d$ $\sim$0.13 (except for the Sagittarius dSph, which extends up to $\sim$30 kpc and has a mean $d_{\mathrm{err}}$/$d$ $\sim$0.16).

We use the 6-D phase space information\footnote{The positions, proper motions, and distances are taken/derived from \textit{Gaia} EDR3 data, whilst the radial velocities are taken from APOGEE DR17.} and convert between astrometric parameters and Galactocentric cylindrical coordinates, assuming a solar velocity combining the proper motion from Sgr~A$^{*}$ \citep{Reid2020} with the determination of the local standard of rest of \citet{Schonrich2010}. This adjustment leads to a 3D velocity of the Sun equal to [U$_{\odot}$, V$_{\odot}$, W$_{\odot}$] = [--11.1, 248.0, 8.5] km s$^{-1}$. We assume the distance between the Sun and the Galactic Centre to be R$_{0}$ = 8.178~kpc \citep{Gravity2019}, and the vertical height of the Sun above the midplane $z_{0}$ = 0.02~kpc \citep{Bennett2019}. Orbital parameters were then determined using the publicly available code \texttt{galpy}\footnote{\href{https:/docs.galpy.org/en/v1.6.0/}{https:/docs.galpy.org/en/v1.6.0/.}} (\citealp[][]{Galpy2015,Galpy2018}), adopting a \citet{McMillan2017} potential and using the St\"ackel approximation of \citet[][]{Binney2012}.

The parent sample employed in this work is comprised of stars that satisfy the following selection criteria:
\begin{itemize}
    \item APOGEE-determined atmospheric parameters: $3500<\mathrm{T}_{\rm eff}<5500$~K and $\log{g}< 3.6$,
    \item APOGEE spectral S/N $>$ 70,
    \item APOGEE \texttt{STARFLAG} = 0,
    \item  astroNN distance accuracy of $d_{\odot, \rm err}/d_{\odot}<0.2$,
\end{itemize}
where $d_{\odot}$ and $d_{\odot\rm err}$ are heliocentric distance and its uncertainty, respectively. The S/N criterion was implemented to maximise the quality of the elemental abundances. The T$_{\rm eff}$ and $\log{g}$ criteria aimed to minimise systematic effects at high/low temperatures, and to minimise contamination by dwarf stars. We also removed stars with \texttt{STARFLAG} flags set, in order to not include any stars with issues in their stellar parameters. 
A further 7,750 globular cluster stars were also removed from consideration using the APOGEE Value Added Catalogue of globular cluster candidate members from Schiavon et al. (2022, in prep.), \citep[building on the method from][using primarily radial velocity and proper motion information]{Horta2020}. Finally, stars belonging to the Large and Small Magellanic clouds were also excluded using the sample from \citet[][]{Hasselquist2021} (removing 3,748 and 1,002 stars, respectively). The resulting parent sample contains 199,030 stars.

\begin{figure*}
\includegraphics[width=\textwidth]{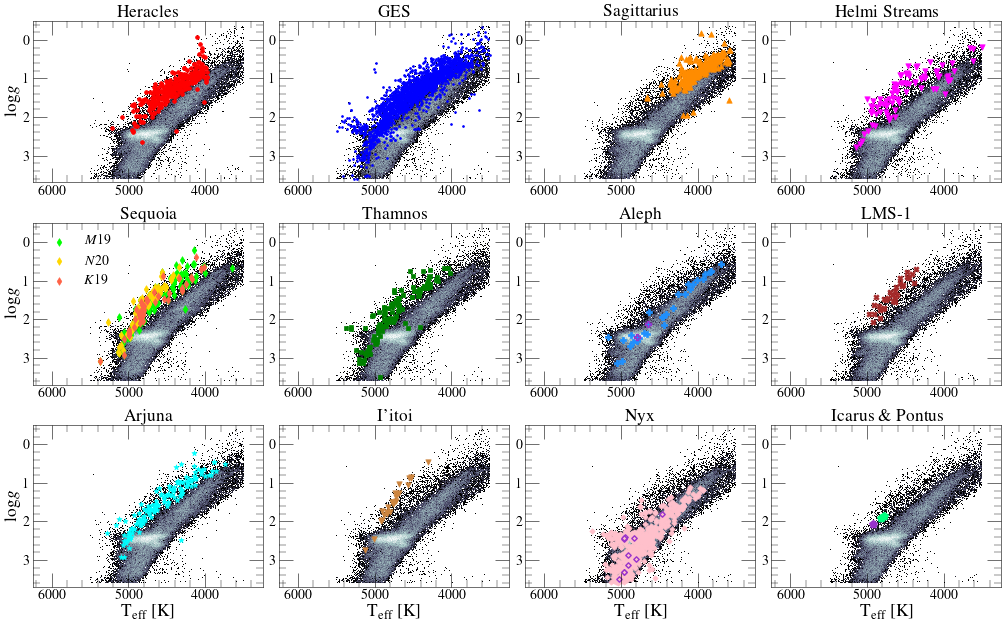}
\caption{Distribution of the identifed substructures in the Kiel diagram. The parent sample, as defined in Section~\ref{data} is plotted as a 2D histogram, where white/black signifies high/low density regions. The coloured markers illustrate the different halo substructures studied in this work. For the bottom right panel, green points correspond to Pontus stars, whereas the purple point is associated with Icarus. Additionally, in the Aleph and Nyx panels, we also highlight with purple edges those stars that overlap between the APOGEE DR17 sample and the samples determined in \citet[][]{Naidu2020} and \citet[][]{Necib2020}, respectively, for these halo substructures.}
\label{fig:hr}
\end{figure*}

In the following subsection we describe the motivation behind the selection criteria employed to select each substructure in the stellar halo of the Milky Way. The criteria are largely built on selections employed in previous works and are summarised in Table \ref{tab:selection} and in Figure~\ref{fig:hr}.

\subsection{Identification of substructures in the stellar halo} 
\label{method}

We now describe the method employed for identifying known substructures in the stellar halo of the Milky Way. We set out to select star members belonging to the various halo substructures.

We strive to identify substructures in the stellar halo by employing solely orbital parameter and phase-space information where possible, with the aim of obtaining star candidates for each substructure population that are unbiased by any chemical composition selection. 

We take a handcrafted approach and select substructures based on simple and reproducible selection criteria that are physically motivated by the data and/or are used in previous works, instead of resorting to clustering software algorithms, which we find cluster the $n$-dimensional space into too many fragments and vary wildly depending on the input parameters to the clustering model. In the following subsections we describe the selection procedure for identifying each substructure independently.

 We note that our samples for the various substructures are defined by a strict application to the APOGEE survey data of the criteria defined by other groups, often on the basis of different data sets.  The latter were per force collected as part of a different observational effort, based on specific target selection criteria.  It is not immediately clear whether or how differences between the APOGEE selection function and those of other catalogues may imprint dissimilarities between our samples and those of the original studies.  We nevertheless do not expect such effects to influence our conclusions, regarding the chemical compositions of these structures, in an important way.

\subsubsection{Sagittarius}
\label{sec_sgr} 
Since its discovery \citep{Ibata1994}, many studies have sought to characterise the nature of the Sagittarius dwarf spheroidal (\citealp[hereafter Sgr dSph; e.g.,][]{Ibata2001,Majewski2003,Johnston2005,Belokurov2006,Yanny2009,Koposov2012,Carlin2018,Antoja2020,Ibata2020,Vasiliev2020}), as well as interpret its effect on the Galaxy using numerical simulations (\citealp[e.g.,][]{Johnston1995,Ibata1997,Law2005,Law2010,Purcell2011,Gomez2013}). More recently, the Sgr~dSph has been the subject of comprehensive studies on the basis of APOGEE data. This has enabled a detailed examination of its chemical compositions, both in the satellite's core and in its tidal tails (\citealp[e.g.,][]{Hasselquist2017,Hasselquist2019,Hayes2020}). Moreover, in a more recent study, \cite{Hasselquist2021} adopted chemical evolution models to infer the history of star formation and chemical evolution of the Sgr dSph.  Therefore, in this paper the Sgr dSph is considered simply as a template massive satellite whose chemical properties can be contrasted to those of the  halo substructures that are the focus of our study.

\begin{figure}
\includegraphics[width=\columnwidth]{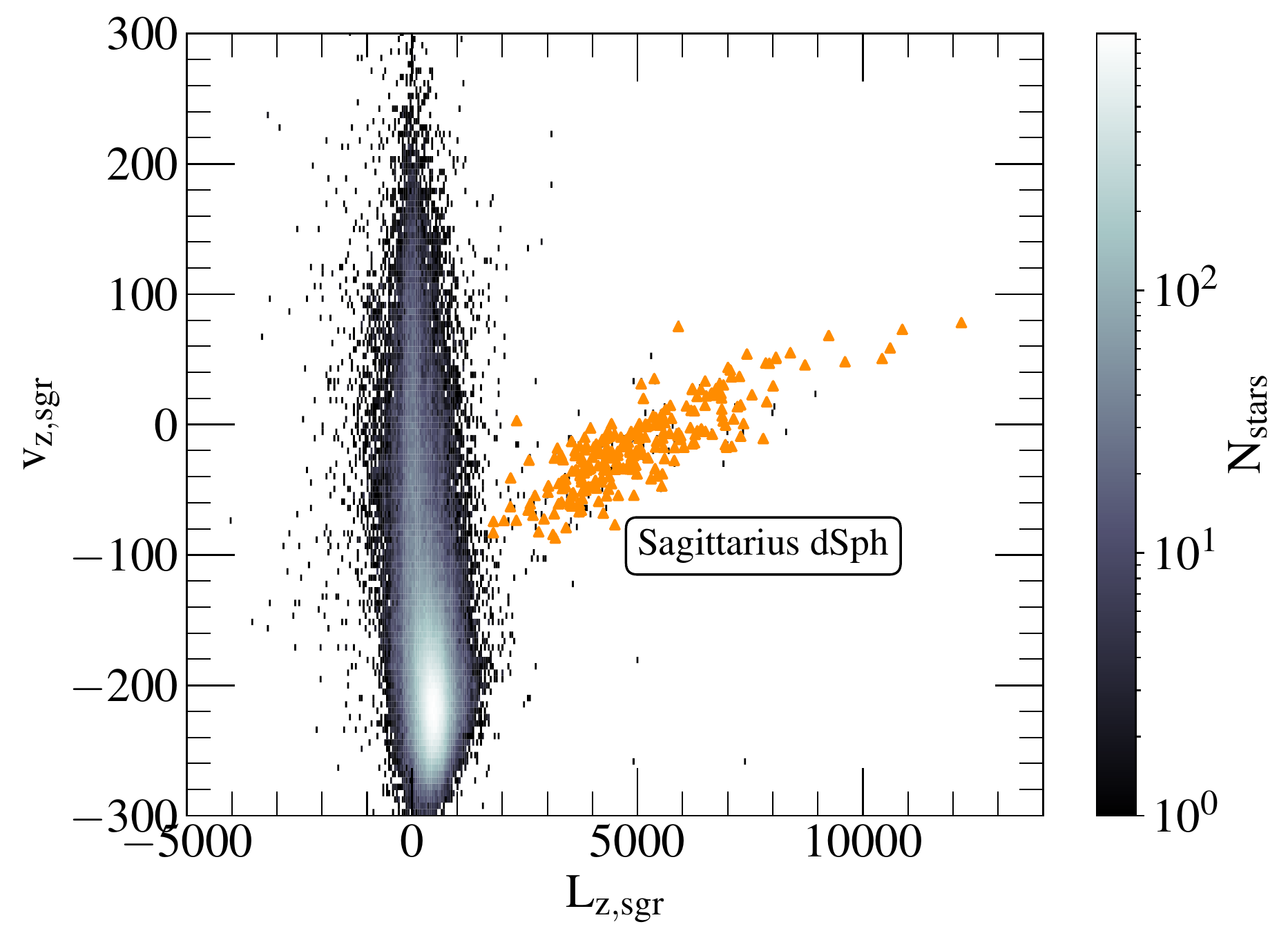}
\caption{Parent sample used in this work in the L$_{\mathrm{z,Sgr}}$ vs $\mathrm{V}_{\mathrm{z,Sgr}}$ plane (see Section~\ref{sec_sgr} for details). Here, Sgr stars clearly depart from the parent sample, and are easily distinguishable by applying the selection criteria from \citet[][]{Hayes2020}, demarked in this illustration by the orange markers.}
    \label{sag_selec}
\end{figure}

While it is possible to select high confidence Sgr~dSph star candidates using Galactocentric positions and velocities \citep{Majewski2003}, \citet{Hayes2020} showed it is possible to make an even more careful selection by considering the motion of stars in a well-defined Sgr orbital plane.  We identify Sgr star members by following the method from \citet{Hayes2020}. Although the method is fully described in their work, we summarise the key steps for clarity and completeness. We take the Galactocentric positions and velocities of stars in our parent sample and rotate them into the Sgr orbital plane according to the transformations described in \citet{Majewski2003}. This yields a set of position and velocity coordinates relative to the Sgr orbital plane, but still centered on the Galactic Centre. As pointed out in \citet{Hayes2020}, Sgr star members should stand out with respect to other halo populations in different Sgr orbital planes. Using this orbital plane transformation, we select from the parent sample Sgr star members if they satisfy the following selection criteria:

\begin{itemize}
\item |$\beta_{\mathrm{GC}}$| < 30 ($^{\circ}$),
\item 18 < L$_{z,\mathrm{Sgr}}$ < 14 ($\times$10$^{3}$ kpc kms$^{-1}$),
\item --150 < $\mathrm{V}_{z,\mathrm{Sgr}}$ < 80 (kms$^{-1}$),
\item  X$_{\mathrm{Sgr}}$ > 0 or X$_{\mathrm{Sgr}}$ < --15 (kpc), 
\item Y$_{\mathrm{Sgr}}$ > --5 (kpc) or Y$_{\mathrm{Sgr}}$ < --20 (kpc),
\item  Z$_{\mathrm{Sgr}}$ > --10 (kpc),
\item pm$_{\alpha}$ > --4 (mas),
\item $d_{\odot}$ > 10 (kpc),
\end{itemize}
where $\beta_{\mathrm{GC}}$ is the angle subtended between the Galactic Centre and the Sgr dSph, L$_{\mathrm{z,Sgr}}$ is the azimuthal component of the angular momentum in the Sgr plane, $\mathrm{V}_{\mathrm{z,Sgr}}$ is the vertical component of the velocity in the Sgr plane, (X$_{\mathrm{Sgr}}$, Y$_{\mathrm{Sgr}}$, Z$_{\mathrm{Sgr}}$) are the cartesian coordinates centred on the Sgr dSph plane, pm$_{\alpha}$ is the right-ascension proper motion, and $d_{\odot}$ is the heliocentric distance, which for Sgr has been shown to be $\sim$ 23 kpc \citep{Vasiliev2020} (although we follow the distance cut from \citet[][]{Hayes2020} for this work, to prevent the exclusion of more nearby parts of the Sgr stream). Our selection yields a sample of 266 Sgr star members, illustrated in the L$_{\mathrm{z,Sgr}}$ vs $\mathrm{V}_{\mathrm{z,Sgr}}$ plane in Fig~\ref{sag_selec}.

\subsubsection{Heracles}
Heracles is a recently discovered metal-poor substructure located in the heart of the Galaxy \citep{Horta2021}. It is characterised by stars on eccentric and low energy orbits. Due to its position in the inner few kpc of the Galaxy, it is highly obscured by dust extinction and vastly outnumbered by its more abundant metal-rich (\textit{in situ}) co-spatial counterpart populations. Only with the aid of chemical compositions has it been possible to unveil this metal-poor substructure, which is discernible in the [Mg/Mn]-[Al/Fe] plane. It is important at this stage that we mention a couple of recent studies which, based chiefly on the properties of the Galactic globular cluster system, proposed the occurrence of an early accretion event whose remnants should have  similar properties to those of Heracles (\citealp[named Kraken and Koala, by][respectively]{Kruijssen2020,Forbes2020}). In the absence of a detailed comparison of the dynamical properties and detailed chemical compositions of Heracles with these putative systems, and recent results by \citet[][]{Pagnini2022}, a definitive association is impossible at the current time.

In this work we define Heracles candidate star members following the work by \citet{Horta2021}, and select stars from our parent sample that satisfy the following selection criteria:  
\begin{itemize}
\item $e$ > 0.6,
\item --2.6 < E < --2 ($\times$10$^{5}$ km$^{2}$s$^{-2}$), 
\item $[\mathrm{Al/Fe}]$ < --0.07 $\&$ [Mg/Mn] $\geqslant$ 0.25, \item $[\mathrm{Al/Fe}]$ $\geqslant$ --0.07 $\&$ [Mg/Mn] $\geqslant$ 4.25$\times$[Al/Fe] + 0.5475.
\end{itemize}
Moreover, we impose a [Fe/H] > --1.7 cut to select Heracles candidate star members in order to select stars from our parent sample that have reliable Mn abundances in APOGEE DR17. Our selection yields a resulting sample of 300 Heracles star members.

\subsubsection{Gaia-Enceladus/Sausage}
Recent studies have shown that there is an abundant population of stars in the nearby stellar halo (namely, R$_{\mathrm{GC}} \lesssim$ 20-25 kpc) belonging to the remnant of an accretion event dubbed the $Gaia$-Enceladus/Sausage (\citealp[GES, e.g.,][]{Belokurov2018,Haywood2018,Helmi2018,Mackereth2019b}). This population is characterised by stars on highly radial/eccentric orbits, which also appear to follow a lower distribution in the $\alpha$-Fe plane, presenting lower [$\alpha$/Fe] values for fixed metallicity than $in$ $situ$ populations.

For this paper, we select GES candidate star members by employing a set of orbital information cuts. Specifically, GES members were selected adopting the following criteria:
\begin{itemize}
\item |L$_{z}$| < 0.5 ($\times$10$^{3}$ kpc km$^{-1}$),
\item --1.6 < E < --1.1 ($\times$10$^{5}$ km$^{2}$s$^{-2}$).
\end{itemize}
This selection is employed in order to select the clump that becomes apparent in the E-L$_{z}$ plane at higher orbital energies and roughly L$_{z}$ $\sim$ 0 (see Fig~\ref{elz}), and to minimise the contamination from high-$\alpha$ disc stars on eccentric orbits (\citealp[namely, the "Splash"][]{Bonaca2017,Belokurov2020}), which sit approximately at E$\sim$--1.8$\times$10$^{5}$ km$^{2}$s$^{-2}$ (see Kisku et al, in prep). The angular momentum restriction ensures we are not including stars on more prograde/retrograde orbits. We find that by selecting the GES substructure in this manner, we obtain a sample of stars with highly radial (J$_{R}$ $\sim$1x10$^{3}$ kpc kms$^{-1}$) and therefore highly eccentric ($e$$\sim$0.9) orbits, in agreement with selections employed in previous studies to identify this halo substructure (\citealp[e.g.,][]{Mackereth2020,Naidu2020,Feuillet2021,Buder2022}). The final GES sample is comprised of 2,353 stars.

We note that in a recent paper, \citet[][]{Hasselquist2021} undertook a thorough investigation into the chemical properties of this halo substructure and compared it to other massive satellites of the Milky Way (namely, the Magellanic Clouds, Sagittarius dSph, and Fornax). Although their selection criteria differs slightly from the one employed in this study, we find that their sample is largely similar to the one employed here, as both studies employed APOGEE DR17 data.

\subsubsection{Retrograde halo: Sequoia, Thamnos, Arjuna, and I'Itoi}\label{sec:retro}
A number of substructures have been identified in the retrograde halo. The first to be discovered was Sequoia (\citealp{barba2019,Matsuno2019,Myeong2019}), which was suggested to be the remnant of an accreted dwarf galaxy. The Sequoia was identified given the retrograde nature of the orbits of its stars, which appear to form an arch in the retrograde wing of the Toomre diagram. Separately, an interesting study by \citet{Koppelman2019b} showed that the retrograde halo can be further divided into two components, separated by their orbital energy values in the E-L$_{z}$ plane. They suggest that the high energy component corresponds to Sequoia, whilst the lower energy population would be linked to a separate accretion event, dubbed Thamnos. In addition, \cite{Naidu2020} proposed the existence of additional retrograde substructure overlapping with Sequoia, characterised by different metallicities, which they named Arjuna and I'Itoi.

As the aim of this paper is to perform a comprehensive study of the chemical abundances of substructures identified in the halo, we utilise all the selection methods used in prior work and select the same postulated substructures in multiple ways, in order to compare their abundances later. Specifically, we build on previous works (\citealp[e.g.,][]{Myeong2019,Koppelman2019b,Naidu2020}) that have aimed to characterise the retrograde halo and select the substructures following a similar selection criteria.

For reference, throughout this work we will refer to the different selection criteria of substructures in the retrograde halo as the "\textit{M19}", "\textit{K19}", and "\textit{N20}" selections, in reference to the \citet{Myeong2019}, \citet{Koppelman_thamnos}, and \citet{Naidu2020} studies that determined the Sequoia substructure, respectively. We will now go through the details of each selection method independently.

The \textit{M19} method (used in \citealt{Myeong2019}) selects Sequoia star candidates by identifying stars that satisfy the following conditions:
\begin{itemize}
\item E > --1.5 ($\times$10$^{5}$ km$^{2}$s$^{-2}$),
\item J$_{\phi}$/J$_{\mathrm{tot}}$ < --0.5,
\item J$_{\mathrm{(J_{z}-J_{R})}}$/J$_{\mathrm{tot}}$ <0.1.
\end{itemize}
Here, J$_{\phi}$, J$_{R}$, and J$_{z}$ are the azimuthal, radial, and vertical actions, and J$_{\mathrm{tot}}$ is the quadrature sum of those components. This selection yields a total of 116 Sequoia star candidates. 

The \textit{K19} method (used in \citealt{Koppelman_thamnos}) identifies Sequoia star members based on the following selection criteria:
\begin{itemize}
\item --0.65 < $\eta$ < --0.4,
\item --1.35 < E < --1 ($\times$10$^{5}$ km$^{2}$s$^{-2}$),
\item L$_{z}$ < 0 (kpc kms$^{-1}$),
\end{itemize}
where $\eta$ is the circularity. These selection criteria yield a total of 45 Sequoia stars.

\begin{figure}
\includegraphics[width=\columnwidth]{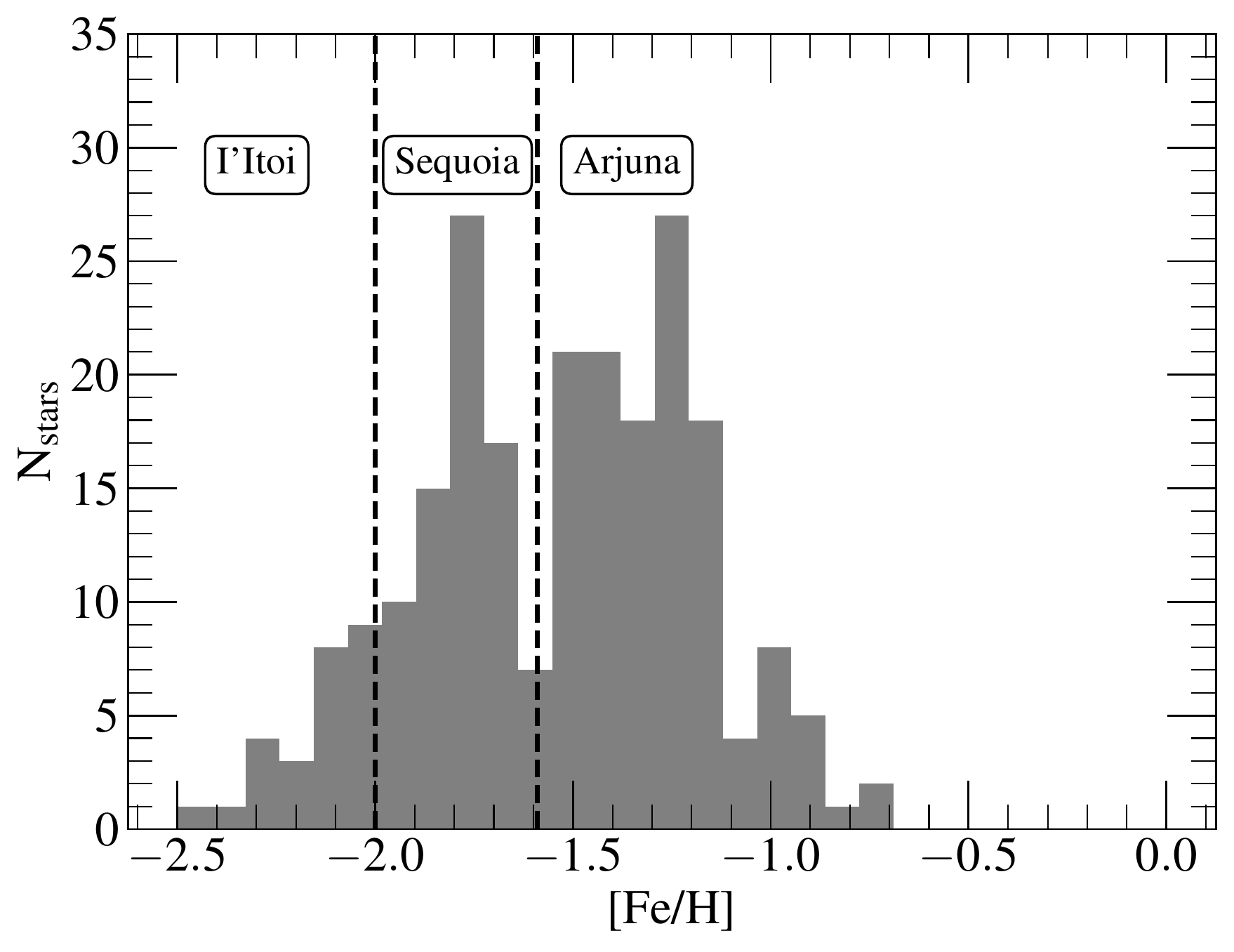}
\caption{Metallicity distribution function of the high-energy retrograde sample determined using the selection criteria from \citet{Naidu2020}. The dashed black vertical lines define the division of this sample used by \citet{Naidu2020} to divide the three high-energy retrograde substructures: Arjuna, Sequoia, and I'itoi. This MDF dissection is based both on the values used in \citet[][]{Naidu2020}, and the distinguishable peaks in this plane (we do not use a replica value of the [Fe/H] used in \citet{Naidu2020} in order to account for any possible metallicity offsets between the APOGEE and H3 surveys). We note that by purely selecting the high-energy retrograde structures in APOGEE DR17-\textit{Gaia}-DR3 as done so in \citet[][]{Naidu2020} leads to some contamination from Magellanic Cloud stars not included in the sample from \citet[][]{Hasselquist2021} that we removed earlier. We exclude these stars by performing an additional [Fe/H]$<$--0.7 and $d_{\odot}$$<$20 kpc cut.}
    \label{mdf_highe}
\end{figure}

Lastly, we select the Sequoia based on the \textit{N20} selection criteria (used in \citealt{Naidu2020}) as follows:
\begin{itemize}
\item $\eta$ < --0.15,
\item E > --1.6 ($\times$10$^{5}$ km$^{2}$s$^{-2}$),
\item L$_{z}$ < --0.7 (10$^{3}$ kpc kms$^{-1}$).
\end{itemize}
This selection produces a preliminary sample comprised of 478 stars. However, we note that by purely selecting the high-energy retrograde structures in APOGEE DR17-\textit{Gaia}-DR3 as done so in \citet[][]{Naidu2020} leads to some contamination from Magellanic Cloud stars not included in the sample from \citet[][]{Hasselquist2021} that we removed earlier. We further exclude these stars by performing an additional [Fe/H]$>$--0.7 and $d_{\odot}$$<$20 kpc cut. This leads to a high-energy retrograde sample comprising of 227 Sequoia stars. Furthermore, \citet{Naidu2020} used this selection to define not only Sequoia, but all the substructures in the high-energy retrograde halo (including the Arjuna and I'itoi substructures). In order to distinguish  Sequoia, I'itoi, and Arjuna, \citet{Naidu2020} suggest performing a metallicity cut, motivated by the observed peaks in the metallicity distribution function (MDF) of their retrograde sample. Thus, we follow this procedure and further refine our Sequoia, Arjuna and I'itoi samples by requiring --1.6 < [Fe/H] cut for Arjuna, --2 < [Fe/H] < --1.6 for Sequoia, and [Fe/H] < --2 for I'itoi, based on the distribution of our initial sample in the MDF (see Fig~\ref{mdf_highe}). This further division yields an \textit{N20} Sequoia sample comprised of 72 stars, an Arjuna sample constituted by 132 stars, and an I'itoi sample comprised of 25 stars.

Following our selection of substructures in the high-energy retrograde halo, we set out to identify stars belonging to the intermediate-energy and retrograde Thamnos 1 and 2 substructures. \citet{Koppelman_thamnos} state that Thamnos 1 and 2 are separate debris from the same progenitor galaxies. For this work we consider Thamnos as one overall structure, given the similarity noted by \citet{Koppelman_thamnos} between the two smaller individual populations in chemistry and kinematic planes. Stars from our parent sample were considered as Thamnos candidate members if they satisfied the following selection criteria:
\begin{itemize}
\item --1.8 < E < --1.6 ($\times$10$^{5}$ km$^{2}$s$^{-2}$),
\item L$_{z}$ < 0 (kpc kms$^{-1}$),
\item $e$ < 0.7,
\end{itemize}
These selection cuts are performed in order to select stars in our parent sample with intermediate orbital energies and retrograde orbits (see Fig~\ref{elz} for the position of Thamnos in the E-L$_{z}$), motivated by the distribution of the Thamnos substructure in the E-L$_{z}$ plane illustrated by \citet{Koppelman_thamnos}. This selection yields a Thamnos sample comprised of 121 stars.

\subsubsection{Helmi stream}
The Helmi stream was initially identified as a substructure in orbital space due to its high V$_{\mathrm{z}}$ velocities \citep{Helmi1999}. More recent work by \citet{Koppelman2019} characterised the Helmi stream in \emph{Gaia} DR2, and found that this stellar population is best defined by adopting a combination of cuts in different angular momentum planes. Specifically, by picking stellar halo stars based on the azimuthal component of the angular momentum ($\mathrm{L}_{z}$), and its perpendicular counterpart ($\mathrm{L}_{\bot}$ = $\sqrt{\mathrm{L}_{x}^{2}+\mathrm{L}_{y}^{2}}$), the authors were able to select a better defined sample of Helmi stream star candidates.
We build on the selection criteria from \citet{Koppelman2019} and define our Helmi stream sample by including stars from our parent population that satisfy the following selection criteria:
\begin{itemize}
\item 0.75 < $\mathrm{L}_{z}$ < 1.7 ($\times$10$^{3}$ kpc kms$^{-1}$),
\item 1.6 < $\mathrm{L}_{\bot}$ < 3.2 ($\times$10$^{3}$ kpc kms$^{-1}$).
\end{itemize}

Our final sample is comprised of 85 Helmi stream stars members.

\subsubsection{Aleph}
Aleph is a newly discovered substructure presented in a detailed study of the Galactic stellar halo by \cite{Naidu2020} on the basis of the H3 survey \citep{Conroy2019}. It was initially identified as a sequence below the high $\alpha$-disc in the $\alpha$-Fe plane. It is comprised by stars on very circular prograde orbits. For this paper, we follow the method described in \citet{Naidu2020} and define Aleph star candidates as any star in our parent sample which satisfies the following selection criteria:
\begin{itemize}
\item 175 < $\mathrm{V}_{\phi}$ < 300 (kms$^{-1}$),
\item  |V$_{R}$| < 75  (kms$^{-1}$),
\item $[\mathrm{Fe/H}]$ > --0.8,
\item $[\mathrm{Mg/Fe}]$ < 0.27,
\end{itemize}
where $\mathrm{V}_{\phi}$ and V$_{R}$ are the azimuthal and radial components of the velocity vector (in Galactocentric cylindrical coordinates), and we use Mg as our $\alpha$ tracer element. The selection criteria yield a sample comprised of 128,578 stars. We find that the initial selection criteria determine a preliminary Aleph sample that is dominated by \textit{in situ} disc stars, likely obtained due to the prograde nature of the velocity cuts employed as well as the chemical cuts. Thus, in order to remove disc contamination and select $true$ Aleph star members, we employ two further cuts in vertical height above the plane (namely, |$z$| > 3 kpc) and in vertical action (i.e., 170 < J$_{z}$ < 210 kpc kms$^{-1}$), which are motivated by the distribution of Aleph in these coordinates (see Section 3.2.2 from \citealt{Naidu2020}). We also note that the vertical height cut was employed in order to mimic the H3 survey selection function. After including these two further cuts, we obtain a final sample of Aleph stars comprised of 28 star members.

\subsubsection{LMS-1}
LMS-1 is a newly identified substructure discovered by \citet{Yuan2020}. It is characterised by metal poor stars that form an overdensity at the foot of the omnipresent GES in the E-L$_{z}$ plane. This substructure was later also studied by \citet[][]{Naidu2020}, who referred to it as Wukong. We identify stars belonging to this substructure adopting a similar selection as \citet{Naidu2020}, however adopting different orbital energy criteria to adjust for the fact that we adopt the \citet{McMillan2017} Galactic potential (see Fig 23 from Appendix B in that study). Stars from our parent sample were deemed LMS-1 members if they satisfied the following selection criteria:
\begin{itemize}
\item 0.2 < $\mathrm{L}_{z}$ < 1 ($\times$10$^{3}$ kpc kms$^{-1}$),
\item  --1.7 < E < --1.2 ($\times$10$^{5}$ km$^{-2}$s$^{-2}$),
\item $[\mathrm{Fe/H}]$ < --1.45,
\item 0.4 < $e$ < 0.7,
\item |$z$| > 3 (kpc).
\end{itemize}
We note that the $e$ and $z$ cuts were added to the selection criteria listed by \citet{Naidu2020}. This is because we conjectured that instead of eliminating GES star members from our selection (as \citet[][]{Naidu2020} do), it is more natural in principle to find additional criteria that distinguishes these two overlapping substructures. Thus, we select stars on less eccentric orbits than those belonging to GES, but still more eccentric than most of the Galactic disc (i.e., 0.4 < $e$ < 0.7). Furthermore, in order to ensure we are observing stars at the same distances above the Galactic plane as in \citet{Naidu2020} (defined by the selection function of the H3 survey), we add a vertical height cut of |$z$| > 3 (kpc). Our selection identifies 31 stars belonging to the LMS-1 substructure.

\subsubsection{Nyx}
Nyx has recently been put forward by \cite{Necib2020}, who identified a stellar stream in the solar neighbourhood, that they suggest to be the remnant of an accreted dwarf galaxy \citep{Necib2020}. Similar to Aleph, it is characterised by stars on very prograde orbits, at relatively small mid-plane distances (|Z| < 2 kpc) and close to the solar neighbourhood (i.e., |Y| < 2 kpc and |X| < 3 kpc). The Nyx structure is also particularly metal-rich (i.e., [Fe/H] $\sim$ --0.5). Based on the selection criteria used in \citet{Necib2020}, we select Nyx star candidates employing the following selection criteria:
\begin{itemize}
\item 110 < V$_{r}$ < 205 (kms$^{-1}$),
\item 90 < V$_{\phi}$ < 195 (kms$^{-1}$),
\item |X| < 3 (kpc), |Y| < 2 (kpc), |Z| < 2 (kpc).
\end{itemize}
The above selection criteria yield a sample comprising of 589 Nyx stars.

\subsubsection{Icarus}
\label{sec_icarus}
Icarus is a substructure identified in the solar vicinity, comprised by stars that are significantly metal-poor ([Fe/H] $\sim$ --1.45) with circular (disc-like) orbits \citep{Refiorentin2021}. In this work, we select Icarus star members using the mean values reported by those authors and adopting a two sigma uncertainty cut around the mean. The selection used is listed as follows:
\begin{itemize}
    \item $\mathrm{[Fe/H]}$ < --1.05,
    \item $\mathrm{[Mg/Fe]}$ < 0.2,
    \item 1.54 < $\mathrm{L}_{z}$ < 2.21 ($\times$10$^{3}$ kpc kms$^{-1}$),
    \item L$_{\bot}$ < 450 (kpc kms$^{-1}$),
    \item $e$ < 0.2,
    \item $z_{\mathrm{max}}$ < 1.5 (kpc).
\end{itemize}
These selection criteria yield an Icarus sample comprised of one star. As we have only been able to identify one star associated with this substructure, we remove it from the main body of this work and focus on discerning why our selection method only identifies 1 star in Appendix~\ref{appen_icarus}. Furthermore, we combine the one Icarus star found in APOGEE DR17 with 41 stars found by \citet[][]{Refiorentin2021} in APOGEE DR16, in order to study the nature of this substructure in further detail. Our results are discussed in Appendix~\ref{appen_icarus}.  

\subsubsection{Pontus}

 Pontus is a halo substructure  recently proposed by \cite{Malhan2022}, on the basis of an analysis of {\it Gaia} EDR3 data for a large sample of Galactic globular clusters and stellar streams.  These authors identified a large number of groupings in action space, associated with known substructures.  \cite{Malhan2022} propose the existence of a previously unknown susbtructure they call {\it Pontus}, characterised by retrograde orbits and intermediate orbital energy.  Pontus is located just below {\it Gaia}-Enceladus/Sausage in the E-Lz plane, but displays less radial orbits (Pontus has an average radial action of J$_{R}$$\sim$500 kpc kms$^{-1}$, whereas \textit{Gaia}-Enceladus/Sausage displays a mean value of J$_{R}$$\sim$1,250 kpc kms$^{-1}$). In this work, we utilise the values listed in Section 4.6 from \citet[][]{Malhan2022} to identify Pontus candidate members in our sample. We note that because both that study and ours use the \citet[][]{McMillan2017} potential to compute the IoM, the orbital energy values will be on the same scale. Our selection criteria for Pontus are the following:

\begin{itemize}
    \item  --1.72 < E < --1.56 ($\times$10$^{5}$ km$^{-2}$s$^{-2}$),
    \item --470 < $\mathrm{L}_{z}$ < 5 ($\times$10$^{3}$ kpc kms$^{-1}$),
    \item 245 < $\mathrm{J}_{R}$ < 725 (kpc kms$^{-1}$),
    \item 115 < $\mathrm{J}_{z}$ < 545 (kpc kms$^{-1}$),
    \item 390 < $\mathrm{L}_{\perp}$ < 865 (kpc kms$^{-1}$),
    \item 0.5 < $e$ < 0.8,
    \item 1 < $R_{\mathrm{peri}}$ < 3 (kpc),
    \item 8 < $R_{\mathrm{apo}}$ < 13 (kpc),
    \item $\mathrm{[Fe/H]}$ < --1.3,
\end{itemize}
 
\noindent where $R_{\mathrm{peri}}$ and $R_{\mathrm{apo}}$ are the perigalacticon and apogalacticon radii, respectively. Using these selection criteria, we identify two Pontus candidate members in our parent sample. As two stars comprise a sample too small to perform any statistical comparison, we refrain from comparing the Pontus stars in the main body of this work. Instead, we display and discuss the chemistry of these two Pontus stars in Appendix~\ref{appen_pontus}, for completeness.

\subsubsection{Cetus}

As a closing remark, we note that we attempted to identify candidate members belonging to the Cetus \citep{Newberg2009} stream. Using the selection criteria defined in Table 3 from \citet[][]{Malhan2022}, we found no stars associated with this halo substructure that satisfied the selection criteria of our parent sample. This is likely due to a combination of two factors: {\it (i)} Cetus is a diffuse stream orbiting at large heliocentric distances ($d_{\odot}$$\gtrsim$30 kpc, \citealp[][]{Newberg2009}), which APOGEE does not cover well; {\it (ii)} it occupies a region of the sky around the southern polar cap, where APOGEE does not have many field pointings, at approximately $l$$\sim$143$^{\circ}$ and $b$$\sim$--70$^{\circ}$ \citep[][]{Newberg2009}.

\setlength{\tabcolsep}{25pt}
\begin{table*}
\centering
\begin{tabular}{ |p{3.25cm}|p{9.25cm}|p{0.3cm}}
\hline
 Name & Selection criteria & N$_{\mathrm{stars}}$\\
\hline
\hline
Heracles & $e$ > 0.6; --2.6 < E < --2 ($\times$10$^{5}$ km$^{2}$s$^{-2}$); $[\mathrm{Al/Fe}]$ < --0.07 $\&$ [Mg/Mn] $\geqslant$ 0.25; $[\mathrm{Al/Fe}]$ $\geqslant$ --0.07; [Mg/Mn] $\geqslant$ 4.25$\times$[Al/Fe] + 0.5475; [Fe/H] > --1.7 & 300 \\
\hline
\textit{Gaia}-Enceladus/Sausage &  |L$_{z}$| < 0.5 ($\times$10$^{3}$ kpc km s$^{-1}$) ; --1.6 < E < --1.1 ($\times$10$^{5}$ km$^{2}$s$^{-2}$) & 2353  \\
\hline
Sagittarius & |$\beta_{\mathrm{GC}}$|
< 30 ($^{\circ}$); 1.8 < L$_{z,\mathrm{Sgr}}$ < 14 ($\times$10$^{3}$ kpc kms$^{-1}$); --150 < $\mathrm{V}_{z,\mathrm{Sgr}}$ < 80 (kms$^{-1}$);  X$_{\mathrm{Sgr}}$ > 0 (kpc) or X$_{\mathrm{Sgr}}$ <--15 0 (kpc); Y$_{\mathrm{Sgr}}$ > --5 (kpc) or Y$_{\mathrm{Sgr}}$ < --20 (kpc); Z$_{\mathrm{Sgr}}$ > --10 (kpc); pm$_{\alpha}$ > --4 (mas); $d_{\odot}$ > 10 (kpc) & 266\\
\hline
Helmi stream & 0.75 < $\mathrm{L}_{z}$ < 1.7 ($\times$10$^{3}$ kpc kms$^{-1}$); 1.6 < $\mathrm{L}_{\bot}$ < 3.2 ($\times$10$^{3}$ kpc kms$^{-1}$)& 85\\
\hline
(\textit{M19}) Sequoia  & E > --1.5 ($\times$10$^{5}$ km$^{2}$s$^{-2}$); J$_{\phi}$/J$_{\mathrm{tot}}$<--0.5; J$_{\mathrm{(J}_{z}-\mathrm{J}_{R})}$/J$_{\mathrm{tot}}$<0.1 & 116\\
\hline
(\textit{K19}) Sequoia & 0.4<$\eta$<0.65; 
--1.35<E<--1 ($\times$10$^{5}$ km$^{2}$s$^{-2}$); L$_{z}$<0 (kpc kms$^{-1}$) &45\\
\hline
(\textit{N20}) Sequoia & $\eta$>0.15; E>--1.6 ($\times$10$^{5}$ km$^{2}$s$^{-2}$); L$_{z}$<--0.7 ($\times$10$^{3}$ kpc kms$^{-1}$); --2 < [Fe/H] < --1.6 & 72\\
\hline
Thamnos & --1.8 < E < --1.6 ($\times$10$^{5}$ km$^{2}$s$^{-2}$); L$_{z}$ < 0 (kpc kms$^{-1}$); $e$ < 0.7 & 121\\
\hline
Aleph & 175 < $\mathrm{V}_{\phi}$ < 300 (kms$^{-1}$); |V$_{R}$| < 75  (kms$^{-1}$); Fe/H > --0.8; Mg/Fe < 0.27; |z| > 3 (kpc); 170 < J$_{z}$ < 210 (kpc kms$^{-1}$) & 28\\
\hline
LMS-1 &  0.2 < $\mathrm{L}_{z}$ < 1 ($\times$10$^{3}$ kpc kms$^{-1}$);  --1.7 < E < --1.2 ($\times$10$^{5}$ km$^{-2}$s$^{-2}$); [Fe/H] < --1.45; 0.4 < $e$ < 0.7; |$z$| > 3 (kpc)& 31\\
\hline
Arjuna & $\eta$ > 0.15; E > --1.6 ($\times$10$^{5}$ km$^{2}$s$^{-2}$); L$_{z}$ < --0.7 ($\times$10$^{3}$ kpc kms$^{-1}$); --1.6 < [Fe/H] & 132\\
\hline
I'itoi & $\eta$ > 0.15; E > --1.6 ($\times$10$^{5}$ km$^{2}$s$^{-2}$); L$_{z}$ < --0.7 ($\times$10$^{3}$ kpc kms$^{-1}$); [Fe/H] < --2 & 25\\ 
\hline
Nyx &  110 < V$_{R}$ < 205 (kms$^{-1}$); 90 V$_{\phi}$ < 195 (kms$^{-1}$); |X| < 3 (kpc), |Y| < 2 (kpc), |Z| < 2 (kpc)& 589\\
\hline
Icarus & $\mathrm{[Fe/H]}$ < --1.45; $\mathrm{[Mg/Fe]}$ < 0.2; 1.54 < $\mathrm{L}_{z}$ < 2.21 ($\times$10$^{3}$ kpc kms$^{-1}$); L$_{\bot}$ < 450 (kpc kms$^{-1}$); $e$ < 0.2; $z_{\mathrm{max}}$ < 1.5 & 1\\
\hline
Pontus &  --1.72 < E < --1.56 ($\times$10$^{5}$ km$^{-2}$s$^{-2}$); --470 < $\mathrm{L}_{z}$ < 5 ($\times$10$^{3}$ kpc kms$^{-1}$); 245 < $\mathrm{J}_{R}$ < 725 (kpc kms$^{-1}$); 115 < $\mathrm{J}_{z}$ < 545 (kpc kms$^{-1}$); 390 < $\mathrm{L}_{\perp}$ < 865 (kpc kms$^{-1}$); 0.5 < $e$ < 0.8; 1 < $R_{\mathrm{peri}}$ < 3 (kpc); 8 < $R_{\mathrm{apo}}$ < 13 (kpc); [Fe/H] < --1.3 & 2\\
\hline
\hline
\end{tabular}
\caption{Summary of the selection criteria employed to identify the halo substructures, and the number of stars obtained for each sample. For a more thorough description of the selection criteria used in this work, see Section~\ref{method}. We note that all the orbital parameter values used are obtained adopting a \citet[][]{McMillan2017} potential.}
\label{tab:selection}
\end{table*}

\section{Kinematic properties}
\label{kinematics}

\begin{figure*}
\includegraphics[width=\textwidth]{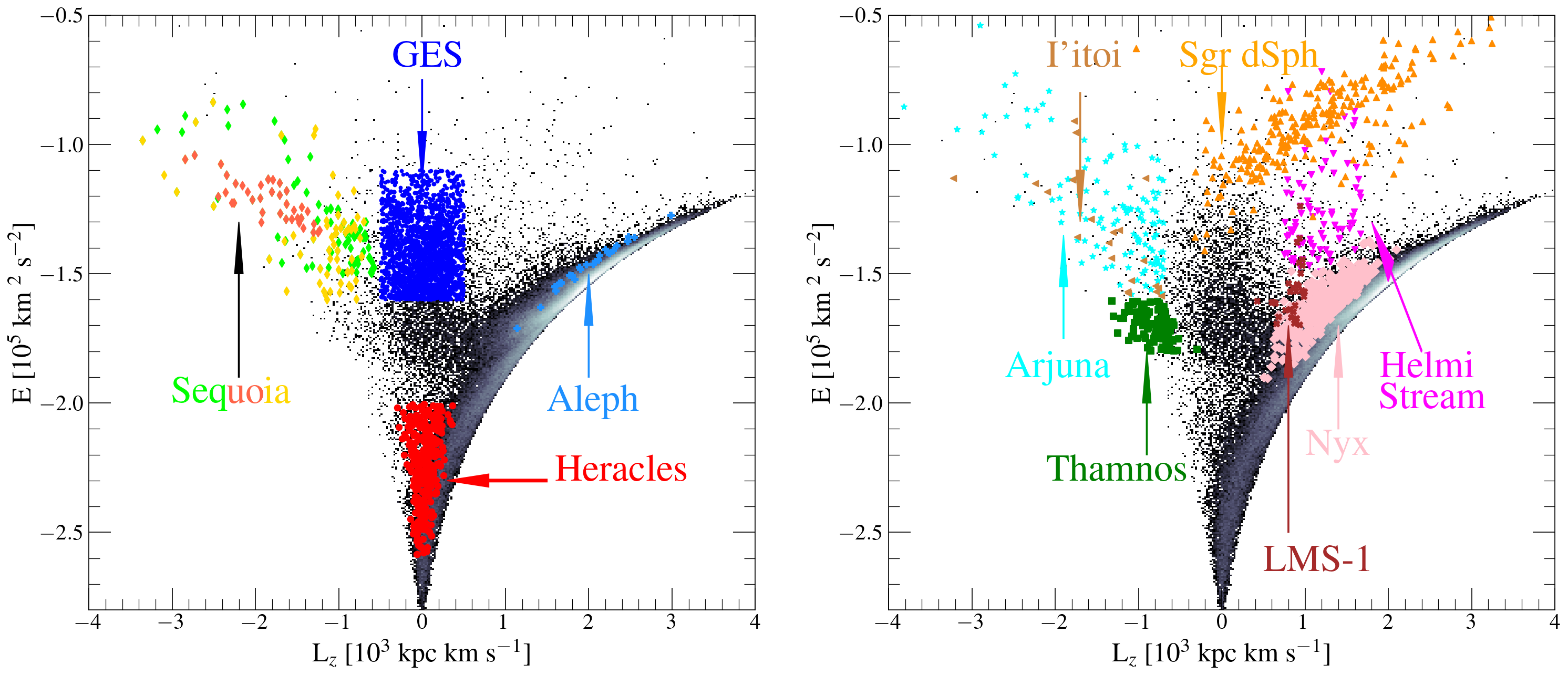}
\caption{Distribution of the identified halo substructures in the orbital energy (E) versus angular momentum w.r.t. the Galactic disc (L$_{z}$) plane. The parent sample is plotted as a 2D histogram, where white/black signifies high/low density regions. The coloured markers illustrate the different structures studied in this work, as denoted by the arrows (we do not display Pontus(Icarus) as we only identify 2(1) stars, respectively). The figure is split into two panels for clarity.}
    \label{elz}
\end{figure*}

In this Section, we present the resulting distributions of the identified halo substructures in the orbital energy (E) versus the azimuthal component of the angular momentum (L$_{z}$) plane in Fig~\ref{elz}. The parent sample is illustrated as a density distribution and the halo substructures are shown using the same colour markers as in Fig~\ref{fig:hr}. By construction, each substructure occupies a different locus in this plane. However, we do notice some small overlap between some of the substructures (for example, between GES and Sequoia, or GES and LMS-1), given their similar selection criteria. More specifically, we find that  Heracles dominates the low energy region (E < --2$\times$10$^{5}$ km$^{-2}$s$^{-2}$), whereas all the other substructures are characterised by higher energies. As shown before (\citealp[e.g.,][]{Koppelman_thamnos,Horta2021}), we find that GES occupies a locus at low L$_{z}$ and relatively high E, which corresponds to very radial/eccentric orbits. We find the retrograde region (i.e., L$_{z}$ < 0 $\times$10$^{3}$ kpc kms$^{-1}$) to be dominated by Thamnos at intermediate energies (E $\sim$ --1.7$\times$10$^{5}$ km$^{-2}$s$^{-2}$), and by Sequoia, Arjuna and I'itoi at higher energies (E > --1.4$\times$10$^{5}$ km$^{-2}$s$^{-2}$); on the other hand, in the prograde region (L$_\mathrm{z}$ > 0 $\times$10$^{3}$ kpc kms$^{-1}$), we find the LMS-1 and Helmi stream structures, which occupy a locus at approximately E $\sim$ --1.5$\times$10$^{5}$ km$^{-2}$s$^{-2}$ and L$_\mathrm{z}$ $\sim$ 500 kpc kms$^{-1}$, and E $\sim$ --1.4$\times$10$^{5}$ km$^{-2}$s$^{-2}$ and L$_\mathrm{z}$ $\sim$ 1,000 kpc kms$^{-1}$, respectively. Furthemore, the loci occupied by the Aleph and Nyx substructures closely mimic the region defined by disc orbits. Lastly, sitting above all other structures we find the Sagittarius dSph, which occupies a position at high energies and spans a range of angular momentum between 0 < L$_\mathrm{z}$ < 2,000 kpc kms$^{-1}$.

\section{Chemical Compositions} 
\label{chemistry}
\begin{figure}
\includegraphics[width=\columnwidth]{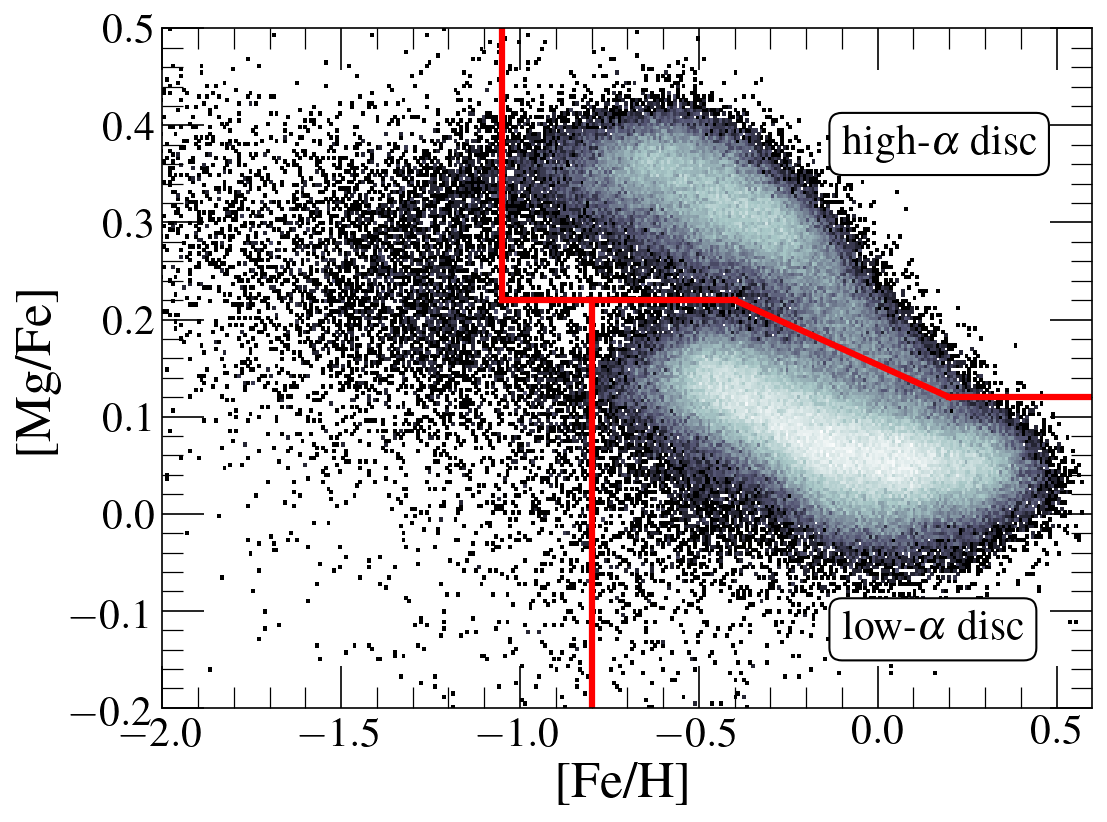}
\caption{Parent sample in the Mg-Fe plane. The solid red lines indicate cut employed to select the \textit{in situ} high- and low-$\alpha$ samples that we use in our $\chi^{2}$ analysis, where the diagonal dividing line is defined as [Mg/Fe] > --0.167[Fe/H] + 0.15.}
    \label{discs}
\end{figure}

In this Section we turn our attention to the main focus of this work: a chemical abundance study of substructures in the stellar halo of the Milky Way. In this Section, we seek to first characterise these substructures qualitatively in multiple chemical abundance planes that probe different nucleosynthetic pathways. In Section~\ref{sec_abundances} we then compare mean chemical compositions across various substructures in a quantitative fashion. Our aim is to utilise the chemistry to further unravel the nature and properties of these halo substructures, and in turn place constraints on their star formation and chemical enrichment histories. We also aim to compare their chemical properties with those from {\it in situ} populations (see Fig~\ref{discs} for how we determine \textit{in situ} populations). By studying the halo substructures using different elemental species we aim to develop an understanding of their chemical evolution contributed by different nucleostynthetic pathways, contributed either by core-collapse supernovae (SNII), supernovae type Ia (SNIa), and/or Asymptotic Giant Stars (AGBs). Furthermore, as our method for identifying these substructures relies mainly on phase space and orbital information, our analysis is not affected by chemical composition biases (except for the case of particular elements in the Heracles, Aleph, LMS-1, Arjuna, I'itoi, and (H3) Sequoia sample).

Our results are presented as follows. In Section \ref{sec_alphas} we present the distribution of the halo substructures in the $\alpha$-Fe plane, using Mg as our $\alpha$ element tracer. In Section \ref{sec_ironpeak}, we show the distribution of these substructures in the Ni-Fe abundance plane, which provides insight into the chemical evolution of the iron-peak elements. Section \ref{sec_oddZ} displays the distribution of the halo substructures in an odd-Z-Fe plane, where we use Al as our tracer element. Furthermore, we also show the C and N abundance distributions in Section \ref{sec_cn}, the Ce abundances (namely, an $s$-process neutron capture element) in Section~\ref{sec_neutron_capture}, and the distribution of the halo substructures in the [Mg/Mn]-[Al/Fe] plane in Section \ref{sec_other_chem} \footnote{For each chemical plane, we impose a further set of cuts of \texttt{X$_{-}$FE$_{-}$FLAG=0} and [X/Fe]$_{\mathrm{error}}$<0.15 to ensure there are no unforeseen issues when determining the abundances for these halo substructures in \texttt{ASPCAP}.}. This last chemical composition plane is interesting to study as it has recently been shown to help distinguish stellar populations from "\textit{in situ}" and accreted origins (\citealp[e.g.,][]{Hawkins2015,Das2020,Horta2021}). Upon studying the distribution of the substructures in different chemical composition planes, we finalise our chemical composition study in Section \ref{sec_abundances} by performing a quantitative comparison between the substructures studied in this work for all the (reliable) elemental abundances available in APOGEE. For simplicity, we henceforth refer to the [X/Fe]-[Fe/H] plane as just the X-Fe plane.

As mentioned in Section~\ref{data}, we exclude the Pontus and Icarus substructures from our quantitative chemical comparisons as the number of candidate members of these substructures in the APOGEE catalogue is too small.  The properties of Icarus and Pontus are briefly discussed in Appendices~\ref{appen_icarus} and \ref{appen_pontus}, respectively.

\subsection{$\alpha$-elements}
\label{sec_alphas}
We first turn our attention to the distribution of the substructures in the $\alpha$-Fe plane. This is possibly the most interesting chemical composition plane to study, as it can provide great insight into the star formation history and chemical enrichment processes of each subtructure (\citealp[e.g.,][]{Matteucci1986,Wheeler1989,McWilliam1997,Tolstoy2009,Nissen2010,Bensby2014}). Specifically, we seek to identify the presence of the $\alpha$-Fe knee. For this work, we resort to magnesium as our primary $\alpha$ element, as this has been shown to be the most reliable $\alpha$ element in previous APOGEE data releases \citep[e.g., DR16;][]{Jonsson2020}. For the distributions of the remaining $\alpha$ elements determined by ASPCAP (namely, O, Si, Ca, S, and Ti), we refer the reader to Fig~\ref{sife}-Fig.~\ref{tife} in Appendix~\ref{app_alphas}.

\begin{figure*}
\includegraphics[width=\textwidth]{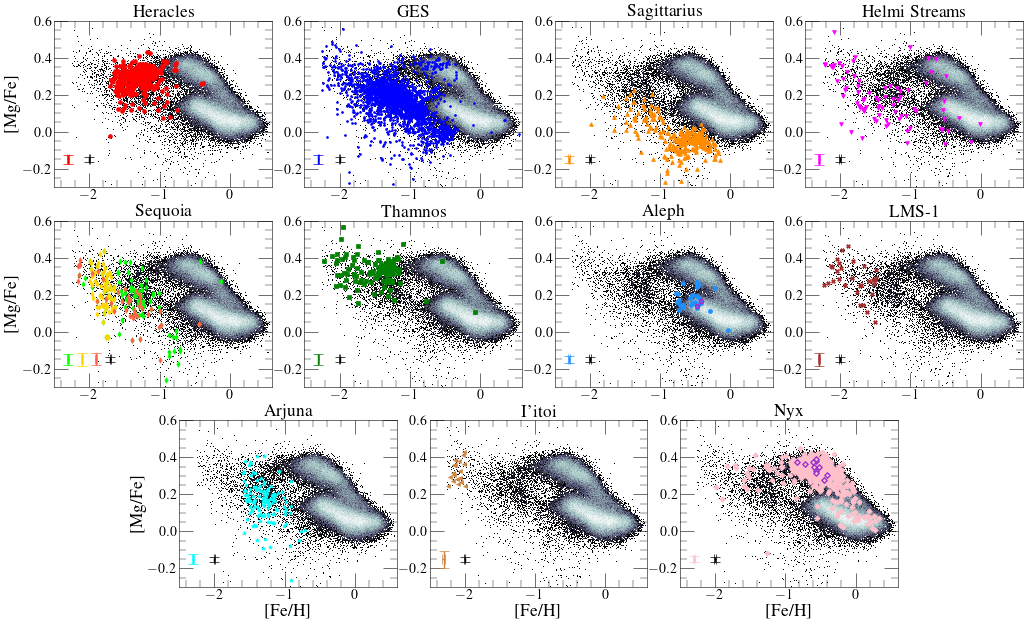}
\caption{The resulting parent sample and identified structures from Fig~\ref{elz} in the Mg-Fe plane. The mean uncertainties in the abundance measurements for halo substructures (colour) and the parent sample (black) are shown in the bottom left corner. Colour coding and marker styles are the same as Fig~\ref{elz}. For the Aleph and Nyx substructures, we also highlight with purple edges stars from our APOGEE DR17 data that are also contained in the Aleph and Nyx samples from the \citet[][]{Naidu2020} and \citet[][]{Necib2020} samples, respectively.}
    \label{mgfes}
\end{figure*}

Figure \ref{mgfes} shows the distribution of each substructure in the Mg-Fe plane (coloured markers) compared to the parent sample (2D density histogram). We find that all the substructures --except for Aleph and Nyx-- occupy a locus in this plane which is typical of low mass satellite galaxies and accreted populations of the Milky Way (\citealp[e.g.,][]{Tolstoy2009,Hayes2018,Mackereth2019b}), characterised by low metallicity and lower [Mg/Fe] at fixed [Fe/H] than {\it in situ} disc populations. Moreover, we find that different substructures display distinct [Mg/Fe] values, implying certain differences despite their overlap in [Fe/H]. However, we also note that at the lowest metallicities ([Fe/H] < --1.8), the overlap between different halo substructures increases. 

Next we discuss the distribution  on the Mg-Fe plane of stars in our sample belonging to each halo substructure.

\begin{itemize}

\item $Sagittarius$: As shown in previous studies ( \citealp[e.g.,][]{Monaco2005,Sbordone2007,Carreta2010,Mcwilliam2013,Hasselquist2017,Hasselquist2019,Hayes2020,Minelli2021}), the stellar populations of the Sgr dSph galaxy are characterised by substantially lower [Mg/Fe] than even the low-$\alpha$ disc at the same [Fe/H] and traces a tail towards higher [Mg/Fe] with decreasing [Fe/H]. Conversely, at the higher [Fe/H] end, we find that Sgr stops decreasing in [Mg/Fe] and shows an upside-down "\textit{knee}", likely caused by a burst in SF at late times that may be due to its interaction with the Milky Way (\citealp[e.g.,][]{Hasselquist2017,Hasselquist2021}). Such a burst of star formation intensifies the incidence of SNe II, with a consequent boost in the ISM enrichment in its nucleosynthetic by-products, such as Mg, over a short timescale ($\sim10^7$~yr).  Because Fe is produced predominantly by SN Ia over a considerably longer timescale ($\sim10^8-10^9$~yr), Fe enrichment lags behind, causing a sudden increase in [Mg/Fe].

\item $Heracles$: The [Mg/Fe] abundances of the Heracles structure occupy a higher locus than that of other halo substructures of similar metallicity, with the exception of Thamnos. As discussed in \citet{Horta2021}, the distribution of Heracles in the $\alpha$-Fe plane is peculiar, differing from that of GES and other systems by the absence of the above-mentioned $\alpha$-knee. As conjectured in \citet{Horta2021}, we suggest that this distribution results from an early quenching of star formation, taking place before SN~Ia could contribute substantially to the enrichment of the interstellar medium (ISM). 

\item $Gaia$-$Enceladus/Sausage$: Taking into account the small yet clear contamination from the high-$\alpha$ disc at higher [Fe/H] (see Fig~\ref{mgfe_GES} for details), the distribution of GES dominates the metal-poor and $\alpha$-poor populations of the Mg-Fe plane (as pointed out in previous studies \citealp[e.g.,][]{Helmi2018,Hayes2018,Mackereth2019b}), making it easily distinguishable from the high-/low-$\alpha$ discs. We find that GES reaches almost solar metallicities, displaying the standard distribution in the Mg-Fe plane with a change of slope --the so-called ``$\alpha$-knee''-- occurring at approximately [Fe/H]$\sim$--1.2 \citep{Mackereth2019b}. The metallicity of the ``knee'' has long been thought to be an indicator of the mass of the system \cite[e.g.,][]{Tolstoy2009}, and indeed it occurs at [Fe/H]$>$--1 for both the high- and low-$\alpha$ discs.  As a result, GES stars in the $shin$ part of the  $\alpha$-$knee$ are characterised by lower [Mg/Fe] at constant metallicity than disc stars. Interestingly, even at the plateau ([Fe/H]$<$--1.2), GES seems to present lower [Mg/Fe] than the high-$\alpha$ disc, although this needs to be better quantified. Furthermore, we note the presence of a minor population of [Mg/Fe] < 0 stars at --1.8 < [Fe/H] < --1.2, which could be contamination from a separate halo substructure, possibly even a satellite of the GES progenitor (see Fig~\ref{mgfe_GES} for details).

Based purely on the distribution of its stellar populations on the Mg-Fe plane, one would expect the progenitor of GES to be a relatively massive system \citep[see][ for details]{Mackereth2019b}. The fact that the distribution of its stellar populations in the Mg-Fe plane covers a wide range in metallicity, bracketing the knee and extending from [Fe/H]<--2 all the way to [Fe/H]$\sim$--0.5 suggests a substantially prolonged history of star formation.

\item $Sequoia$: The distribution of the \textit{M19}, \textit{K19}, and \textit{N20} selected samples (selected on the criteria described in \citet{Myeong2019}, \citet{Koppelman_thamnos}, and \citet{Naidu2020}, respectively) occupy similar [Mg/Fe] values ([Mg/Fe]$\sim$0.1) at lower metallicities ([Fe/H] $\lesssim$--1). More specifically, we find that all three Sequoia samples occupy a similar position in the Mg-Fe plane, one that overlaps with that of GES and other substructures at similar [Fe/H] values. Along those lines, we find that the \textit{K19} and \textit{M19} Sequoia samples seem to connect with the \textit{N20} Sequoia sample, where the \textit{N20} sample comprises the lower metallicity component of the \textit{K19}/\textit{M19} samples. We examine in more detail the distribution of the Sequoia stars in the $\alpha$-Fe plane in Section~\ref{sec:Sequoia}.

\begin{figure}
\includegraphics[width=\columnwidth]{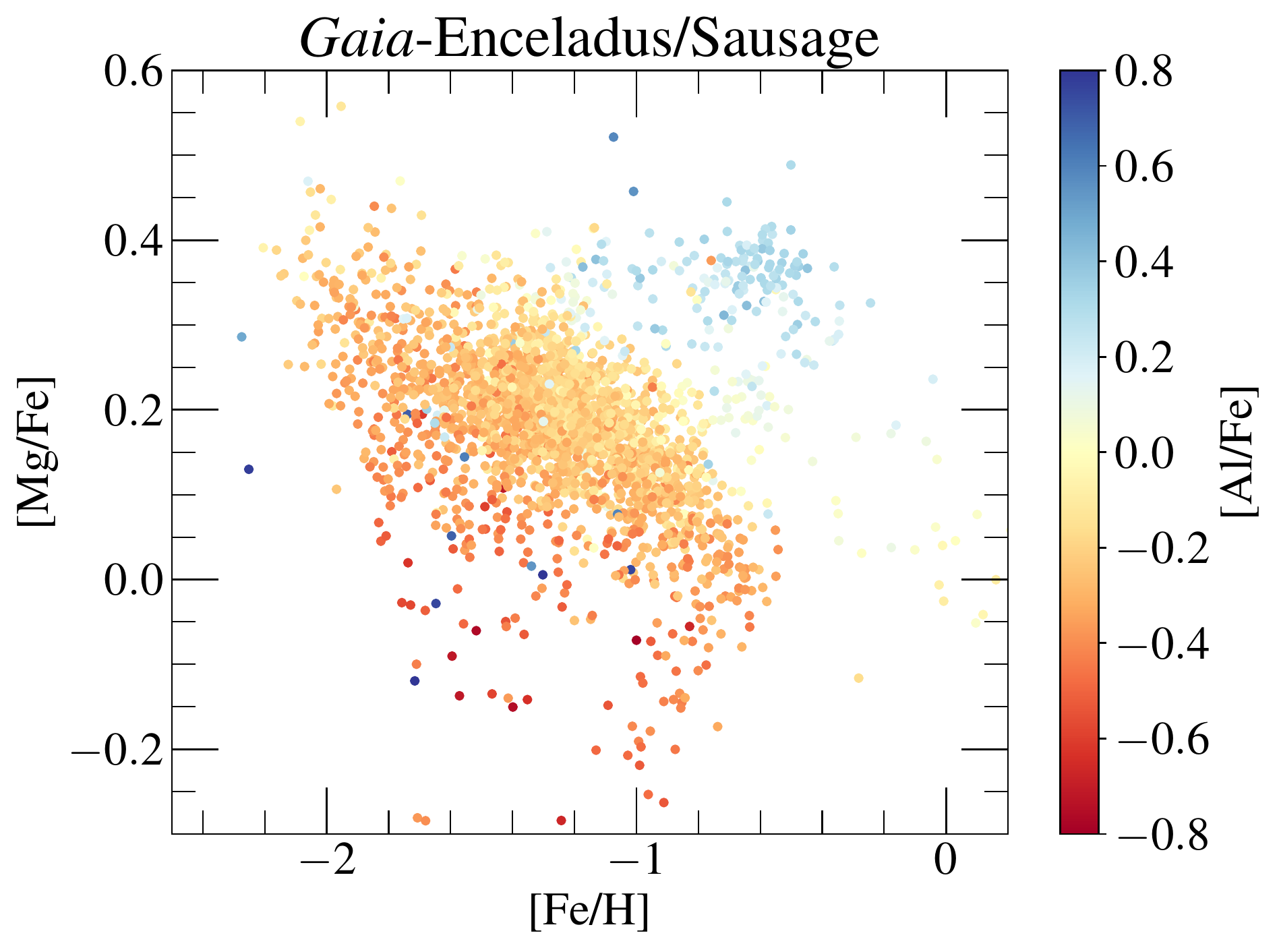}
\caption{\textit{Gaia}-Enceladus/Sausage (GES) sample in the Mg-Fe plane, colour coded by the [Al/Fe] abundance values. The low [Al/Fe] stars are true GES star candidates, which display the expected low [Al/Fe] abundances observed in accreted populations (see Section~\ref{sec_oddZ} for details). Conversely, the high [Al/Fe] stars are clear contamination from the high-$\alpha$ disc, likely associated with disc stars on very eccentric and high energy orbits (\citealp{Bonaca2017,Belokurov2020}). A striking feature becomes apparent in this plane: at --1.8 < [Fe/H] < --0.8, there is a population of very [Mg/Fe]-poor stars (i.e., [Mg/Fe] below $\sim$0), that could possibly be contamination from a separate halo substructure (although these could also be due to unforeseen problems in their abundance determination).}
    \label{mgfe_GES}
\end{figure}

\item $Helmi$ $stream$: Despite the lower number of members associated to this substructure, its chemical composition in the Mg-Fe plane appears to follow a single sequence, with slope and scatter similar to that of GES, with the addition of a small number of likely contaminants overlapping with the low- and high-alpha disk. Furthermore, it is confined to low metallicities ([Fe/H]<--1.2) and intermediate magnesium values ([Mg/Fe]$\sim$0.2). However, we do note that the stars identified for this substructure appear to be scattered across a wide range of [Fe/H] values. Specifically, the Helmi stream occupies a locus that overlaps with the GES for fixed [Fe/H]. In Fig~\ref{fig:knees_subs} we show that the best-fitting piece-wise linear model prefers a knee that is "inverted" similar to, although less extreme than, the case of Sgr dSph. We discuss this in more detail in Section~\ref{sec:helmi}.

\item $Thamnos$: The magnesium abundances of Thamnos suggest that this structure is clearly different from other substructures in the retrograde halo (namely, Sequoia, Arjuna, and I'itoi). It presents a much higher mean [Mg/Fe] for fixed metallicity than the other retrograde substructures, and appears to follow the Mg-Fe relation of the high-$\alpha$ disc. We find that Thamnos presents no $\alpha$-knee feature, and occupies a similar locus in the Mg-Fe plane to that of Heracles. The distribution of this substructure in this plane with the absence of an $\alpha$-Fe "knee" suggests that this substructure likely quenched star formation before the onset of SN~Ia. 

\item $Aleph$: By construction, Aleph occupies a locus in the Mg-Fe plane that overlaps with the metal-poor component of the low-$\alpha$ disc. Given the distribution of this substructure in this chemical composition plane, and its very disc-like orbits, we suggest it is possible that Aleph is constituted by warped/flared disc populations. Because the data upon which our work and that by \cite{Naidu2020} are based come from different surveys, it is important however to ascertain that selection function differences between APOGEE and H3 are not responsible for our samples to have very different properties, even though they are selected adopting the same kinematic criteria.  In an attempt to rule out that hypothesis we cross-matched the \citet[][]{Naidu2020} catalogue with that of APOGEE DR17 to look for Aleph stars in common to the two surveys. We find only two such stars\footnote{We find that these two stars have a \texttt{STARFLAG} set with \texttt{PERSIST$_{-}$LOW} and \texttt{BRIGHT$_{-}$NEIGHBOUR}, and thus do not survive our initial parent selection criteria. However, these warnings are not critical, and should not have an effect on their abundance determinations.}, which we highlight in Fig~\ref{mgfes} with purple edges. While this is a very small number, the two stars seem to be representative of the chemical composition of the APOGEE sample of Aleph stars, which is encouraging.

\item $LMS-1$: Our results from Fig~\ref{mgfes} show that the LMS-1 occupies a locus in the Mg-Fe plane which appears to form a single sequence with the GES at the lower metallicity end. Based on its Mg and Fe abundances, and the overlap in kinematic planes of LMS-1 and GES, we suggest it is possible that these two substructures could be linked, where LMS-1 constitutes the more metal-poor component of the GES. We investigate this possible association further in Section~\ref{sec_abundances}.

\item $Arjuna$: This substructure occupies a distribution in the Mg-Fe plane that follows that of the GES. Despite the sample being lower in numbers than that for GES, we still find that across --1.5 < [Fe/H] < --0.8 this halo substructure overlaps in the Mg-Fe plane with that of the GES substructure. Given the strong overlap between these two systems, as well as their proximity in the E-L$_{z}$ plane (see Fig~\ref{elz}), we suggest it is possible that Arjuna could be part of the GES substructure, and further investigate this possible association in Section~\ref{sec_abundances}.

\item $I'itoi$: Despite the small sample size, we find I'itoi presents high [Mg/Fe] and low [Fe/H] values (the latter by construction), and occupies a locus in the Mg-Fe plane that appears to follow a single sequence with the Sequoia (all three samples) and the GES sample. 

\item $Nyx$: The position of this substructure in the Mg-Fe strongly overlaps with that of the high-$\alpha$ disc. Given this result, and the disc-like orbits of stars comprising this substructure, we conjecture that Nyx is constituted by high-$\alpha$ disc populations, and further investigate this association in Section~\ref{sec_abundances}. Furthermore, in a similar fashion as done for the Aleph substructure, we highlight in Fig~\ref{mgfes} with purple edges those stars in APOGEE DR17 that are also contained in the Nyx sample from \citet[][]{Necib2020}, in order to ensure that our results are not biased by the APOGEE selection function. We find that the overlapping stars occupy a locus in this plane that overlaps with the Nyx sample determined in this study, and the high-$\alpha$ disc.

\end{itemize}

\subsection{Iron-peak elements}
\label{sec_ironpeak}

\begin{figure*}
\includegraphics[width=\textwidth]{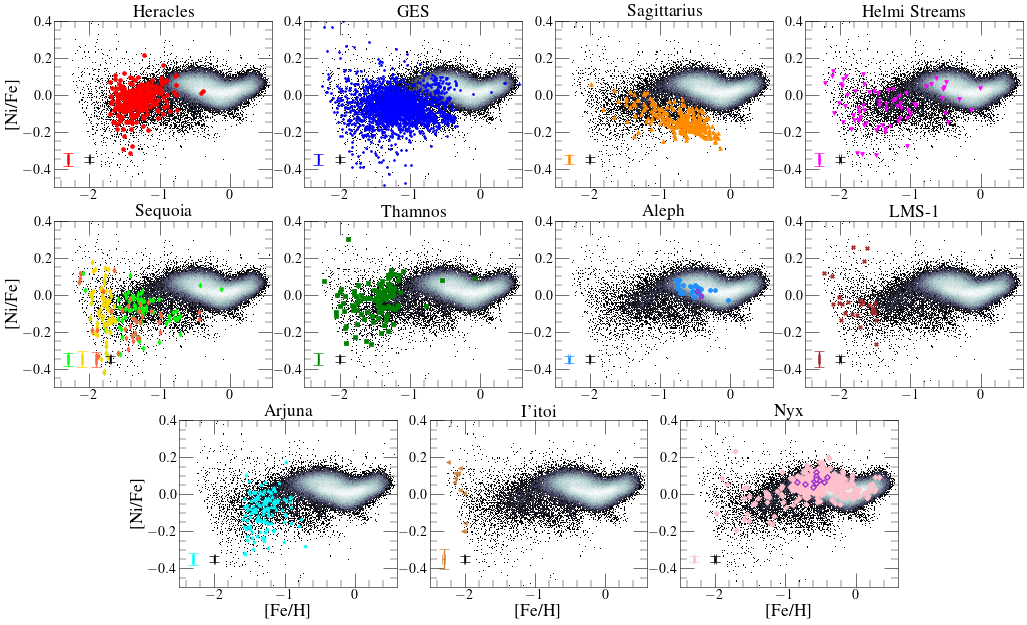}
\caption{The same illustration as in Fig~\ref{mgfes} in the Ni-Fe space. We note that the grid limit of appears clearly in this plane at the lowest [Fe/H] values. }
    \label{nifes}
\end{figure*}

Following our analysis of the various substructures in the Mg-Fe plane, we now focus on studying their distributions in chemical abundance planes that probe nucleosynthetic pathways contributed importantly by Type Ia supernovae. We focus on nickel (Ni), which is the Fe-peak element that is determined the most reliably by ASPCAP, besides Fe itself.  For the distribution of the structures in other iron-peak element planes traced by ASPCAP (e.g., Mn, Co, and Cr), we refer the reader to Fig.~\ref{mnfe}-- Fig.~\ref{crfe} in Appendix~\ref{app_ironpeak}. 

The distributions of the halo substructures in the Ni-Fe plane are shown in Fig.~\ref{nifes}. We find that the distributions of GES, Sgr dSph, the Helmi stream, Arjuna, LMS-1, and the three Sequoia samples occupy a locus in this plane that is characteristic of low mass satellite galaxies and/or accreted populations of the Milky Way, displaying lower [Ni/Fe] abundances than the low- and high-$\alpha$ disc populations (\citealp[e.g.,][Shetrone et al. 2022, in prep.]{Shetrone2003,Mackereth2019b,Horta2021}). In contrast, the data for Heracles and Thamnos display a slight correlation between [Ni/Fe] and [Fe/H], connecting with the high-$\alpha$ disc at  ${\rm [Fe/H]}\sim-1$ (despite the differences of these substructures in the other chemical composition planes with \textit{in situ} disc populations). Conversely, we find that the Aleph and Nyx structures clearly overlap with \textit{in situ} disc populations at higher [Fe/H] values, agreeing with our result for these substructures on the Mg-Fe plane. The distribution of I'itoi shows a spread in [Ni/Fe] for a small range in [Fe/H], that is likely due to observational error at such low metallicities. 

Interpretation of these results depends crucially on an understanding of the sources of nickel enrichment.  Like other Fe-peak elements, nickel is contributed by a combination of SNIa and SNe~II \cite[e.g.,][]{Weinberg2019,Kobayashi2020}.  The disc populations display a bimodal distribution in Figure~\ref{nifes}, which is far less pronounced than in the case of Mg.  This result suggests that the contribution by SNe~II to nickel enrichment may be more important than previously thought (but see below).  It is thus possible that the relatively low [Ni/Fe] observed in MW satellites and halo substructures has the same physical reason as their low [$\alpha$/Fe] ratio, namely, a low star formation rate \cite[e.g.,][]{Hasselquist2021}. This hypothesis can be checked by examining the locus occupied by halo substructures in a chemical plane involving an Fe-peak element with a smaller contribution by SNe~II, such as manganese \cite[e.g.,][]{Kobayashi2020}.  If indeed the [Ni/Fe] depression is caused by a decreased contribution by SNe~II, one would expect [Mn/Fe] to display a different behaviour.  Figure~\ref{mnfe} confirms that expectation, with substructures falling on the same locus as disc populations on the Mn-Fe plane.  

Another possible interpretation of the reduced [Ni/Fe] towards the low metallicity characteristic of halo substructures is a metallicity dependence of nickel yields \citep{Weinberg2021}.  We may need to entertain this hypothesis since, in contrast to the results presented in Figure~\ref{nifes}, no [Ni/Fe] bimodality is present in the solar neighbourhood disc sample studied by \cite{Bensby2014}, which may call into question our conclusion that SNe~II contribute relevantly to nickel enrichment.  It is not clear whether the apparent discrepancy between the data for nickel in \cite{Bensby2014} and this work is due to lower precision in the former, sample differences, or systematics in the APOGEE data.

Given the distribution of the substructures in the Ni-Fe plane, our results suggest that: {\it i)} Sgr dSph, GES, Sequoia (all three samples), and the Helmi stream substructures show a slightly lower mean [Ni/Fe] than \textit{in situ} populations at fixed [Fe/H], as expected for accreted populations in the Milky Way on the basis of previous work \citep[e.g.,][]{Nissen1997,Shetrone2003}; {\it ii)} Heracles and Thamnos fall on the same locus on the Ni-Fe plane, presenting a slight correlation between [Ni/Fe] and [Fe/H]; {\it iii)} as in the case of the Mg-Fe plane, Arjuna and LMS-1 occupy a similar locus in the Ni-Fe plane to that of GE/S, further supporting the suggestion that these substructures may be associated; {\it iv)} Aleph/Nyx mimic the behaviour of {in situ} low-/high-$\alpha$ disc populations, respectively. 

\subsection{Odd-Z elements}
\label{sec_oddZ}

\begin{figure*}
\includegraphics[width=\textwidth]{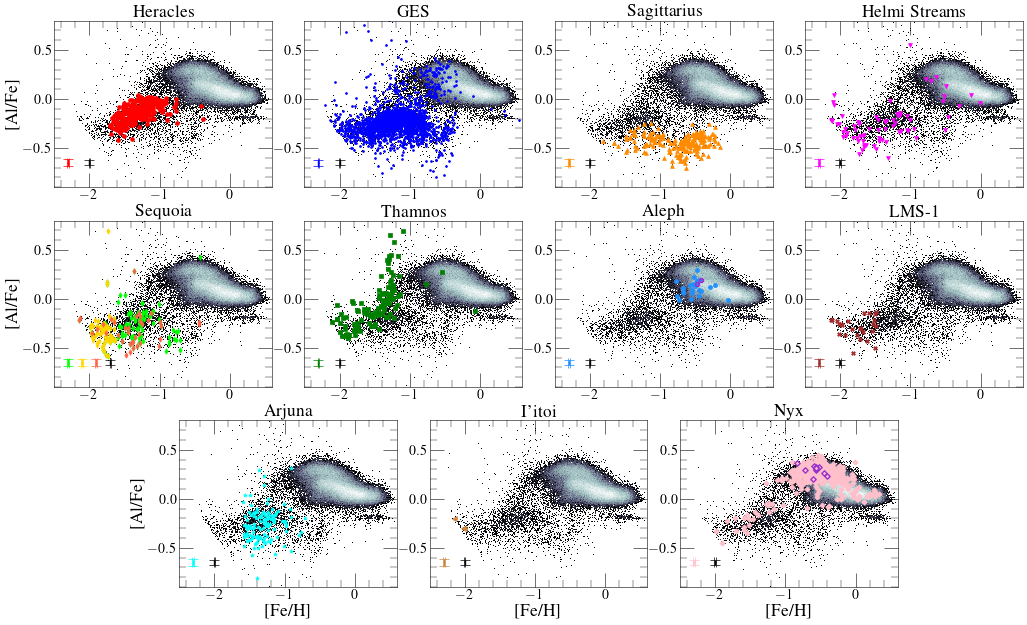}
\caption{The same illustration as in Fig~\ref{mgfes} in Al-Fe space. We note that the grid limit appears clearly in this plane at the lowest [Fe/H] values.}
    \label{alfes}
\end{figure*}

Aside from $\alpha$ and iron-peak elements, other chemical abundances provided by ASPCAP/APOGEE that are interesting to study are the odd-Z elements. These elements have been shown in recent work to be depleted in satellite galaxies of the MW and accreted systems relative to populations formed \textit{in situ} (\citealp[e.g.,][]{Hawkins2015,Das2020,Horta2021,Hasselquist2021}). For this paper, we primarily focus on the most reliable odd-Z element delivered by ASPCAP: aluminium. For the distribution of the structures in other odd-Z chemical abundance planes yielded by APOGEE (namely, Na, and K), we refer the reader to Fig.~\ref{nafe} and Fig.~\ref{kfe} in Appendix~\ref{app_oddZ}.

Fig.~\ref{alfes} displays the distribution of the substructures and parent sample in the Al-Fe plane, using the same symbol convention as adopted in Fig.~\ref{mgfes}. We note that the parent sample shows a high density region at higher metallicities, displaying a bimodality at approximately [Fe/H] $\sim$ --0.5, where the high-/low-[Al/Fe] sequences correspond to the high-/low-$\alpha$ discs, respectively. In addition, there is a sizeable population of aluminium-poor stars with ${\rm -0.5\simless[Al/Fe]\simless0}$ ranging from the most metal-poor stars in the sample all the way to ${\rm [Fe/H]\sim-0.5}$\footnote{The clump located at {$\rm [Al/Fe]\sim-0.1$} and [Fe/H] > 0 is not real, but rather an artifact due to systematics in the abundance analysis which does not affect the bulk of the data.}.  This is the locus occupied by MW satellites and most accreted substructures, with the exception of Aleph and Nyx.  Note also that the upper limit of the distribution of the Heracles population on this plane is determined by the definition of our sample \cite[see][for details]{Horta2021}.

The majority of the substructures studied occupy a similar locus in this plane, which agrees qualitatively with the region where the populations from MW satellites are usually found \cite[e.g.,][]{Hasselquist2021}.  There is strong overlap between stars associated with the GES, the Helmi stream, Arjuna, Sequoia (all three samples), and LMS-1 substructures. More specifically, we find that GES dominates the parent population sample at [Fe/H] < --1, being located at approximately [Al/Fe] $\sim$ --0.3. At a slightly higher value of [Al/Fe] $\sim$ --0.15 and similar metallicities, we find Heracles and Thamnos. In contrast, Sgr dSph is characterised by an overall  lower [Al/Fe] $\sim$ --0.5 value, which extends below the parent disc population towards higher [Fe/H], reaching almost solar metallicity.  Within the ${\rm -2 \simless [Fe/H] \simless -1}$ interval, Heracles, GES, Helmi streams, Thamnos, Nyx, and, to a lesser extent, Sequoia, show some degree of correlation between [Al/Fe] and [Fe/H].   Towards the metal-poor end, we find the LMS-1 located at [Al/Fe]$\sim$--0.3, which is consistent with the value found for I'itoi, although the sample of aluminium abundances for this latter structure is very small and close to the detection limit. As in the case of magnesium and nickel, all three Sequoia samples occupy the same locus in the Al-Fe plane as Arjuna, which strongly overlap with GES.  Again in the case of the Al-Fe plane, we find that the case for Nyx and Aleph follow closely the trend established by {\it in situ} disc populations.

\subsection{Carbon and Nitrogen}
\label{sec_cn}

\begin{figure*}
\includegraphics[width=\textwidth]{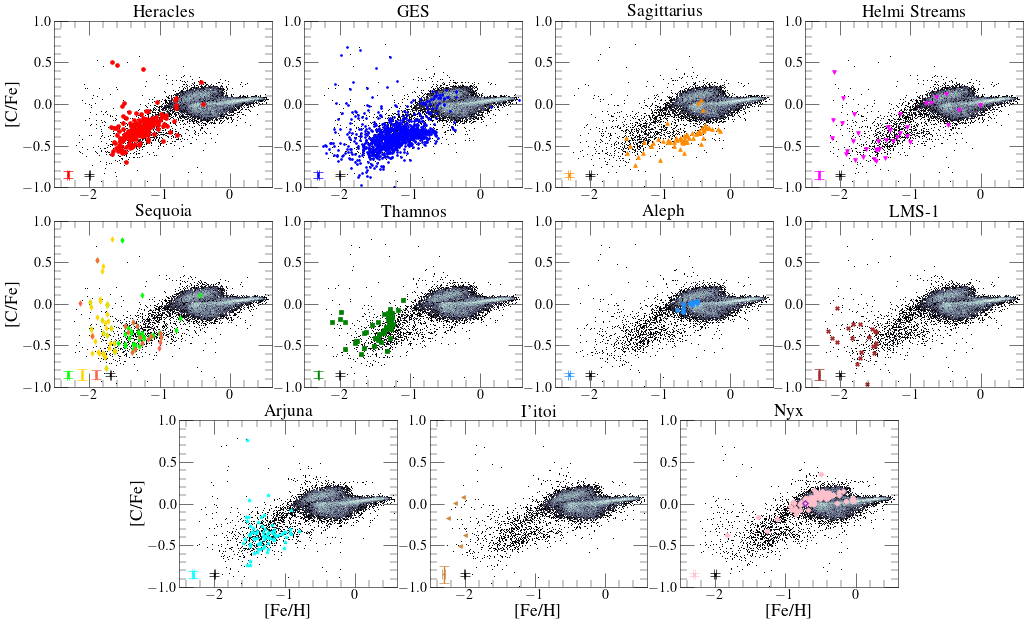}
\caption{The same illustration as Fig.~\ref{mgfes} in the C-Fe plane. For this chemical plane we restrict our sample to a surface gravity range of 1 < log$g$ < 2 in order to minimise the effect of internal mixing in red giant stars.} 
    \label{cfe}
    
\end{figure*}

\begin{figure*}
\includegraphics[width=\textwidth]{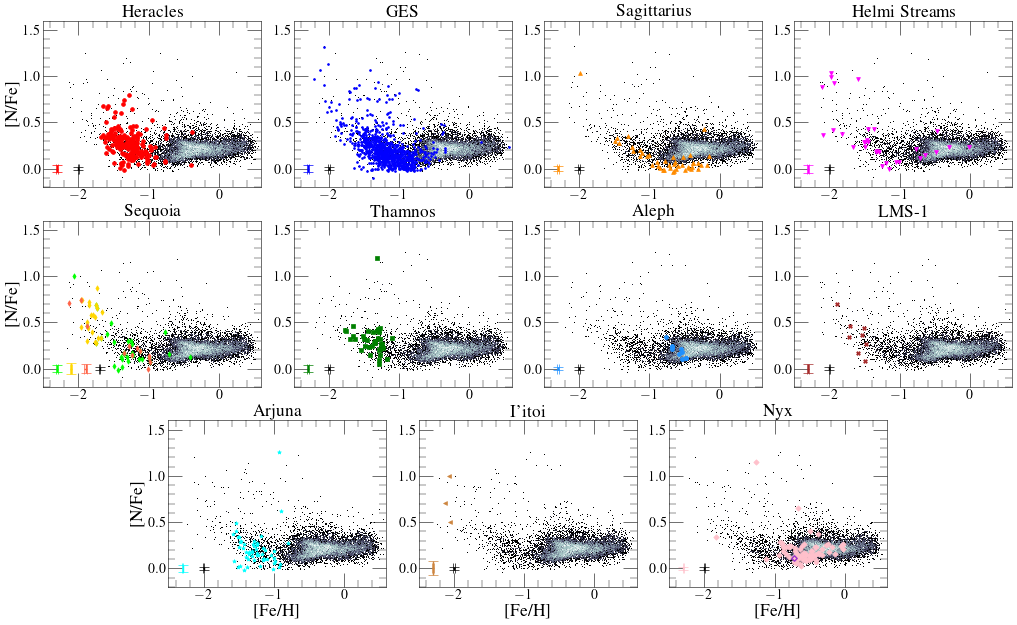}
\caption{The same illustration as Fig.~\ref{mgfes} in the N-Fe plane. As done in Fig~\ref{cfe}, for this chemical plane we restrict our sample to a surface gravity range of 1 < log$g$ < 2 in order to minimise the effect of internal mixing in red giant stars. We note that the grid limit appears clearly in this plane at the lowest [Fe/H] values.}
    \label{nfe}
    
\end{figure*}

In this subsection, we examine the distribution of stars belonging to various substructures in the C-Fe and N-Fe abundance planes, shown in Figures~\ref{cfe} and ~\ref{nfe}, respectively. We note that in these chemical planes, we impose an additional surface gravity constraint of 1 < log$g$ < 2 in order to minimise the effect of internal mixing along the giant branch.

 In the C-Fe plane, most substructures are characterised by sub-solar [C/Fe], displaying a clear correlation between that abundance ratio and metallicity.  The exceptions, as in all previous cases, are Aleph and Nyx, which again follow the same trends as {\it in situ} populations. Interestingly, the Sgr dSph presents the lowest values of [C/Fe] at fixed [Fe/H], tracing a tight sequence at approximately [C/Fe]$\sim$--0.5, spanning from --1.4 < [Fe/H] < --0.2, approximately $\sim$0.5 dex below that of the Galactic disc. In the case of I'itoi, due to the low numbers of stars in this sample we are unable to draw any conclusions.

The distribution of substructures in the N-Fe plane follows a different behaviour than seen in all other chemical planes.  Again, except for Aleph and Nyx, all systems display a trend of increasing [N/Fe] towards lower metallicities, starting at ${\rm [Fe/H]\simless-1}$.  This trend cannot be ascribed to systematics in the ASPCAP abundances or evolutionary effects, as the abundances are corrected for variations with $\log g$.  Nitrogen abundances are notoriously uncertain, particularly in the low metallicity regime. In their analysis of DR16 elemental abundances, \citet[][]{Jonsson2020} report the presence of a trend of nitrogen abundances with $T_{\rm eff}$.  They suggest that such trend is likely caused by systematics in the comparison samples, which are based on rather uncertain analyses of optical spectra.  They find that the relation between [C/N] and $\log g$ is in good agreement with theoretical expectations. The compilation by \cite{Kobayashi2020} shows that the [N/Fe] trend at low metallicity is strongly dependent on the analysis methods.  Discerning the source of systematics in the ASPCAP abundances at ${\rm [Fe/H]\simless-1}$ is beyond the scope of this paper, which focuses on a strictly differential analysis of the data, within a metallicity regime where ASPCAP elemental abundances attain exceedingly high precision and accuracy (Section~\ref{sec_abundances}).

For completeness, data covering the whole range of $\log g$ for all substructures are displayed in the (C+N)-Fe plane in Fig~\ref{cnfe} in Appendix~\ref{app_carbon_nitrogen}. By combining carbon and nitrogen abundances, we minimise the effect of CNO mixing along the giant branch. In this plane, MW satellites and accreted populations typically display a lower [(C+N)/Fe] chemical composition than their \textit{in situ} counterparts \citep[e.g.,][]{Horta2021,Hasselquist2021}. This is in fact what we observe for all the structures identified, again with the exception of Aleph and Nyx, whose locus overlaps with that of \textit{in situ} disc populations.

\subsection{Cerium}
\label{sec_neutron_capture}

\begin{figure*}
\includegraphics[width=\textwidth]{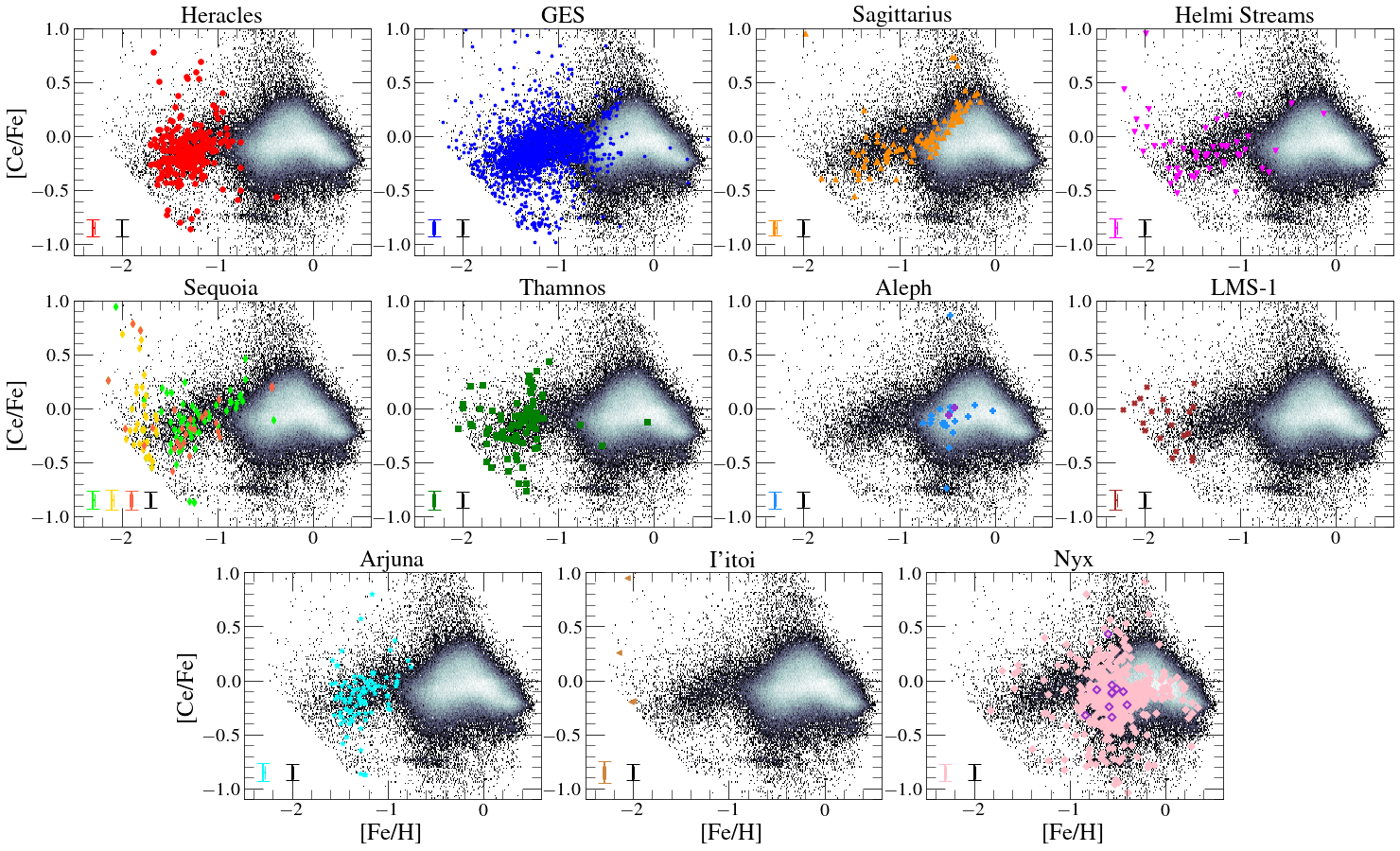}
\caption{The same illustration as Fig.~\ref{mgfes} in the Ce-Fe plane. We note that the grid limit appears clearly in this plane at the lowest [Fe/H] values. }
    \label{cefes}
\end{figure*}

Cerium is a neutron capture element of the $s$-process family, with a large enrichment contribution from AGB stars \citep{Sneden2008,Jonsson2020,Kobayashi2020}. In Fig.~\ref{cefes}, the disc sample at [Fe/H]$>-1$ has a roughly horizontal locus at [Ce/Fe]$\approx-0.1$ dominated by stars in the {\it high}-$\alpha$ population and an upward-pointing triangular locus dominated by stars in the {\it low}-$\alpha$ sequence, reaching [C/Fe]$\approx+0.4$ at [Fe/H]$\approx-0.2$.  The scatter within each of these components is large and may be partly observational.  The presence of substantial Ce in high-$\alpha$ stars suggests that massive stars with short lifetimes make a significant, prompt contribution.  The rising-then-falling trend in the low-$\alpha$ population is expected from the metallicity-dependent yield of intermediate mass AGB nucleosynthesis: at low [Fe/H] the number of seeds available for neutron capture increases with increasing metallicity, but at high [Fe/H] the number of neutrons per seed becomes to low to produce the heavier $s$-process elements \citep{Gallino1998}. See \cite{Weinberg2021} for plots of [Ce/Mg] vs. [Mg/H] and further discussion of the disc trends.

In this chemical composition plane, we find that all the identified substructures, with the exception of Aleph and Nyx, present [Ce/Fe] abundances that follow the mean trend with [Fe/H] of the parent population until [Fe/H]~$\sim-1$.  Aleph and Nyx have higher [Fe/H] stars that generally lie within the broad disc locus.  Interestingly, the Sgr stars with [Fe/H]~$>-1$ show a rising [Ce/Fe] trend that tracks the behaviour of the low-$\alpha$ disc population.  This trend is not obvious in the other substructures, though with the exception of Aleph and Nyx they have few stars at [Fe/H]~$>-1$.  We interpret this upturn in both Sgr and the low-$\alpha$ disc as the signature of an AGB contribution with a metallicity dependent yield.

At [Fe/H]~$<-1$, the parent halo population and most substructure stars exhibit mildly sub-solar [Ce/Fe] with substantial scatter, which may have a significant observational component.  The [Ce/Fe] is similar to that of typical high-$\alpha$ disc stars at [Fe/H]~$>-1$.  However, these stars have lower [$\alpha$/Fe] than the high-$\alpha$ disc, the signature of Fe enrichment from Type Ia SNe, so if the Ce in these populations is a prompt contribution from massive stars one might have naively expected them to have depressed [Ce/Fe].  It is difficult to disentangle the effects of metallicity-dependent Ce yields, differences in the relative contributions of high-mass and intermediate-mass stars, and the impact of Type Ia SN enrichment on the Fe abundance; further observational investigation and theoretical modeling will be needed to do so. The [Ce/Fe] locus of substructure stars at $-2 < {\rm [Fe/H]} < -1$ is similar to that in the dwarf satellites studied by \cite{Hasselquist2021}.

\subsection{ The [Al/Fe] vs [Mg/Mn] plane}
\label{sec_other_chem}

\begin{figure*}
\includegraphics[width=\textwidth]{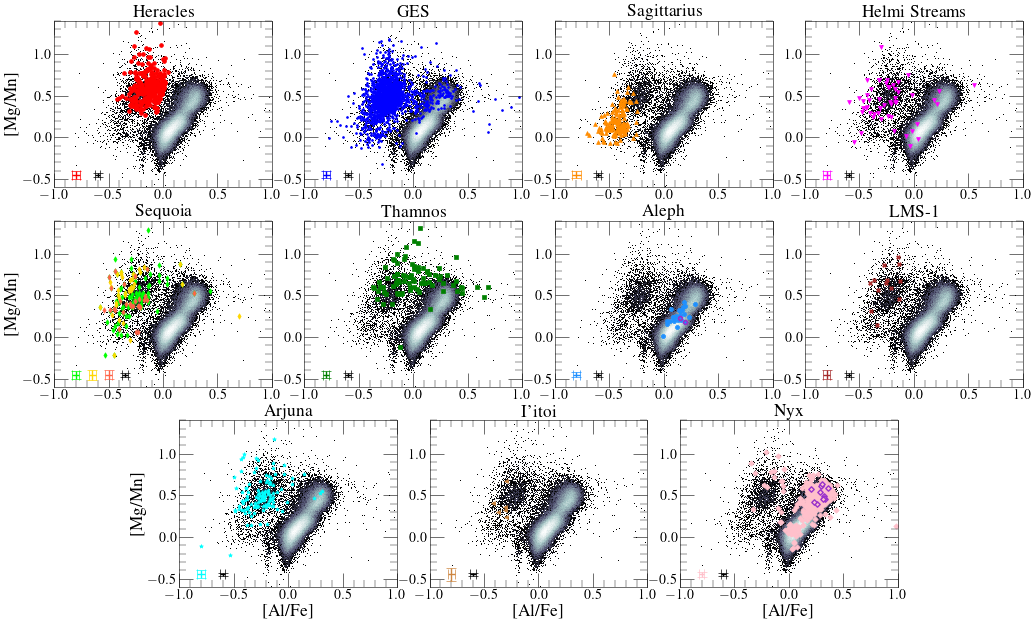}
\caption{The same illustration as Fig.~\ref{mgfes} in the [Mg/Mn]-[Al/Fe] plane. }
    \label{mgmn}
    
\end{figure*}

Having studied the distribution of the identified structures in chemical abundance planes that aimed to give us an insight into the different nucleosynthetic pathways, contributed either by core-collapse, type Ia supernovae, and AGB stars, we now focus our attention on analysing the distribution of substructures in the stellar halo in an abundance plane that lends insights into the accreted or \textit{in situ} nature of Galactic stellar populations: namely, the [Mg/Mn]-[Al/Fe] plane.

Fig.~\ref{mgmn} shows the resulting distribution of the various structures in the [Mg/Mn]-[Al/Fe] plane. This chemical plane has been proposed by \cite{Das2020} as a means to distinguishing accreted populations from those formed {\it in situ}.  \cite{Horta2021} showed that {\it in situ} stellar populations with a small degree of chemical evolution occupy the same locus in that plane as accreted populations. By construction, the Heracles substructure falls in the accreted locus of the diagram (see Fig~1 in \citet{Horta2021} for reference). However, our results show that all the other structures, except for Aleph and Nyx, also occupy the accreted locus of this plane. Interestingly, we find that although the GES, Sgr dSph, the Helmi stream, Sequoia (all three samples), Thamnos, LMS-1, Arjuna, and I'itoi substructures occupy the same locus, they appear to show some small differences. Specifically, we find that Sgr dSph occupies a locus in this plane positioned at lower mean [Mg/Mn] than the other structures. This is likely due to Sgr being more recently accreted by the Milky Way, and thus had more time to develop stellar populations with enriched Mn abundances that have been contributed on a longer timescale by type Ia supernovae. This feature is also seen to a lesser extent for I'itoi. Conversely, at higher [Mg/Mn] values (but still low [Al/Fe]) we find GES, Heracles (by construction), Thamnos, LMS-1, the Helmi stream, Sequoia (all three samples), and Arjuna. The distribution of these substructures in this chemical plane reinforces the hypothesis of these halo substructures arising from an accreted origin. 

In a similar fashion to the other chemical composition planes, we find that Aleph and Nyx overlap with \textit{in situ} (disc) populations at higher [Al/Fe], suggestive that Aleph and Nyx are likely substructures comprised of \textit{in situ} disc populations.

\section{A quantitative comparison between halo substructure abundances}
\label{sec_abundances}
After qualitatively examining the chemical compositions of the previously identified halo substructures in a range of chemical abundance planes, we now focus on comparing the abundances in a quantitative fashion using a $\chi^{2}$ method. To do so, we compare the mean value of thirteen different elemental abundances, manufactured across different nucleosynthetic channels, for each substructure at a fixed metallicity that is well covered by the data. The set of elemental abundances chosen to run this quantitative test was determined based on the distribution of the parent sample in the respective chemical composition plane, where we only chose those elements that did not display a large scatter towards low metallicity due to increased abundance uncertainties (on the order of $\sigma$ $\sim$0.15 dex). Out of the initial 20 elemental abundances available in ASPCAP (excluding Fe), we utilise the following thirteen elements: C, N, O, Mg, Al, Si, S, K, Ca, Ti, Mn, Ni, and Ce. We note that Na, P, V, Cr, and Co were removed due to the large scatter at low metallicity, whereas Cu and Nd were not considered due to \texttt{ASPCAP} not being able to determine abundances for these elements in APOGEE DR17. 


We employ in our quantiative comparison a methodology that is immune to, or very minimally affected by, selection effects. For that reason, the metallicity distribution function does not enter our analysis.  Our quantitative comparisons between various substructures are instead based on the abundance ratios at a reference [Fe/H], and the slope of the relation between abundance ratios and [Fe/H]. The method proceeds as follows:

i) We select a high- and low-$\alpha$ (disc/halo) population based on Fig~\ref{discs} for reference, and utilise these as representative  samples for any comparison between \textit{in situ} populations and halo substructures. In order to account for any distance selection function effects, we restrict our high-/low-$\alpha$ samples to stars within $d < 2$ kpc, and also determine an "inner high-$\alpha$" sample (restricted to R$_{\mathrm{GC}} < 4$ kpc) which we use to compare to Heracles (which has a spatial distribution that is largely contained within $\sim$4 kpc from the Galactic Centre).

ii) Before performing any chemical composition comparisons, we correct the abundances for systematic trends with surface gravity. Systematic abundance variations trends with $\log g$ can be caused, one one hand, by real physical chemical composition variations as a function of evolutionary stage and/or, on the other, by systematic errors in elemental abundances as a function of stellar parameters.  The former chiefly impact elements such as C, N, and O, whose atmospheric abundances are altered by mixing during evolution along the giant branch.  The latter impact various elements in distinct, though more subtle, ways \citep[see discussion in][]{Weinberg2021}.  Surface gravity distributions of various substructures differ in important ways (Figure~\ref{fig:hr}), so that such systematic abundance trends with $\log g$ can induce spurious artificial chemical composition differences between substructures.  We thus follow a procedure similar to that outlined by \cite{Weinberg2021} to correct each elemental abundance using the full parent sample. As systematic trends with $\log g$ are more important towards the low and high ends of the $\log g$ distribution, we restrict our sample to stars within the $1 < \log g  < 2$ range. We then fit a second order polynomial to the [X/H]-$\log g$ relation, and calculate the difference between that fit and the overall [X/H] median. The difference between these two quantities for any given $\log g$ is then added to the original [X/H] values so as to produce a flat relation between [X/H]$_{\rm corrected}$ and $\log g$. We then use these values to determine corrected [X/Fe] abundances. In a recent study, \cite{Eilers2021} pointed out that simple corrections for abundance trends as a function of $\log g$ could erase real differences associated with abundance gradients within the Galaxy.  That is because a magnitude limited survey may cause an artificial dependence of $\log g$ on distance. Our study aims at contrasting the chemical compositions of substructures that are in principle associated with spatially self-contained progenitors.  Thus, systematic differences linked to spatial abundance variations within each structure are irrelevant for our purposes, so a straightforward correction for abundance variations with $\log g$ are perfectly acceptable for our goals.

iii) Upon obtaining corrected abundances for every halo substructure, we determined the uncertainties in the abundances using a bootstrapping resampling with replacement method (utilising the \texttt{astropy.stats.bootstrap} routine by \citealp[][]{astropy:2018}).  We generated 1,000 realisations of the X-Fe chemical composition planes for every element and every halo substructure sample in order to assess the scatter in the abundance distribution. For example, we generated 1,000 realisations of the C-Fe distribution for GES by drawing 2,353 values from the observed distribution, with replacement (where 2,353 is the size of our GES sample).

iv) For each one of the 1,000 bootstrapped realisations of a chemical composition plane of a halo substructure, we determine the [X/Fe] value at a given metallicity ([Fe/H]$_{\mathrm{comp}}$) that is covered by both halo substructures being compared by taking a 0.05 dex slice in [Fe/H] around [Fe/H]$_{\mathrm{comp}}$ and determining the median value for stars in that [Fe/H] interval. This yields 1,000 [X/Fe] median values for each of the thirteen elements studied for every halo substructure (and disc sample) compared. We note that the [Fe/H]$_{\mathrm{comp}}$ values selected to perform this comparison are safely within the metallicity (and stellar parameter regime) where ASPCAP analysis is very reliable.

v) We take the mean and standard deviation of the medians distribution for every [X/Fe] as our representative [X/Fe] and uncertainty value, respectively, and use these to quantitatively compare the chemical abundances between two populations. This sample median from the mean of the bootstrap medians is always close to the full sample median itself.

vi) Upon obtaining the mean and uncertainty chemical abundance values for every halo substructure at [Fe/H]$_{\mathrm{comp}}$ (for a range of thirteen reliable elemental abundances in APOGEE), we quantitatively compare the chemical compositions across halo substructures adopting a $\chi^{2}$ statistic to assess the chemical similarities between different substructures. This quantity was computed by using the following relation:
\begin{equation}
    \chi^{2} =  \sum_{i} \frac{\Big(\mathrm{[X/Fe]}_{i,\mathrm{sub}} - \mathrm{[X/Fe]}_{i,\mathrm{ref}}\Big)^{2}}{\Big(\sigma_{\mathrm{[X/Fe]}_{i,\mathrm{sub}}}^{2} + \sigma_{\mathrm{[X/Fe]}_{i,\mathrm{ref}}}^{2}\Big)},
    \label{eq:chi2}
\end{equation}
where [X/Fe]$_{\mathrm{sub}}$ and [X/Fe]$_{\mathrm{ref}}$ are the abundances of the halo substructure and the compared reference stellar population, respectively, and $\sigma_{\mathrm{[X/Fe],sub}}$ and $\sigma_{\mathrm{[X/Fe],ref}}$ are the corresponding uncertainties to those abundance values. Since GES is the halo substructure for which our sample is the largest, we use this substructure as our main reference against which all other substructures, as well as \textit{in situ} disc populations are contrasted. 

vii) Lastly, in order to infer if two stellar populations present consistent chemical abundances in a statistical manner, we determine the probability value of the $\chi^{2}$ result for twelve degrees of freedom using the \texttt{scipy} \citep[][]{Scipy2020} \texttt{stats.chi2.cdf} routine. In addition to the $\chi^{2}$ value, we also compute a metric of separation (defined as $\Sigma_{\mathrm{[X/Fe]}}$), that is calculated by setting the denominator of Eq~\ref{eq:chi2} equal to 1. This separation metric provides an additional way to quantify how similar the chemical compositions of two halo substructures are that is unaffected by the sample size (as smaller halo substructure samples will have larger uncertainties on their mean abundance values).

viii) For the case of Heracles and Aleph, as the selection of these substructures relies heavily on the use of [Al/Fe] and [Mg/Fe], respectively, we remove these elements when comparing these substructures, and reduce the numbers of degrees of freedom to eleven when calculating the $\chi^{2}$ probability value.

 In order to develop a more clear notion of the meaning of the resulting $\chi^{2}$ values resulting from the above comparisons we perform an additional exercise aimed at gauging the expected $\chi^2$ values for the cases where two samples are identical to each other, or very different.  To accomplish this, we draw, for each substructure, three $N_{\rm sub}$-sized random samples, two from the high-$\alpha$ and one from the low-$\alpha$ disc samples, where $N_{\rm sub}$ is the size of the sample of that substructure.  We then calculate, for each substructure, two $\chi^2$ values, one resulting from the comparison of the high-$\alpha$ disc against itself, and the other from the comparison of the high-$\alpha$ disc against the low-$\alpha$ disc samples. As the high- and low-$\alpha$ disc both cover a similar range in [Fe/H], and their abundances vary with [Fe/H], we select a narrow bin in [Fe/H] from which to draw our high- and low-$\alpha$ disc samples (namely, between --0.45 < [Fe/H] < --0.35). This enables us to obtain random samples of high- and low-$\alpha$ disc populations at the same [Fe/H], and allows us to directly compare the mean and scatter values of the chemical abundances using the $\chi^{2}$ method.

The reader may inspect the resulting $\chi^{2}$ and associated probability ($p_{\chi^{2}}$) values obtained in Table~\ref{tab:chi2}. Here, a $p_{\chi^{2}}$ $\sim$ 1 signifies that two populations are statistically equal, and $p_{\chi^{2}}$ $\sim$ 0 means that they are statistically different. For the quantitative comparisons, we employ a threshold of $p_{\chi^{2}}$ = 0.1 as our benchmark, where we will deem two substructures to be statistically similar if their associated probability value is higher than $p_{\chi^{2}}$ > 0.1, and different if below $p_{\chi^{2}}$ < 0.1. We note that this definition is arbritary, and chosen based on testing the similarity between two substructures in chemical space.

The resulting quantitative comparison between the halo substructures indicates that approximately half of the halo substructures are statistically equal with regards to their chemical abundances. For example, the $\chi^{2}$ comparison between the GES and two of the Sequoia samples implies that these three substructures are statistically the same, as we find that the \textit{M19} and \textit{N20} Sequoia samples all have a high probability of being statistically similar to the GES substructure ($p_{\chi^{2}}$ > 0.4). Conversely, for the \textit{K19} Sequoia shows clear differences with the GES substructure, given its $p_{\chi^{2}}$=0 and $\chi^{2}$=47.9 values. Moreover, we also find striking similarities between GES and other halo substructures. For example, when comparing LMS-1 to GES we obtain a probability value of 0.46. Similarly, an ever closer match is found for the GES-Arjuna comparison, yielding a probability value of 0.92. Along similar lines, we find that when comparing  Nyx to the high-$\alpha$ disc, we obtain a probability value of 0.78, which confirms our initial hypothesis that the Nyx is not an accreted substructure, but instead is a stellar population constituted of high-$\alpha$ disc stars. Moreover, we find that when comparing Heracles, Sgr dSph, and Thamnos with the GES that the probability of these substructures being statistically equal is $\sim$0. This result is not entirely surprising, as all these substructures are postulated to be debris from separate accretion events, and thus their chemical compositions are likely to differ. Interestingly, we find two surprising results: i) despite the Aleph substructure presenting qualitatively the same chemistry as the low-$\alpha$ disc, its $\chi^{2}$ value yields a probability of 0.04, suggesting that these are not as similar as initially hypothesised. These differences are manifested most prominently in [Ti/Fe] and [Ce/Fe]; ii) although Heracles occupies a position in several chemical composition planes that appears to follow a single sequence with the inner high-$\alpha$ populations, the $\chi^{2}$ yielded by that comparison indicates that these are statistically different (with a probability value of $p_{\chi^{2}}$=0). For this comparison, we find that [O/Fe], [Mg/Fe], and [Si/Fe] (i.e., the lighter $\alpha$ elements) are the main culprits for this difference, and to a lesser degree [S/Fe].

For the complete resulting probability values obtained when comparing the chemistry between every halo substructure, as well as the high- and low-$\alpha$ disc, we refer the reader to Fig~\ref{confusion_matrix}.

\begin{figure*}
\includegraphics[width=\textwidth]{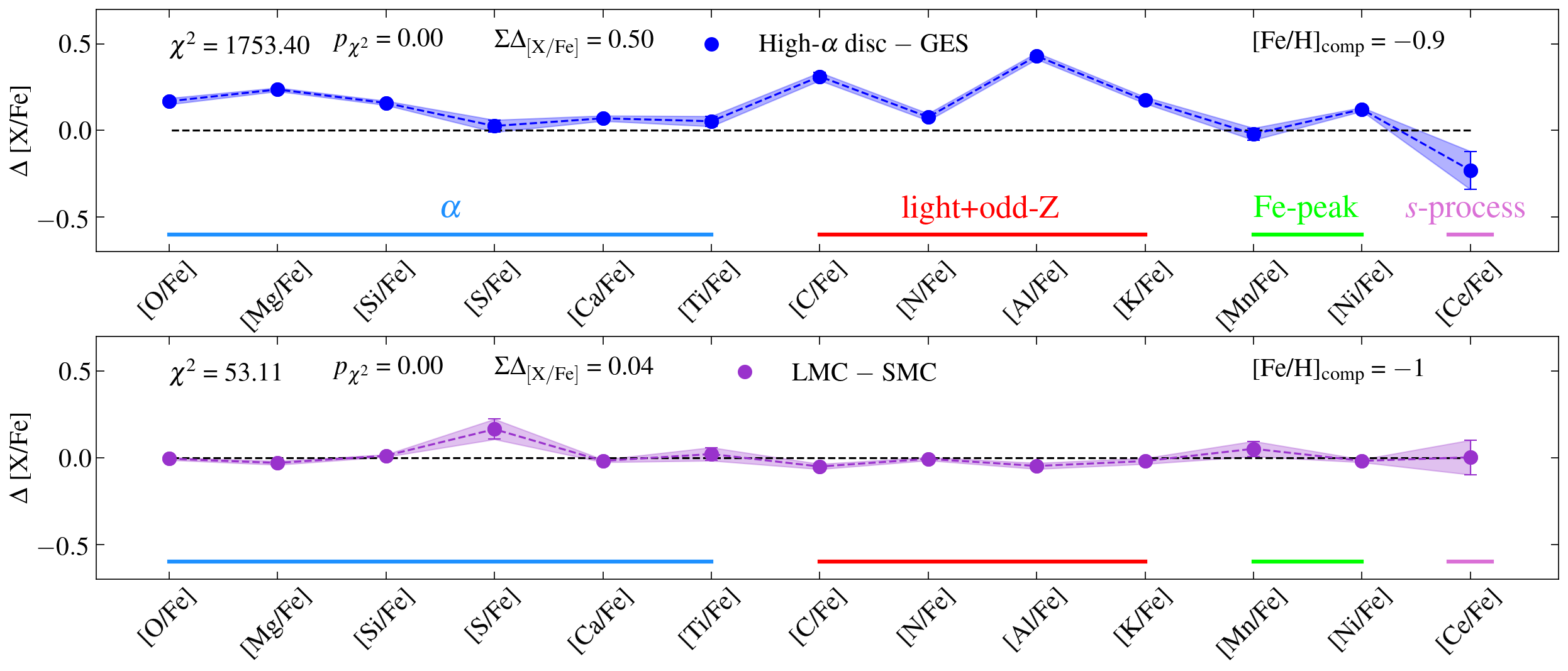}
\caption{$\Delta$[X/Fe] differences between the resulting mean values obtained using the procedure outlined in Section~\ref{sec_abundances} (at [Fe/H]$_{\mathrm{comp}}$) for the \textit{Gaia}-Enceladus/Sausage substructure and the high-$\alpha$ disc stars (top) and for the Large and Small Magellanic Cloud (LMC/SMC) samples from \citet[][]{Hasselquist2021} (bottom) in thirteen different chemical abundance planes, grouped by their nucleosynthetic source channel. The shaded regions illustrate the uncertainty on this $\Delta$[X/Fe] value. Also illustrated in the top right/left are the $\chi^{2}$/$p_{\chi^{2}}$/[Fe/H]$_{\mathrm{comp}}$ values for the comparison between these two populations. As can be seen from the abundance values, the $\chi^{2}$ value, and the $p_{\chi^{2}}$ value, it is evident that the GES/high-$\alpha$ disc and the LMC/SMC are quantitatively different given their chemical compositions.}
    \label{ges_disc}
\end{figure*}

\begin{figure*}
\includegraphics[width=\textwidth]{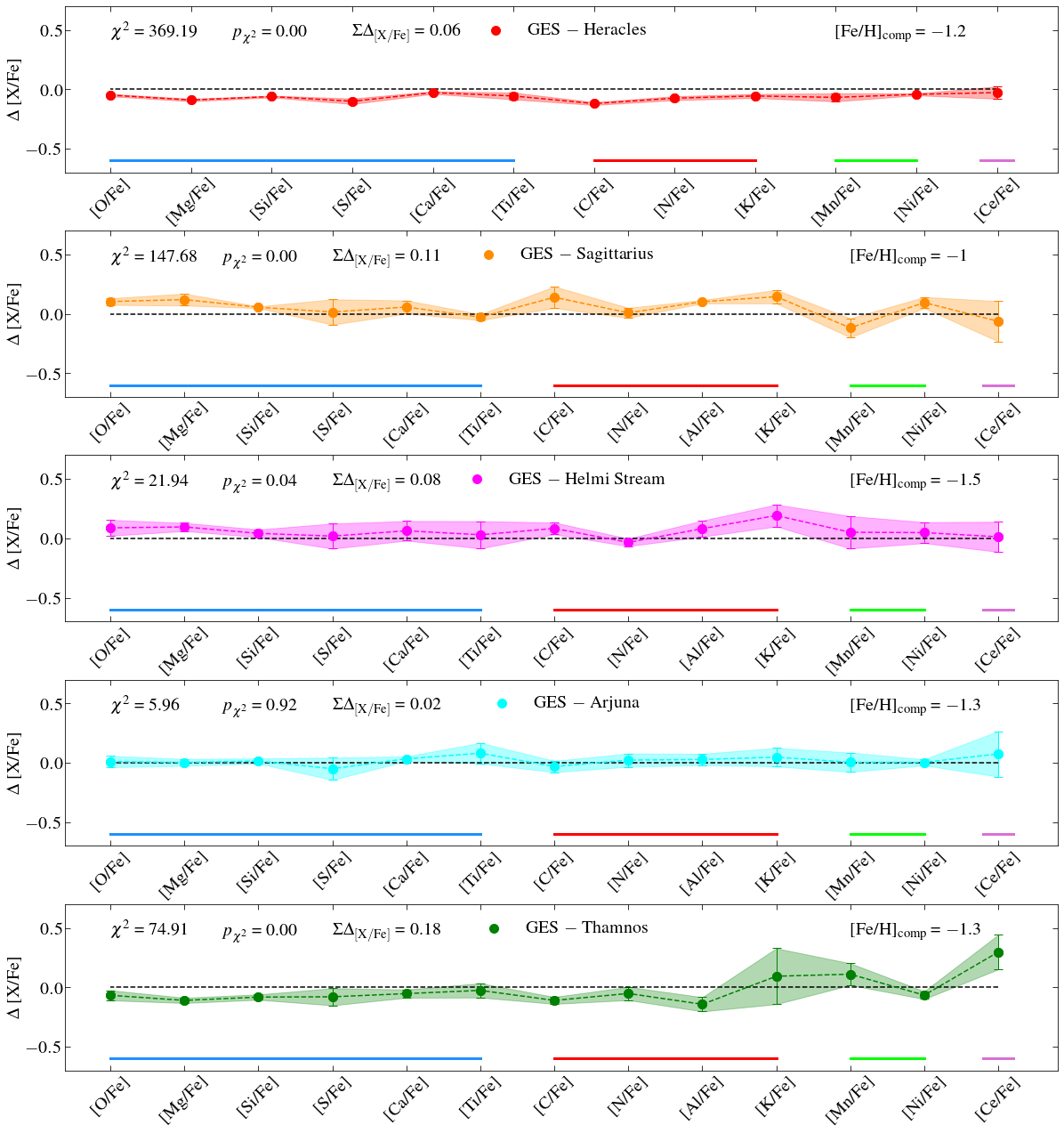}
\caption{The same mean and mean error abundance values as shown in Fig~\ref{ges_disc} but comparing the Heracles, Sagittarius dSph, Helmi stream, Arjuna, and Thamnos substructures with the \textit{Gaia}-Enceladus/Sausage substructure. We note that those substructures with fewer stars present larger uncertainties in their $\Delta$[X/Fe] value.}
    \label{ges_combined1}
\end{figure*}

\begin{figure*}
\includegraphics[width=\textwidth]{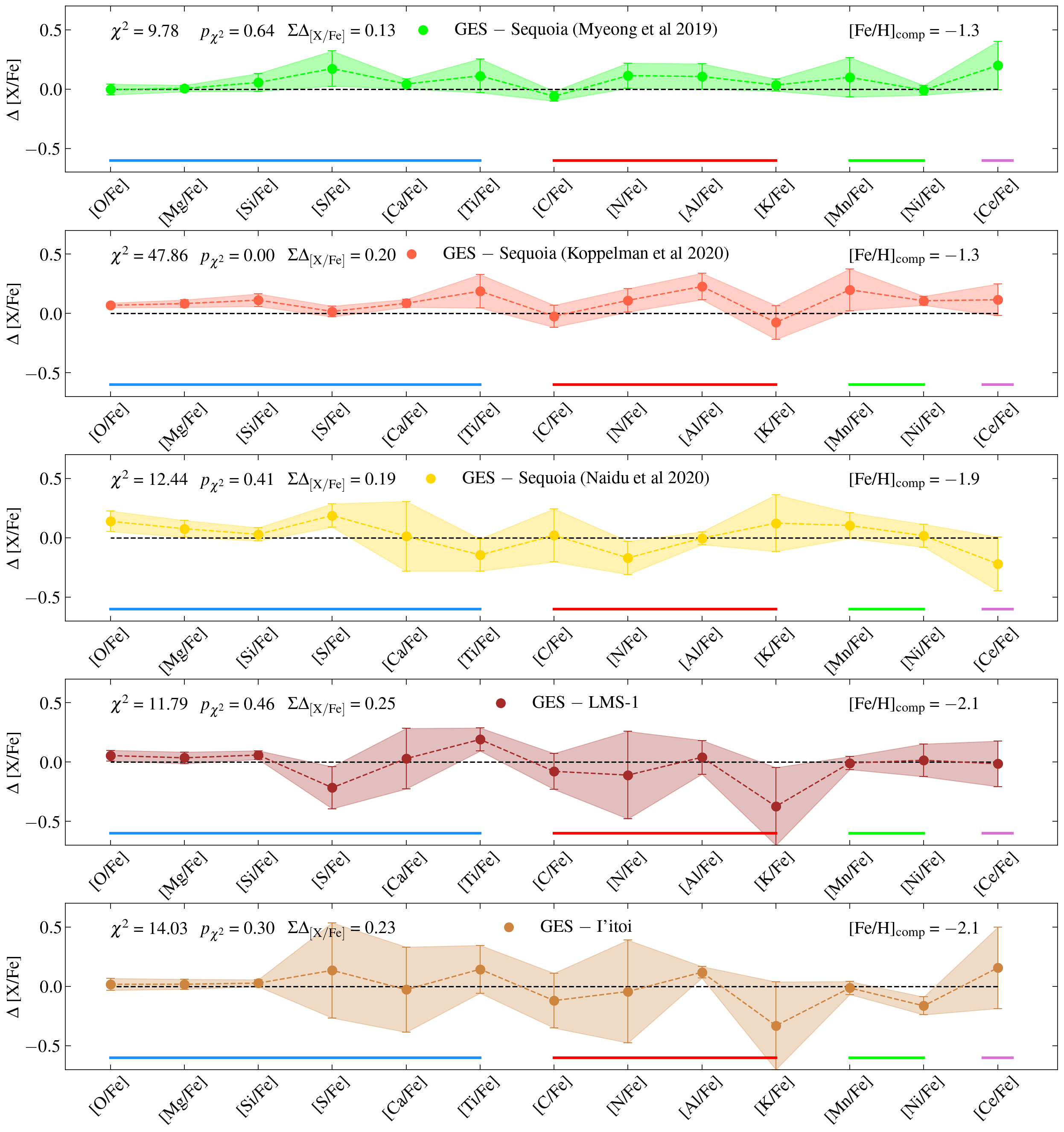}
\caption{The same as Fig~\ref{ges_combined1} but comparing the three Sequoia samples, LMS-1, and I'itoi substructures with the \textit{Gaia}-Enceladus/Sausage substructure.}    \label{ges_combined2}
\end{figure*}

\begin{figure*}
\includegraphics[width=\textwidth]{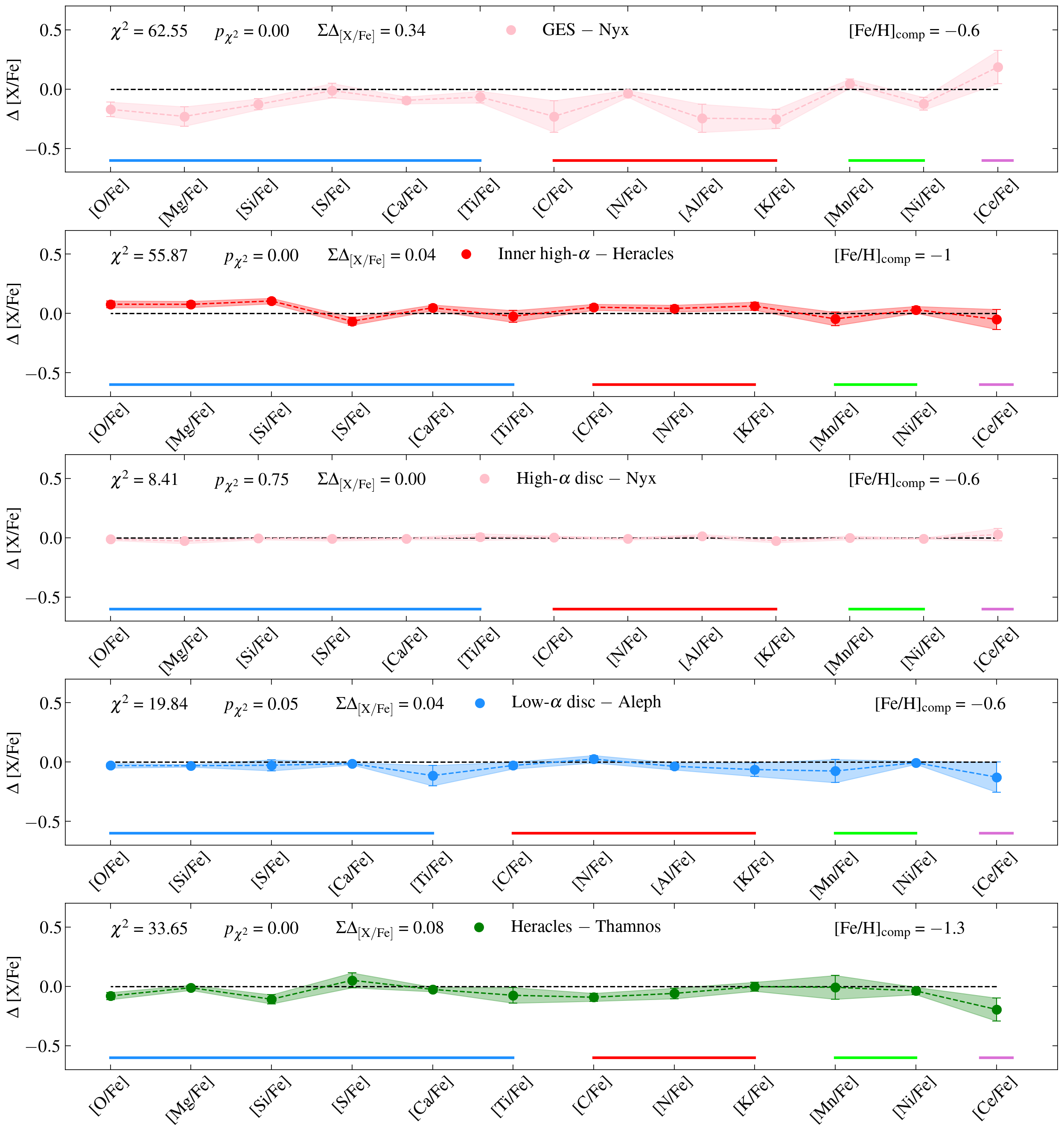}
\caption{The same as Fig~\ref{ges_disc} but comparing the Nyx substructures with the \textit{Gaia}-Enceladus/Sausage substructure and the high-$\alpha$ discs, as well as a comparison between the Heracles and inner high-$\alpha$ disc, Heracles and Thamnos, and Aleph and the low-$\alpha$ disc.}    \label{ges_combined3}
\end{figure*}

\setlength{\tabcolsep}{8pt}
\begin{table*}
\centering
\begin{tabular}{ p{3.5cm}|p{1cm}|p{1cm}|p{1cm}|p{1cm}|p{2cm}|p{2cm}}
\hline
Compared samples & [Fe/H]$_{\mathrm{comp}}$& $\chi^{2}$& $p_{\chi^{2}}$ & $\Sigma \Delta_{\mathrm{[X/Fe]}}$ & high$\alpha$-high$\alpha$ $\chi^{2}$ & high$\alpha$-low$\alpha$ $\chi^{2}$ \\
\hline
\hline
High$\alpha$ disc-GES & --0.9&  1753.4 & 0.00 & 0.50 & 9.78& 4622.96 \\ 
\hline
LMC-SMC & --1.1&  53.1 & 0.00 & 0.04& 10.05 & 1517.67 \\ 
\hline
GES-Heracles & --1.3& 369.2& 0.00& 0.06& 10.74&507.43  \\ 
\hline
GES-Sgr dSph& --1.0 & 147.7& 0.00& 0.11& 8.55  & 542.27\\
\hline
GES-Helmi stream& --1.5 & 21.9 & 0.08 & 0.08& 6.61 &153.93 \\ 
\hline
GES-Sequoia (\textit{M19})& --1.3 & 9.8& 0.64& 0.13& 3.40& 207.94 \\
\hline
GES-Sequoia (\textit{K19})& --1.3 & 47.9& 0.00& 0.2& 4.04&210.68\\ 
\hline
GES-Sequoia (\textit{N20})& --1.9 & 12.4& 0.41& 0.19& 3.42&229.83 \\
\hline
GES-Thamnos& --1.3 & 74.9& 0.00& 0.18& 8.82 &162.81\\
\hline
GES-LMS-1& --2.1& 11.8& 0.46& 0.25& 6.27&371.65 \\
\hline
GES-Arjuna& --1.3 &5.96& 0.92& 0.02&5.48 &230.82\\
\hline
GES-I'itoi& --2.1& 14.0&0.3& 0.23& 6.75& 123.99\\
\hline
GES-Nyx& --0.6& 62.6&0.00& 0.34& 6.45&1246.71  \\
\hline
high$\alpha$ disc-Nyx& --0.6 & 8.4 & 0.75& $<$0.01& 6.45&1246.71\\
\hline
low$\alpha$ disc-Aleph& --0.6& 19.9& 0.05& 0.04& 8.82&195.69\\
\hline
Inner high-$\alpha$-Heracles& --1&55.9&0.00& 0.04 & 10.74&507.43\\
\hline
Heracles-Thamnos&--1.3 & 33.6& 0.0& 0.08& 8.82 &162.81 \\
\hline
\hline
\end{tabular}
\caption{From left to right: compared halo substructures, [Fe/H] value used to compare the two compared substructures, resulting $\chi^{2}$ value from the comparison between the listed halo substructures, the probability value the $\chi^{2}$ result falls upon for a $\chi^{2}$ test with twelve (or eleven for the case of Heracles and Aleph) degrees of freedom, the metric separation $\Sigma \Delta_{\mathrm{[X/Fe]}}$, $\chi^{2}$ value between two randomly chosen high-$\alpha$ disc samples of the same size as the smallest substructure compared, $\chi^{2}$ value between a randomly chosen high-$\alpha$ and low-$\alpha$ disc sample of the same size as the smallest substructure compared. The LMC/SMC samples were taken from \citet[][]{Hasselquist2021}.}
\label{tab:chi2}
\end{table*}

\begin{figure*}
\includegraphics[width=\textwidth]{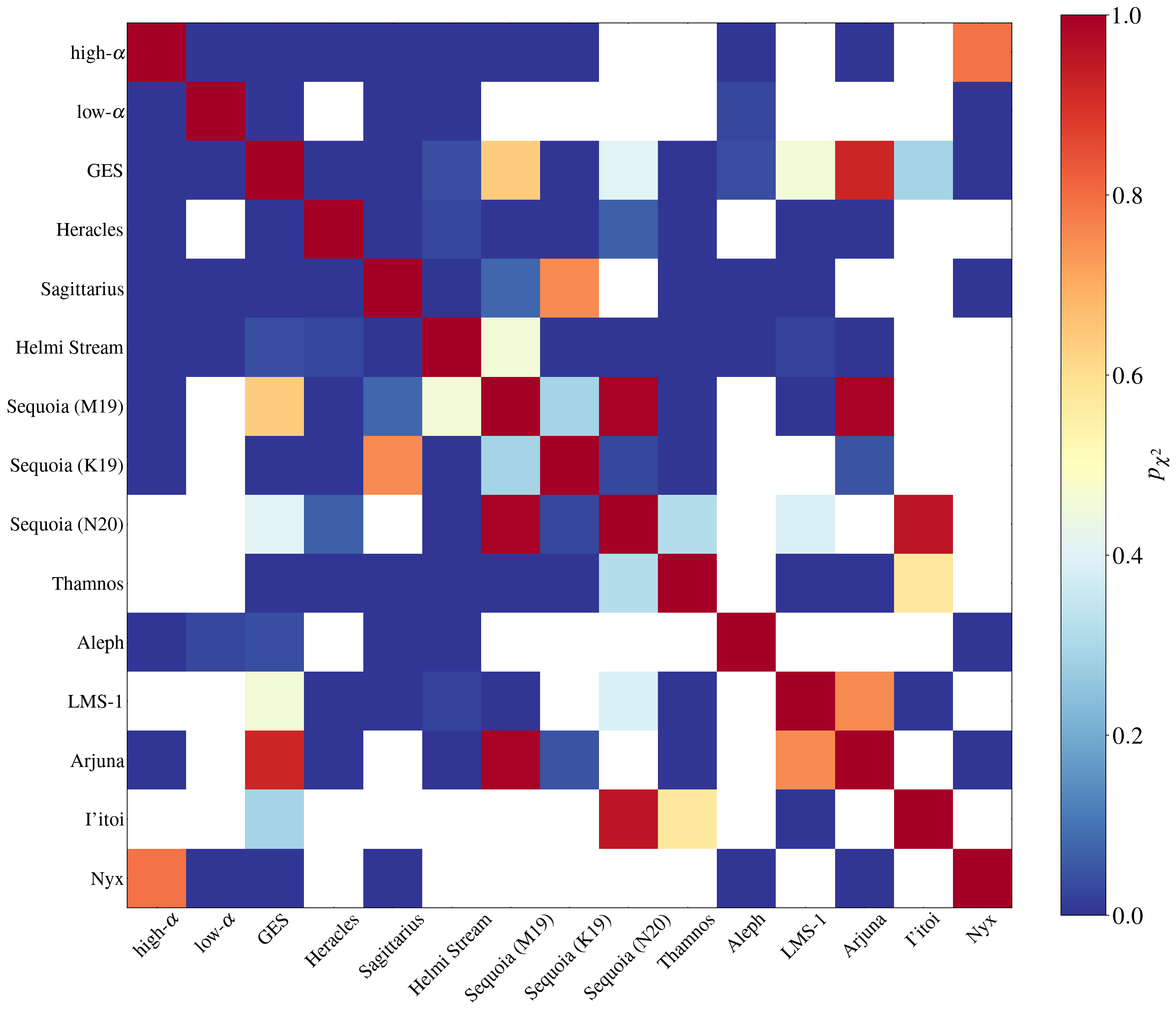}
\caption{Confusion matrix of the probability values (estimated using the $\chi^{2}$ calculated using Eq~\ref{eq:chi2}) obtained when comparing the chemical compositions of all the halo substructures with each other and with a high-/low-$\alpha$ discs. Here, each substructure is compared with its counterpart using a [Fe/H] value that is well covered by the data (see Fig~\ref{confused_matrix_feh} in Appendix~\ref{appen_confusion} for further details), where red(blue) signifies a high(low) probability of two systems being statistically similar given their chemical compositions. Comparisons with blank values are due to the two substructures being compared not having any overlap in [Fe/H].}    \label{confusion_matrix}
\end{figure*}

\section{Discussion}
\label{discussion}

\subsection{Summary of substructure in the stellar halo}
\label{discussion_subs}

Having qualitatively and quantitatively compared the chemical abundances of all the halo substructures under study, in this Section we discuss the results obtained for each substructure in the context of previous work. 

\subsubsection{Heracles}

Stars from this substructure follow low energy, often eccentric orbits with low L$_{z}$, being largely phase-mixed in velocity and action space. All these features are to be expected in the scenario where Heracles was a massive system that merged early in the history of the Milky Way.  Under this hypothesis, both the fact that the young Milky Way system was smaller in size, and that dynamical friction would have driven the system quickly into low energy orbits, it would lead to this stellar population sinking it into the heart of the Galaxy (\citealp{Horta2021,Pfeffer2021})

The chemical compositions of the stars associated with Heracles are in broad agreement with this scenario. The distribution of Heracles in the $\alpha$-Fe plane does not display the $\alpha$-Fe knee or shin components of chemically evolved systems \citep{McWilliam1997}, which suggests that its star formation was quenched before core collapse supernovae contributed significantly to chemical enrichment. In \cite{Horta2021} we checked that this result was not an effect of the chemical composition criteria adopted in the selection of Heracles stars by comparing them with a similarly selected GES sample, which did display a clear $\alpha$-Fe knee signature. Furthermore,  Heracles occupies a locus in different chemical planes that resembles that of low mass galaxies of the Milky Way and/or accreted populations. In particular \cite{Horta2021} show that the stars associated with Heracles make up a clump in the [Mg/Mn]-[Al/Fe] plane in the region occupied by accreted and/or chemically unevolved populations, characterised by low [Al/Fe] and high [Mg/Mn].

Recent work by \citet[][]{Lane2021} suggests that the concentration associated by \cite{Horta2021} with Heracles in E-L$_z$ space could be an artifact of the APOGEE selection function, so the authors caution that the reality of this halo substructure should be further tested. As discussed by \cite{Horta2021}, phase mixing makes it especially hard to discriminate accreted systems from their {\it in situ} counterparts co-located in the inner few kpc of Milky Way on the sole basis of kinematics.  Nonetheless, numerical simulations predict their existence \cite[e.g.][Horta et al., 2022, in prep.]{Fragkoudi2020,Kruijssen2020,Pfeffer2021}.  Detailed chemistry is thus crucial to tease out the remnants of accreted systems from the maze of {\it in situ} populations overlapping in the inner few kpc of the Galaxy.

For that reason, we examine closely the comparison between the abundance patterns of Heracles data and the inner high-$\alpha$ {\it in situ} population at the same [Fe/H] (Fig.~\ref{ges_combined3}). Our $\chi^2$ analysis shows that the two populations differ chemically with high statistical significance ($\chi^2=55.9$, $p_{\chi^{2}}$=0, similar for instance to the value obtained when comparing the LMC to the SMC, see Table~\ref{tab:chi2}). To check whether this result is sensitive to the choice of [Fe/H]$_{\mathrm{comp}}$, we reran the analysis adopting [Fe/H]$_{\mathrm{comp}}$=--0.9 and [Fe/H]$_{\mathrm{comp}}$=--0.95 obtaining $\chi^{2}$ value of 36.28 and 43.40, respectively, which corresponds to a probability value of $p_{\chi^{2}}$=0 in both cases.

In order to further test this claim, and to ensure that our [Al/Fe] cut employed to select Heracles does not bias our sample and results, we repeat the statistical comparison between Heracles and the inner high-$\alpha$ stars by selecting both populations in the [Mg/Mn]-[Al/Fe] plane adopting various [Al/Fe] criteria (namely, by shifting the diagonal line selection restricting the [Al/Fe] values\footnote{In the [Mg/Mn]-[Al/Fe] plane the high-$\alpha$ population is the clump centred at [Mg/Mn]$\sim$0.55 and [Al/Fe]$\sim$0.3}). The motivation behind this is two-fold: $i$) if the diagonal [Al/Fe] selection is the sole reason for the differences in the chemical abundance properties between Heracles and the \textit{in situ} high-$\alpha$ stellar population, then varying the position of this line should affect the resulting $\chi^{2}$ and associated $p_{\chi^{2}}$ at a significant level, and should lead to Heracles and the inner high-$\alpha$ disc becoming chemically indistinguishable when loosening the [Al/Fe] restriction; $ii$) if Heracles is a distinct population from the high-$\alpha$ stars, this test should provide a deeper understanding of the level of contamination in our Heracles sample by \textit{in situ}  populations which might be caused by this selection in the [Mg/Mn]-[Al/Fe] plane.

In order to perform this test, we select Heracles and inner high-$\alpha$ populations in the [Mg/Mn]-[Al/Fe] plane by shifting the diagonal line by \pm0.05 dex and \pm0.1 dex, respectively. Upon repeating the quantitative analysis (at [Fe/H]=--1), we find that Heracles and the inner high-$\alpha$ stars are still statistically distinct, and that the resulting $\chi^{2}$ and $p_{\chi^{2}}$ vary by small amounts. Specifically, the resulting $\chi^{2}$ values we obtain when shifting the diagonal line in the [Mg/Mn]-[Al/Fe] plane by (+0.10,+0.05,--0.05,--0.10) are $\chi^{2}$=(51.32,58.47,58.93,59.43), leading to a $p_{\chi^{2}}$=0 in all cases. As expected, when increasing (decreasing) the sample of Heracles stars to include more (less) chemically evolved populations, the resulting $\chi^{2}$ value decreases (increases) slightly, despite these systems still being statistically different. This is likely due to the fact that we are selecting more (fewer) high-$\alpha$ disc contaminants in our Heracles sample. While the overlap between Heracles and inner Galaxy high-$\alpha$ stars is small at [Fe/H]=--1, given our results it is evident that their chemical compositions are statistically, given our methodology, distinct.

\begin{figure*}
\includegraphics[width=\textwidth]{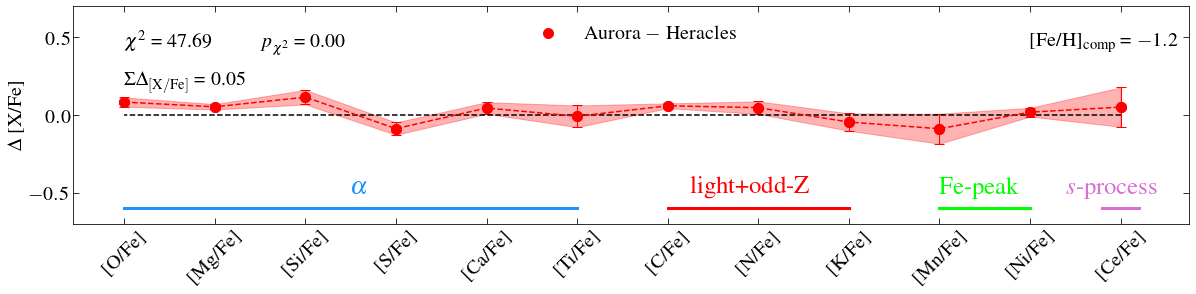}
\caption{The same as Fig~\ref{ges_combined1} but comparing the Aurora APOGEE DR17 targets from \citet[][]{Myeong2022} with the Heracles sample from this work.}    \label{fig_auroravsheracles}
\end{figure*}

In a recent study \citet[][]{Myeong2022} state that the chemical properties of Heracles are consistent with the recently reported Aurora population \citep[see also][]{Rix2022}. Aurora was originally proposed by \citet{Belokurov2022} as an \textit{in situ} component that is kinematically hot, with an approximately  isotropic velocity ellipsoid, and very modest net rotation \citep[see also][]{Conroy2022}. In a follow up study, \cite{Myeong2022} adopt a Gaussian Mixture Model (GMM) approach to describe the distribution of solar neighbourhood populations in chemo-dynamical space, including elemental abundances from APOGEE DR17 and GALAH DR3 \citep{Buder2021}, as well as \textit{Gaia}-based orbital energy.
A key aspect of the \cite{Myeong2022} methodology is that it refrains from adopting any selection criteria in chemical space.  As a result, one of the components emanating from their GMM fit is a highly eccentric (i.e., $e>$0.85) population of APOGEE-\textit{Gaia} stars deemed to be associated with Aurora, which occupy a locus in chemical abundance space that is qualitatively similar to that inhabited by the inner Galaxy Heracles population. 

We examine this claim by applying our statistical method to compare the chemical compositions of Heracles and Aurora, adopting the \cite{Myeong2022} sample for the latter population.  To maximise the statistical robustness of the test, the comparison is performed at [Fe/H]=--1.2, where the samples for both substructures are the densest. As shown in Fig~\ref{fig_auroravsheracles}, we find a resulting $\chi^{2}$=47.69 and $p_{\chi^{2}}$=0, which is similar to the value we have obtained when performing the chemical comparison between Heracles and inner high-$\alpha$  {\it in situ} stars. Moreover, we repeat this test restricting the Heracles sample to the stars with $e>$0.85, for consistency with the Aurora definition by \citet[][]{Myeong2022}. Similarly, we obtain a $\chi^{2}$=40.40 and $p_{\chi^{2}}$=0. Interestingly, the chemical compositions of Aurora stars and those of the inner Galaxy high-$\alpha$ {\it in situ} populations, are statistically similar, yielding $\chi^{2}$=10.17 and $p_{\chi^{2}}$=0.52 when compared at [Fe/H]=--1. These quantitative results support the notion that Heracles and Aurora are chemically distinct. They also attest for the abundance pattern homogeneity of {\it in situ} stellar populations in the inner Galaxy and solar neighbourhood. 

All in all, the overarching conclusion that should be taken from all these additional tests is that Heracles is a stellar population that is distinct, given its chemical abundance values available in APOGEE and the methodology employed in this work, to neighbouring \textit{in situ} populations at the same [Fe/H].

Examining more closely the contrast between the abundance patterns of Heracles and  {\it in situ} populations, including Aurora and their inner high-$\alpha$ counterparts, we find that they differ in interesting ways. By far the elements displaying the largest differences are oxygen, magnesium, and silicon, whose abundances are lower in Heracles than in {\it in situ} populations. At face value, this difference implies less SNII enrichment in Heracles than in the inner high-$\alpha$ disc, possibly reflecting a lower star formation rate. 

We stress that it is possible that the chemical properties ascribed to Heracles may be to some extent influenced by our selection method, which is partly based on chemistry. That selection, however, is far from arbitrary. It is rather informed by the fact that the distribution of inner Galaxy stellar populations form a clear clump in the accreted/chemically unevolved region of the [Mg/Mn]-[Al/Fe] plane \citep[see Figure 1 of][]{Horta2021}.
Our $\chi^2$ analysis shows that Heracles presents an almost unique abundance pattern, differing in a (sometimes small, yet) statistically significant way from most stellar populations under study. For instance, we find that the abundance patterns of Heracles and GES are different (i.e., $p_{\chi^{2}}$ = 0). This is also the case when comparing Heracles to the Sequoia (all three samples), Arjuna, Thamnos, and Nyx.

We conclude by stating that our data are consistent with Heracles being the remnant of a major building block of the halo that merged with the Milky Way in its early history. A possible scenario that could accommodate our quantitative analysis with the claims by \cite{Myeong2022} and \cite{Rix2022} is the following: as the local manifestation of a large {\it in situ} halo, Aurora extends all the way to the inner Galaxy, where it overlaps with Heracles, a likely more compact remnant of an early accretion event. The two systems originally have distinct chemical compositions, with Aurora being characterised by a higher [$\alpha$/Fe] ratio, likely due to a higher early star formation rate. Our measurements most likely underestimate the real chemical differences between Heracles and the {\it in situ} halo.  That is because our Heracles sample is likely contaminated by inner stellar halo counterparts at a level that may be as high as $\sim 40$\%, according to the estimate by \cite{Horta2021}. Further studies based on an expanded set of elemental abundances for a larger sample, as well as detailed modelling, based both on cosmological numerical simulations and standard chemical evolution prescriptions, are required to definitively establish the origin of Heracles and the nature of early {\it in situ} populations.

\subsubsection{\textit{Gaia}-Enceladus/Sausage} \label{sec:GES}

Since its discovery (\citealp[][]{Belokurov2018,Helmi2018}), the \textit{Gaia}-Enceladus/Sausage (GES) substructure has been extensively studied, both from an orbital and chemical compositions perspective (\citealp[e.g.,][]{Hayes2018,Mackereth2019b,Koppelman2019b,Vincenzo2019,Aguado2020,Feuillet2020,Monty2020,Simpson2020,Deokkeun2021,Horta2021,Hasselquist2021,Buder2022, Carrillo2022}). In this work, we have identified a large sample of GES stars, and have shown that stars belonging to this population are characterized by intermediate-to-high orbital energies and high eccentricity, displaying no significant systemic disc-like rotation. 

The chemical compositions of the GES substructure are characterised by lower [$\alpha$/Fe] at [Fe/H]~$\simgreater-1.6$, than high-$\alpha$ disc for most $\alpha$ elements (namely Mg, O, Si, Ca, S), in agreement with previous work (\citealp[e.g.,][]{Hayes2018,Haywood2018,Helmi2018,Mackereth2019b,Horta2021,Buder2022}), and displays an $\alpha$-knee at [Fe/H]$\sim$--1.1 (see Fig~\ref{fig:knees_subs}). The stellar populations of GES are also characterised by lower Al, C, and Ni than \textit{in situ} populations, resembling the abundance patterns of stars from satellites of the Milky Way \citep{Horta2021,Hasselquist2021}. They also occupy the accreted/unevolved region of the [Mg/Mn]-[Al/Fe] plane. In summary, the chemical compositions of GES confirm the results from previous studies, and reinforce the idea that this halo substructure is the remnant of an accreted satellite whose debris dominate the local/inner regions of the stellar halo. 

\subsubsection{Sequoia} 
\label{sec:Sequoia}

The Sequoia substructure was initially discovered due to the highly unbound and retrograde orbits of its constituent stars and associated globular clusters (\citealp[e.g.,][]{barba2019,Matsuno2019,Myeong2019}), which made it easily distinguishable from \textit{in situ} populations. Since its discovery, several groups have sought to identify Sequoia within data from different surveys. Based on an analysis of the \citet{Villalobos2008} $N$-body simulations,  \citet{Koppelman2020} suggested that, rather than a separate system, Sequoia may constitute a fringe GES population comprised of stars in low eccentricity retrograde orbits left over after the merger of that system. A different scenario was proposed by \cite{Naidu2021} who suggest that Sequoia was instead a satellite of GES.  

Whether Sequoia is the remnant of a distinct accretion event or a component of GES, or even its satellite is obviously a very important question, which in principle one should like to address by contrasting the chemical compositions of the two systems.  In order to achieve that goal theoretical predictions for the differences in chemical composition under the various scenarios are required.  \cite{Myeong2019} found that the peak [Fe/H] of their Sequoia sample is lower by $\sim$0.3~dex than that of their GES counterparts.  They argue that the mass-metallicity relation at the relevant redshift implies a 1:10 mass ratio, which in turn is consistent with a Sequoia mass estimate based on the total mass in globular clusters presumably associated with the system.  \cite{Koppelman2020}, in turn, argue that the metallicity gradient expected for a galaxy with the pre-merger size and mass of GES is consistent with the mean metallicity estimates for GES and Sequoia.  However, their metallicity gradient expectation is based on measurements made by \cite{Ho2015} on the gas component of nearby star forming galaxies with $M_{\star}<10^{9.6}$M$_{\odot}$.  It is unclear how accurately those are representative of typical metallicity gradients in dwarf galaxies around 10~Gyr ago.

As discussed in Section~\ref{sec:retro}, our study of the Sequoia system is based on three independent definitions, derived from the \citet[][``\textit{M19}'']{Myeong2019}, \citet[][``\textit{K19}'']{Koppelman_thamnos}, and \citet[][``\textit{N20}'']{Naidu2020} studies.  Because the {\it N20} sample is by definition limited to fairly metal-poor stars, it is not considered in our quantitative analysis.  The mean metallicities we infer when adopting either the {\it M19} or the {\it K19} Sequoia samples are ${\rm \langle[Fe/H]\rangle = -1.41}$ and --1.53, respectively.  In contrast, we obtain ${\rm \langle[Fe/H]\rangle = -1.18}$ for our GES sample, suggesting a difference of $\sim$0.2-0.3~dex, which is  roughly consistent with the estimate by \cite{Myeong2019}.  However, as explained in Section~\ref{Introduction}, our sample is not corrected for selection effects, so we do not ascribe much weight to estimates of any moments of the MDFs of either system.  We thus focus on the differences in abundance pattern, for which unfortunately there are no theoretical predictions that can be confronted by our data.  Consequently the scope of our discussion is limited to comparing the abundance patterns of the two systems, in hopes that the ramifications of that comparison for the nature of Sequoia will be explored in the future on the basis of a more solid theoretical framework.

Qualitatively, we find that the three Sequoia samples occupy a similar locus in all chemical composition planes, resembling that of low mass satelite galaxies of the MW and/or accreted populations, displaying low Mg, Al, C, and Ni, that overlap with the GES substructure. As in the case of the systems discussed above, Sequoia occupies the accreted/unevolved region of the [Mg/Mn]-[Al/Fe] plane, which is encouraging given that two of those samples were selected purely on the basis of orbital parameters. Perhaps not unexpectedly, we find that the \textit{N20} sample appears to be simply the metal-poor tail of the \textit{M19} and \textit{K19} samples.

When running a quantitative $\chi^{2}$ comparison between the {\it M19} and {\it K19} Sequoia samples, we find that their chemical compositions are technically indistinguishable from each other, yielding a probability value of   $p_{\chi^{2}}$=0.23. Therefore we conclude, that the chemical composition we obtain for the Sequoia system  at the reference metallicity of [Fe/H]$_{\mathrm{comp}}=-1.3$, does not depend on the selection criterion adopted.
Despite this similarity, a quantitative comparison between Sequoia and GES leads to conflicting results.  When contrasting GES with the {\it M19} sample, the probability obtained is $p_{\chi^{2}}$=0.64, suggesting that Sequoia and GES are chemically indistinguishable. Conversely, when comparing the {\it K19} Sequoia sample with GES, the resulting probability value is $p_{\chi^{2}}$=0.0 (with a $\chi^{2}$=47.9), which implies a statistically significant difference.

Our samples for GES and Sequoia cover in a statistically meaningful way a wide range of metallicities, so that they lend themselves nicely to a more detailed comparison of the distributions of those two systems in the $\alpha$-Fe plane. A crucial diagnostic is the metallicity of the ``$\alpha$-knee'', [Fe/H]$_{\rm knee}$ (Section~\ref{sec_alphas}), which is strongly sensitive to the details of the star formation history of the system. In recent studies, \cite{Matsuno2019}, \cite{Monty2020}, and \cite{Aguado2020} have obtained different values for [Fe/H]$_{\rm knee}$, ranging from $\sim-2.5$ to $\sim-1.5$, depending on the sample and/or $\alpha$ element considered.  We take advantage of our homogeneous sample in order to revisit this measurement. We perform a piece-wise linear fit to our GES and Sequoia samples in order to accurately determine the position of the [Fe/H]$_{\rm knee}$ in those two systems, in a manner similar to the approach followed by \cite{Mackereth2019b}.

\begin{figure}
\includegraphics[width=\columnwidth]{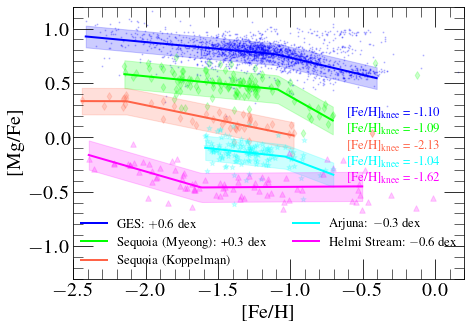}
\caption{Piece-wise polynomial fit (solid line) and 1-$\sigma$ dispersion (shaded region) for  GES, Sequoia, Arjuna, and Helmi stream samples. Data and fits are displaced vertically for clarity. The resulting [Fe/H]$_{\rm knee}$ values are shown. The Mg-Fe knee of GES and the (\textit{M19}) Sequoia are within 0.01~dex from each other, with the largest difference being found between the two Sequoia samples. By the same token, [Fe/H]$_{\rm knee}$ for Arjuna differs from that GES by only 0.06~dex. The star formation efficiencies of these systems, as indicated by [Fe/H]$_{\rm knee}$, seem not to have been substantially different. Conversely, for the Helmi stream we find an "inverted" knee, that occurs at [Fe/H]$_{\rm knee}$$\sim$--1.6, suggestive of a very different star formation history when compared to GES, Sequoia, and Arjuna. For the case of the \textit{K19} Sequoia,the Mg-Fe knee is much more metal-poor ([Fe/H]$_{\rm knee}$=--2.13), again suggestive a lower star formation rate when compared to GES, the other Sequoia samples, and Arjuna.}
    \label{fig:knees_subs}
\end{figure}

The resulting fits (solid lines) and 1-$\sigma$ dispersions (shaded regions) are displayed along with  samples for GES (blue), Sequoia {\it M19} (green) and {\it K19} (red) in Figure~\ref{fig:knees_subs}. The piece-wise function was determined using the \texttt{PiecewiseLinFit} function included as part of the \texttt{pwlf} package \citep[][]{pwlf}.  The [Fe/H]$_{\rm knee}$ values for GES and the \textit{M19} Sequoia sample are within $\sim$0.02 dex from each other. Conversely, the fit to the {\it K19} sample yields a much lower [Fe/H]$_{\rm knee}\simless-2$. Although this result is in relatively good agreement with some of the previous work \citep[e.g.,][]{Matsuno2019,Monty2020}, we deem it quite uncertain, because our sample has very few stars with [Fe/H]$\simless-2$, so that the slope of the ``plateau'' part of the distribution is poorly constrained.  On the other hand, the slope of the Sequoia sequence on the Mg-Fe plane according to the {\it M19} sample is significantly different from that of the {\it K19} sample, suggesting a genuine difference in the star formation history inferred from each sample.

In summary, the situation regarding the nature of the Sequoia system remains somewhat inconclusive.  The fundamental question one would like to answer is whether Sequoia is the remnant of a distinct system, or a component of GES.  While chemistry can provide clues, a definitive answer depends on knowledge of typical or expected pre-merger internal chemical composition variations of dwarf satellites of the Milky Way.  To our knowledge, such information is not available either in the form of observational measurements or theoretical predictions.  When turning to the lower level question of whether Sequoia and GES have similar chemical compositions, we find additional uncertainty, as the result depends on how the Sequoia system is defined.  Out of three definitions adopted in this paper, one delivers a Sequoia sample that differs from GES both in terms of abundance pattern and [Fe/H]$_{\mathrm{knee}}$ \citep{Koppelman2020} \footnote{This result also corroborates the recent findings from \citet[][]{Matsuno2022_seq} using high-resolution spectra of 12 Sequoia stars obtained with the Subaru High Dispersion Spectrograph.}.  Conversely, the sample defined as in \cite{Myeong2019} results in a Sequoia system that is much more similar to GES. To reach resolution of this impasse one will need a bigger and better defined observed sample of the Sequoia system, and preferrably one that is not affected importantly by   selection effects.

\subsubsection{Helmi stream} 
\label{sec:helmi}
The Helmi stream is a halo substructure that appears to jut out of the Galactic disc. It is characterised by stars on highly perpendicular (i.e., high L$_{\perp}$ and v$_{\mathrm{z}}$) and prograde orbits (\citealp{Helmi1999,Koppelman2019}), which appear to form a pillar at high orbital energies in the prograde wing of the E-L$_{z}$ plane (see Fig~\ref{elz}). Despite this substructure being discovered decades ago, its chemical compositions have not been studied in great detail until recently, largely due to the difficulty of obtaining a high confidence sample with reliable chemical composition information. Such studies suggest that the chemical abundance ratios of the Helmi stream resemble that of the rest of halo stars (\citealp[][]{Roederer2010,Limberg2021,Gull2021,Nissen2021}). However, lower abundance ratios in Na and $\alpha$ elements \citep[][]{Matsuno2022}, as well as heavier neutron capture elements \citep[][]{Aguado2021}, have been recorded, suggestive that the Helmi streams have a smaller contribution from massive star nucleosynthesis.

Our results on the chemical compositions of the Helmi stream imply that, unsurprisingly, this substructure presents chemistry that is typical of accreted populations and/or dwarf satellites (i.e., low Mg, Al, C, and Ni). We also find that, despite it being selected purely on a kinematic and position basis, it occupies the accreted/unevolved region of the [Mg/Mn]-[Al/Fe] which, combined with its orbital properties, confirms its accreted nature. Furthermore, we find that, when comparing this halo substructure with the others studied in this work, the Helmi stream differs statistically from all other halo substructures.  In Figure~\ref{fig:knees_subs} we display the data for the Helmi stream alongside a piecewise polynomial fit performed in the same way as described for GES and the Sequoia samples. Interestingly, the best fit for the Helmi stream indicate the occurrence of an ``inverted knee'', whereby the slope of the relation between Mg-Fe becomes less negative. As discussed above for the case of the Sgr dSph, this is the signature of a burst of star formation. This is a very interesting result that is in line with recent work by \citet[][]{Ruiz2022}, who looked at the colour-magnitude diagram of Helmi Stream star candidates in the \textit{Gaia} DR3 data and found hints that this halo substructure experienced a star formation burst approximately $\sim$8Gyr ago. We encourage this finding merits further investigation on the basis of a larger sample. 

We note that in a recent paper \citet[][]{Aguado2021} examined the chemical compositions of the metal-poor part of the Helmi stream (which they call S2). Their sample is comprised of stars within --3 $<$ [Fe/H] $<$ --1.5, and thus only overlap with our Helmi stream sample in the metal-poor regime. \citet[][]{Aguado2021} find an $\alpha$-knee for this substructure at [Fe/H]$\sim$--2, which is different in both shape and value to the one shown in this work. We attribute this difference to the lack of Helmi stream stars in our sample below [Fe/H] $<$ --2. Encouragingly however, we find that at [Fe/H] = --2 our Helmi stream sample has [Mg/Fe]$\sim$0.3 dex, which corroborates the findings from \citet[][]{Aguado2021}.

\subsubsection{Arjuna}
The existence of this substructure was proposed by \cite{Naidu2020} as part of the H3 survey \citep{Conroy2019}.  \cite{Naidu2020} show that the MDF of the retrograde component of the halo displays three peaks, which they ascribe to Arjuna (the most metal-rich), Sequoia, and I'itoi (the most metal-poor, see Fig~\ref{mdf_highe}). \cite{Naidu2021} argue that Arjuna corresponds to the outer parts of the GES progenitor, which, according to their fiducial numerical simulation was stripped early in the accretion process, thus preserving the highly retrograde nature of the GES approaching orbit. In additional support to that proposition, \cite{Naidu2021} point out that the peak [Fe/H] and mean [$\alpha$/Fe] of Arjuna are in excellent agreement with those of GES, which in turn should be consistent with a much lower metallicity gradient in the GES progenitor than suggested by \cite{Koppelman2020}.

The detailed quantitative comparison of the chemical compositions of Arjuna and GES (Figure~\ref{ges_combined1}) shows that the similarity of these two systems indeed encompasses a broader range of elemental abundances, leading to a $p_{\chi^{2}}$ = 0.92. In addition, the distributions of Arjuna and GES stars in the $\alpha$-Fe plane are also very similar, with the two values for [Fe/H]$_{\rm knee}$ agreeing within 0.06~dex (see Fig~\ref{fig:knees_subs}).

Therefore, our results are at face value in agreement with the suggestion by \cite{Naidu2021} that the stars associated with the Arjuna substructure were originally part of GES. That association predicts a very low metallicity gradient for GES at the time of the merger with the Milky Way. Further theoretical and observational work is required to ascertain the reality of that prediction \cite[see discussion in, e.g.,][]{Horta2021,Naidu2021}.

\subsubsection{I'itoi}

Similarly to Arjuna, the I'itoi substructure is a high-energy retrograde substructure identified by \citet{Naidu2020}. However, it is comprised by more metal-poor stars (see Fig~\ref{mdf_highe}) than its high-energy retrograde counterparts, Arjuna and Sequoia.  \cite{Naidu2021} propose that I'itoi was in fact a satellite of GES, based on its low metallicity and high energy retrograde orbit.  Our detailed comparison of the chemistry of GES and I'itoi suggests that their abundance patterns are consistent with a $p_{\chi^{2}}$ = 0.3 (Figure~\ref{ges_combined2}). We point out, however, that this result is highly uncertain, given the relatively small size of our I'itoi sample and its low metallicity ([Fe/H]$_{\rm comp}=-2.1$), which places its stars in a regime where ASPCAP abundances are relatively uncertain. The matter needs revisiting on the basis of more detailed chemical composition studies applied to a larger sample.

\subsubsection{Thamnos}

Initially conjectured by its discoverers to be the amalgamation of two smaller systems \citep{Koppelman_thamnos}, Thamnos is a substructure that occupies a locus in the retrograde wing of the velocity and IoM planes. It is comprised of stars with intermediate orbital energy (i.e., E$\sim$--1.8$\times$10$^{5}$ km$^{2}$ s$^{-2}$) and fairly eccentric orbits ($e$$\sim$0.5), that occupies a position at the foot of GES in the Toomre diagram. 

As in the case of most orbital substructures in this study, we find that the locus occupied by Thamnos in the Ni-Fe, C-Fe, and [Mg/Mn]-[Al/Fe] chemical  planes resembles that of low-mass satellite galaxies and accreted populations of the Milky Way.  However, Thamnos distinguishes itself from other substructures by showing a relatively high [$\alpha$/Fe] ratio, although not as high as Heracles. In fact, Thamnos does not match the abundance pattern of any other substructure in this study.

\subsubsection{Aleph}

Aleph was identified in a study of the stellar halo based on the H3 survey \citep{Naidu2020}. In this work we identify Aleph members by selecting from our parent sample stars that satisfy the selection criteria outlined in \citet{Naidu2020}. We have also searched for stars that are included in both the sample by \citet[][]{Naidu2020} and in APOGEE DR17 (highlighted in the chemical abundance figures with purple edges).
 As described in their work, this substructure is comprised of stars on very strongly prograde orbits with low eccentricity, which appear to follow the same distribution as the Galactic disc component, although at higher J$_{z}$ values. 

 We find that the locus occupied by Aleph in various chemical planes is placed somewhere in between low- and high-$\alpha$ disc stars, while sitting squarely within the \textit{in situ} region of the [Mg/Mn]-[Al/Fe] plane. This is also the case for the two stars common with the sample by \citet[][]{Naidu2020}. Our quantitative comparison of the Aleph chemistry with that of the low-$\alpha$ disc suggests a statistically significant difference, albeit on the borderline ($\chi^{2}$ = 19.8 and $p_{\chi^{2}}$ = 0.05). This is in fact not surprising, seeing as the distribution of Aleph stars on the Mg-Fe plane straddles both the high and low-$\alpha$ discs (Figure~\ref{mgfes}). These results suggest that Aleph may be a stellar population comprised of a mix of warped/flared low-$\alpha$ disc, and high-$\alpha$ disc stars, which also explains its location within the locus of {\it in situ} stellar populations in the [Mg/Mn]-[Al/Fe] plane. 
 
In addition, the chemical abundances of Aleph are in good qualitative agreement with the chemical abundances of the Anticenter Stream (ACS) \citep[][]{Grillmair2006}, as reported by \citet[][]{Laporte2020}. \citet[][]{Laporte2020} argue that the ACS is the remains of a tidal tail of the Galactic disc excited during the first pericentric passage of the Sagittarius dwarf galaxy. If this scenario is correct, the fact that Aleph presents chemical compositions that are barely distinguishable from the low-$\alpha$ disc suggests that the low-$\alpha$ disc of the Galaxy was in place sometime before this satellite interaction. We encourage future work to test this hypothesis.

\subsubsection{LMS-1}

LMS-1 is a metal-poor substructure comprised of stars on mildly prograde orbits at intermediate/high orbital energies. \citet{Yuan2020} identified this substructure by applying a clustering algorithm to SDSS and LAMOST data in the E-L$_{z}$ plane. Although it presents great overlap with the GES in IoM space, it is suggested to be an independent substructure based on the detection of a metallicity peak in the MDF of the stars included within the E-L$_{z}$ box defining this system \citep{Naidu2020}. \cite{Yuan2020} also note that there are potentially several globular clusters with similar metallicity and orbital properties. In this work, we identified LMS-1 candidates adopting the same selection criteria as \citet{Naidu2020}'s, obtaining a relatively small sample of only 31 stars. 

We examine the distribution of LMS-1 stars in various chemical planes, concluding that its chemistry is consistent with those of other accreted systems, with all its stars falling in the "accreted/unevolved" region of the [Mg/Mn]-[Al/Fe] plane. Furthermore, the $\sim$0.5 probability value obtained when comparing this halo substructure with GES suggests that {\it at face value} LMS-1 could be part of either of the omnipresent GES substructure. However, these comparisons are made at [Fe/H]$_{\rm comp}=-2.1$, where our samples are small and ASPCAP abundances are relatively more uncertain and {\it in situ} and accreted structures tend to converge towards the same locus in the regions of chemical space sampled by APOGEE \citep[e.g.,][]{Horta2021}.  Moreover, given its location in IoM space, it is possible that our LMS-1 sample is importantly contaminated by GES stars.

As mentioned in Section~\ref{sec:helmi}, \citet[][]{Jean2017} show that a single accretion event can lead to multiple overdensities in orbital space \citep[see also][]{Koppelman2020}. Due to the chemical similarity between LMS-1 and GES, and the close proximity between these two halo substructures in orbital planes (see Fig~\ref{elz}), an association between these two systems seems tempting. However, we defer a firmer conclusion to a future when more elemental abundances are obtained for a larger sample of both LMS-1 and GES candidates.

\subsubsection{Nyx}
The Nyx substructure is conjectured to be a stellar stream in the solar vicinity of the Milky Way \citep{Necib2020}. Given the chemical compositions obtained for this substructure in this work and its strong overlap with the high-$\alpha$ disc in all the chemical planes studied, we suggest that the Nyx is likely comprised by \textit{in situ} high-$\alpha$ disc populations.  Our quantitative estimate of the similarity between Nyx and the stars from the high-$\alpha$ disc yields a very low $\chi^{2}$ with associated likelihood probability of 0.75.  We note that the stars in common between our sample and that of \citet{Necib2020} seem to boldly confirm this result. Along these lines, we note that our result is in agreement with a recent study focused on studying the chemical compositions of the Nyx substructure \citep[][]{Zucker2021}, who also conjecture Nyx to be comprised of Galactic (high-$\alpha$) disc populations.

\section{Conclusions}
\label{conclusion}
The unequivocal association with accretion events of halo substructures identified on the basis of orbital information alone is extremely difficult, as demonstrated by recent numerical simulations \citep[e.g.,][]{Jean2017,Koppelman2020,Naidu2021}. Substructure in integrals of motion space can also be influenced or even artificially created by survey selection effects \citep[][]{Lane2021}. Because the chemical compositions of halo substructures contain a fossilised record of the  evolutionary histories of their parent galaxies, abundance pattern information can help linking substructure in orbital space to their progenitor systems.

In this work we have utilised a cross-matched catalogue of the latest APOGEE (DR17) and \textit{Gaia} (EDR3) data releases in order to study the chemo-dynamic properties of substructures previously identified in the stellar halo of the Milky Way. We have successfully distinguished stars in the APOGEE DR17 catalogue that are likely associated with the following substructures: \textit{Gaia}-Enceladus/Sausage, Sagittarius dSph, Heracles, Helmi stream, Sequoia, Thamnos, Aleph, LMS-1, Arjuna, I'itoi, Nyx, Icarus, and Pontus. Using the wealth of chemical composition information provided by APOGEE, we have examined the distributions of the stellar populations associated with these substructures in a range of abundance planes, probing different nucleosynthetic channels. We performed a quantitative comparison of the abundance patterns of all the substructures studied in order to evaluate their mutual associations, or lack thereof.  Below we summarise our main conclusions:

\begin{itemize}

\item We show that the chemical compositions of the majority of the halo substructures so far identified in the Galactic stellar halo (namely, \textit{Gaia}-Enceladus/Sausage, Heracles, Sagittarius dSph, Helmi stream, Sequoia, Thamnos, LMS-1, Arjuna, I'itoi) present chemical compositions which resemble those of low-mass satellites of the MW and/or accreted populations. There are however a couple of exceptions, namely Nyx and Aleph, that do not follow this pattern and instead present chemical properties similar to those of populations formed \textit{in situ}. Furthermore, in Appendix~\ref{appen_icarus} we discuss the nature of Icarus, concluding that this substructure is likely comprised of stars formed \textit{in situ}, for which the ASPCAP abundances are unreliable. 

\item The chemical properties of the inner-Galaxy structure, Heracles, differ from those of its {\it in situ} metal-poor counterparts, both in the inner halo and in the solar neighbourhood (Aurora).  In particular, Heracles displays lower abundances of $\alpha$ elements oxygen, magnesium, and silicon than the  {\it in situ} populations.  A possible interpretation of this result is that the star formation rate of Heracles was lower than that of the early {\it in situ} halo. By the same token, Heracles is found to have higher [$\alpha$/Fe] ratios than GES which, as suggested by \cite{Horta2021}, is an indication of an early truncation of star formation and the resulting absence of an $\alpha$-knee in the former system. The abundance pattern of Heracles is indeed found to differ from all of the other substructures studied in this work.  We propose a scenario according to which Heracles and Aurora overlap spatially in the innermost Galaxy. Further studies based on additional elemental abundances for larger samples, as well as detailed numerical modelling, are required to disentangle the {\it in situ} and accreted metal-poor populations cohabiting the innermost Galactic halo (i.e., R$_{\mathrm{GC}}$$<$ 4 kpc).

\item We show that a large fraction of the substructures studied (namely, (\textit{M19/N20}) Sequoia, Arjuna, LMS-1, I'itoi) present chemistry indistinguishable from that of the omnipresent \textit{Gaia}-Enceladus/Sausage. These findings bring into question the independence of these substructures, which are at least partially overlapping with GES in kinematic planes (see Fig~\ref{elz}). In view of these similarities, claims in the literature about the nature of Sequoia as being originally a higher angular momentum component located in the outskirts of GES \citep[][]{Koppelman2020}, a satellite of GES \citep{Naidu2021}, or an altogether unrelated system \citep{Myeong2019} may need to be reexamined. The possibility that Sequoia (selected as done in \citet[][]{Myeong2019} and \citet[][]{Naidu2020}) was a detached, but much less massive galaxy than GES is likely challenged by their chemical similarity.  On the other hand, confirmation that it, or Arjuna, might be the remains of populations originally located in the outskirts of GES depends crucially on the magnitude of chemical composition gradients one should expect for dwarf galaxies at $z\approx2$, and on whether that is a sufficiently discriminating criterion.

\item We found that the halo hosts substructures which differ from GES in a statistically significant way. Three among those susbstructures display chemistry that is genuinely suggestive of an accreted nature, namely Heracles, Thamnos, and the Helmi stream (although for the latter this conclusion is not firm due to uncertainties in the chemistry and small sample size). 

\item Conversely, the chemistry of the two remaining substructures, Nyx and Aleph, is indistinguishable from that of {\it in situ} populations. We conjecture that Nyx forms part of the high-$\alpha$ disc. For the case of Aleph, we suggest that it could likely comprised of stars both from the low-$\alpha$ (flared/warped) disc as well as stars from the high-$\alpha$ disc, or be part of the Anti-Centre stream.

\item The situation regarding the nature of the Sequoia
system remains somewhat inconclusive given our results. Out of three definitions adopted in this paper, one delivers a Sequoia sample that differs from GES both in terms of abundance pattern and [Fe/H]$_{\mathrm{knee}}$ \citep[][]{Koppelman_thamnos}. Conversely, the sample defined as in \citet[][]{Myeong2019} results in a Sequoia system that is much more similar to GES. For the case of the Sequoia as defined in \citet[][]{Naidu2020}, the hard [Fe/H] cut prevents us from performing any quantitative comparison.

\item Our results suggest that the local/inner (R$_{GC}$ $\lesssim$ 20kpc) halo is comprised of the debris from at least three massive accretion events (Heracles, GES, and Sagittarius) and two/three lower mass debris (Thamnos, Helmi stream, Sequoia). Upcoming large spectroscopic surveys probing deeper into the outer regions of the stellar halo (beyond R$_{GC}$ $\sim$20kpc) will likely uncover the debris from predicted lower-mass/more recent accretions (\citealp[e.g.,][Horta et al. 2022, in prep]{Bosch2016}), and will enable the full characterisation of those already known. Conversely the precise contribution of massive accretion to the stellar populations content of the inner few kpc of the halo is still to be fully gauged. Heracles is likely the result of a real building block of the halo, likely the most massive (in relative terms) to have ever happened in the history of the Milky Way, but we are still scratching its surface.

\end{itemize}

This paper presents a chemical composition study of substructures identified (primarily) on the basis of phase-space and orbital information in the stellar halo of the Milky Way. Current and upcoming surveys will continue to map the stellar halo and will provide further clues to the nature and reality of phase-space substructures discovered in recent years. For the inner regions of the Galaxy and the local halo, the Milky Way Mapper \citep[][]{sdssv} and the Galactic component of the MOONS GTO survey \citep{moons} will provide revolutionising chemo-dynamical information for massive samples. For the outer regions of the stellar halo, the WEAVE \citep[][]{Dalton2012}, 4MOST \citep[][]{DeJong2019}, and DESI \citep[][]{DESI2016} surveys will provide spectroscopic data for millions of stars in both high and low resolution. In addition, in this work we have only studied the chemistry of the most massive accretion events. However, there is a plethora of lower-mass halo substructures in the form of coherent stellar streams that has not been studied in this work and also require to be fully examined (see \citealt[][]{Li2021} for a recent example). The advent of surveys like $S^{5}$ \citep[][]{Li2019} will aid in such endeavours. All this information, when coupled with the exquisite astrometry and upcoming spectroscopic information from the \textit{Gaia} satellite, will provide the necessary information for further discoveries and examinations of substructure in the stellar halo of the Milky Way.

\section*{Acknowledgements}
The authors thank GyuChul Myeong and Vasily Belokurov for making the IDs of the Aurora stars available, Vasily Belokurov for helpful comments on earlier versions of the manuscript, Rohan Naidu and Charlie Conroy for making available the H3 catalogue in digital format, and Paola Re Fiorentin and Alessandro Spagna for sharing their Icarus star's APOGEE IDs. DH thanks Tadafumi Matsuno, Melissa Ness, Holger Baumgardt, and Cullan Howlett for helpful scientific discussions, and Sue, Alex, and Debra for their constant support. D.G. gratefully acknowledges financial support from the Direcci\'on de Investigaci\'on y Desarrollo de la Universidad de La Serena through the Programa de Incentivo a la Investigaci\'on de Acad\'emicos (PIA-DIDULS). Funding for the Sloan Digital Sky Survey IV has been provided by
the Alfred P. Sloan Foundation, the U.S. Department of Energy Office
of Science, and the Participating Institutions. SDSS acknowledges
support and resources from the Center for High-Performance Computing
at the University of Utah. The SDSS web site is www.sdss.org. SDSS
is managed by the Astrophysical Research Consortium for the
Participating Institutions of the SDSS Collaboration including the
Brazilian Participation Group, the Carnegie Institution for Science,
Carnegie Mellon University, the chilean Participation Group, the
French Participation Group, Harvard-Smithsonian Center for Astrophysics,
Instituto de Astrof\'{i}sica de Canarias, The Johns Hopkins University,
Kavli Institute for the Physics and Mathematics of the Universe
(IPMU) / University of Tokyo, the Korean Participation Group,
Lawrence Berkeley National Laboratory, Leibniz Institut f\"{u}r Astrophysik
Potsdam (AIP), Max-Planck-Institut f\"{u}r Astronomie (MPIA Heidelberg),
Max-Planck-Institut f\"{u}r Astrophysik (MPA Garching), Max-Planck-Institut
f\"{u}r Extraterrestrische Physik (MPE), National Astronomical Observatories
of china, New Mexico State University, New York University, University
of Notre Dame, Observatório Nacional / MCTI, The Ohio State University,
Pennsylvania State University, Shanghai Astronomical Observatory,
United Kingdom Participation Group, Universidad Nacional Autónoma
de México, University of Arizona, University of Colorado Boulder,
University of Oxford, University of Portsmouth, University of Utah,
University of Virginia, University of Washington, University of
Wisconsin, Vanderbilt University, and Yale University.

This work presents results from the European Space Agency (ESA) space mission Gaia. Gaia data are being processed by the Gaia Data Processing and Analysis Consortium (DPAC). Funding for the DPAC is provided by national institutions, in particular the institutions participating in the Gaia MultiLateral Agreement (MLA). The Gaia mission website is \href{https://www.cosmos.esa.int/gaia}{https://www.cosmos.esa.int/gaia}. The Gaia archive website is \href{https://archives.esac.esa.int/gaia}{https://archives.esac.esa.int/gaia}.

{\it Software:} Astropy \citep{astropy:2013,astropy:2018}, SciPy
\citep{Scipy2020}, NumPy \citep{NumPy}, Matplotlib \citep{Hunter:2007},
Galpy \citep{Galpy2015,Galpy2018}, TOPCAT \citep{Taylor2005}.

{\it Facilities:} Sloan Foundation 2.5m Telescope of Apache Point Observatory (APOGEE-North), Ir\'en\'ee du Pont 2.5m Telescope of Las Campanas Observatory (APOGEE-South), \textit{Gaia} satellite/European Space Agency (\textit{Gaia}).

\section*{Data availability}
All APOGEE DR17 data used in this study is publicly available and can be found at: https: /www.sdss.org/dr17/


\bibliographystyle{mnras}
\bibliography{refs}
\vspace{0.2cm}
$^{1}$ \textit{Astrophysics Research Institute, 146 Brownlow Hill, Liverpool, L3 5RF, UK}\\
$^{2}$ \textit{School of Mathematics and Physics, The University of Queensland, St. Lucia, QLD 4072, Australia}\\
$^{3}$ \textit{Canadian Institute for Theoretical Astrophysics, University of Toronto, Toronto, ON M5S 3H8, Canada}\\
$^{4}$\textit{Dunlap Institute for Astronomy and Astrophysics, University of Toronto, Toronto, ON M5S 3H4, Canada}\\
$^{5}$ \textit{Department of Astronomy and Center for Cosmology and AstroParticle Physics,
Ohio State University, Columbus, OH, 43210, USA}\\
$^{6}$\textit{Space Telescope Science Institute, 3700 San Martin Drive, Baltimore, MD 21218, USA}\\
$^{7}$ \textit{Lund Observatory, Department of Astronomy and Theoretical Physics, Box 43, SE-221\,00 Lund, Sweden}\\
$^{8}$ \textit{Department of Astronomy, University of Virginia, Charlottesville, VA 22904-4325, USA}\\
$^{9}$ \textit{Instituto de Astrof\'isica de Canarias (IAC), E-38205 La Laguna, Tenerife, Spain}\\
$^{10}$ \textit{Universidad de La Laguna (ULL), Departamento de Astrof\'isica, E-38206 La Laguna, Tenerife, Spain}\\
$^{11}$\textit{The Observatories of the Carnegie Institution for Science, Pasadena, CA 91101, USA}\\
$^{12}$\textit{New Mexico State University, Las Cruces, NM 88003, USA}\\
$^{13}$ \textit{University of Arizona, Tucson, AZ 85719, USA}\\
$^{14}$ \textit{Observat\'orio Nacional, S\~ao Crist\'ov\~ao, Rio de Janeiro, Brazil}\\
$^{15}$ \textit{Departamento de Astronom\'{i}a, Casilla 160-C, Universidad de Concepci\'{o}n, Chile}\\
$^{16}$ \textit{Instituto de Investigaci\'{o}n Multidisciplinario en Ciencia y Tecnolog\'{i}a, Universidad de La Serena. Avenida Ra\'{u}l Bitr\'{a}n S/N, La Serena, Chile}\\
$^{17}$ \textit{Departamento de Astronom\'{i}a, Facultad de Ciencias, Universidad de La Serena. Av. Juan Cisternas 1200, La Serena, Chile}\\
$^{18}$ \textit{Materials Science and Applied Mathematics, Malm\"o University, SE-205 06 Malm\"o, Sweden}\\
$^{19}$ \textit{Centro de Investigaci\'on en Astronom\'ia, Universidad Bernardo O'Higgins, Avenida Viel 1497, Santiago, Chile}\\
$^{20}$ \textit{Gothard Astrophysical Observatory, ELTE Eotv\"os Lor\'and University, 9700 Szombathely, Szent Imre H. st. 112, Hungary}\\
$^{21}$ \textit{MTA-ELTE Lend{\"u}let "Momentum" Milky Way Research Group, Hungary}\\
$^{22}$ \textit{Departamento de Ciencias Fisicas, Facultad de Ciencias Exactas, Universidad Andres Bello, Av. Fernandez Concha 700, Santiago, Chile}\\
$^{23}$ \textit{Vatican Observatory, V00120 Vatican City State, Italy}\\
$^{24}$\textit{Centro de Astronom{\'i}a (CITEVA), Universidad de Antofagasta, Avenida Angamos 601, Antofagasta 1270300, Chile}\\
$^{25}$ \textit{University of Texas at Austin, McDonald Observatory, TX 79734-3005, USA}\\
$^{26}$\textit{National Optical Astronomy Observatories, Tucson, AZ 85719, USA}\\
$^{27}$ \textit{Department of Physics $\&$ Astronomy, University of Utah, Salt Lake City, UT 84105, USA}

\appendix

\section{The reality of Icarus}
\label{appen_icarus}
The Icarus substructure is a metal-poor, [Mg/Fe]-poor structure that was identified by \citet{Refiorentin2021} using the APOGEE DR16 and GALAH DR3 data. We have attempted to identify Icarus star members by employing the criteria listed by \citet{Refiorentin2021}, and identified only one potential member within the parent sample described in Section~\ref{sec_icarus}.  We thus decided to retrieve the DR17 properties of the 41 Icarus stars presented by \citet[][]{Refiorentin2021} on the basis of APOGEE DR16 data.

\begin{figure}
\includegraphics[width=\columnwidth]{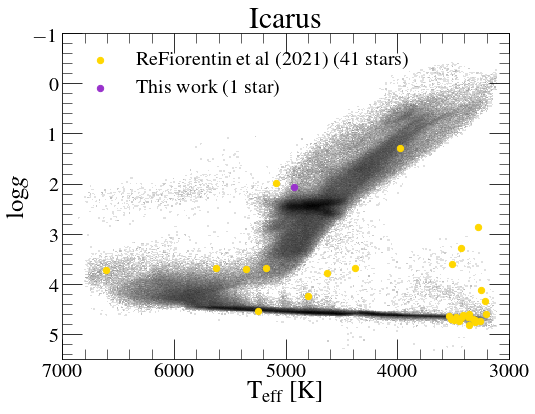}
\caption{Kiel diagram of the parent sample used in this work, the Icarus star identified in this paper (purple), and the Icarus stars identified by \citet[][]{Refiorentin2021} (yellow), using APOGEE DR17 data.}
    \label{icarus_loggteff}
\end{figure}

\begin{figure}
\includegraphics[width=\columnwidth]{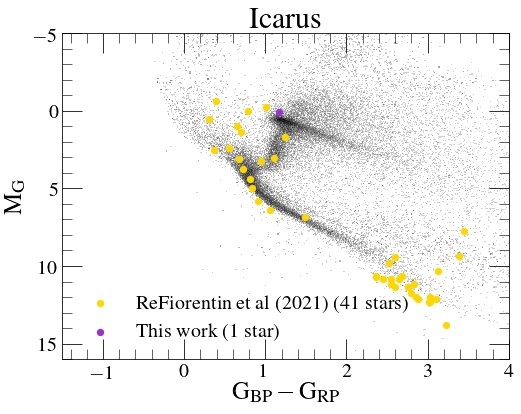}
\caption{The same samples as in Fig~\ref{icarus_loggteff}, now in the \textit{Gaia} colour magnitude diagram using \textit{Gaia} EDR3 data.}
    \label{icarus_cmd}
\end{figure}

We found that only 3 out of the 41 Icarus candidates have T$_{\mathrm{eff}}$ and $\log g$ characteristic of red giant stars. The remaining 38 stars have stellar parameters consistent with a main sequence or subgiant nature (Figure~\ref{icarus_loggteff}), which is the reason why the selection criteria described in Section~\ref{data} excluded them from our parent sample. This finding is confirmed by the positions of the Icarus member sample in the \textit{Gaia} colour-magnitude diagram shown in Figure~\ref{icarus_cmd}. Icarus was conjectured to be an accreted substructure in the Galactic disc based largely on chemical abundance information.  However elemental abundances for sub-giant and main-sequence, and particularly for M~dwarfs, were not very reliable in APOGEE DR16 \citep[although see][]{Birky_2020,Souto2020}.  Therefore, we decided to perform some checks to see if the resulting abundances for these (primarily) M~dwarf stars are reliable in APOGEE DR17.

We checked the \texttt{STAR$_{-}$FLAG} and  \texttt{ASPCAP$_{-}$FLAG} flags of the 41 Icarus stars determined by \citet[][]{Refiorentin2021}. The results from this examination showed that: {\it i)} 25/41 stars had a "\texttt{BRIGHT$_{-}$NEIGHBOUR}", "\texttt{PERSIST$_{-}$HIGH}", "\texttt{PERSIST$_{-}$LOW}", "\texttt{LOW$_{-}$SNR}", \texttt{SUSPECT$_{-}$BROAD$_{-}$LINES}, \texttt{SUSPECT$_{-}$ROTATION}", \texttt{MULTIPLE$_{-}$SUSPECT}", flag set in their \texttt{STAR$_{-}$FLAG}, suggesting the occurrence of various issues with the observed spectra of these stars, which can lead up to uncertainties in stellar parameters and elemental abundances; {\it ii)} 40/41 stars had one or more of the following ASPCAP\_FLAGs raised: "\texttt{STAR$_{-}$WARN}", "\texttt{COLOUR$_{-}$TEMPERATURE$_{-}$WARN}", "\texttt{SN$_{-}$WARN}", "\texttt{VSINI$_{-}$WARN}", "\texttt{VMICRO$_{-}$WARN}", "\texttt{N$_{-}$M$_{-}$WARN}", "\texttt{TEFF$_{-}$WARN}", "\texttt{LOGG$_{-}$WARN}", suggesting that the results from the \texttt{ASPCAP} analysis for these stars are likely uncertain. 

In light of these results, we chose to exclude Icarus from this study, and suggest that this substructure is not the debris from a cannibalised dwarf galaxy, but is rather likely comprised of (primarily) M-dwarf stars with unreliable APOGEE abundances in the disc of the Milky Way.

\section{The chemical compositions of Pontus according to APOGEE}
\label{appen_pontus}

As we have only been able to identify two star members belonging to the Pontus halo substructure, we will refrain from making any strong statement about its chemistry. However, based solely on the chemical compositions of the two member stars we have identified in this work, we find that the Pontus substructure presents elevated abundances in $\alpha$-elements (namely, in O, Mg, Si), low C, Al, Ti, and Ce, and approximately solar Mn and Ni. The two stars we find associated with the Pontus substructure display similar chemical compositions to that of other halo substructures and/or satellite galaxies of the Milky Way. Although it is impossible to attribute any similarities or differences on the basis of two stars alone, given the close proximity between these two Pontus stars and Thamnos in the E-L$_{z}$ plane, it is tempting to associate the former to the later. However, a larger sample of Pontus stars with reliable chemistry is required to ascertain this speculation.

\begin{figure*}
\includegraphics[width=\textwidth]{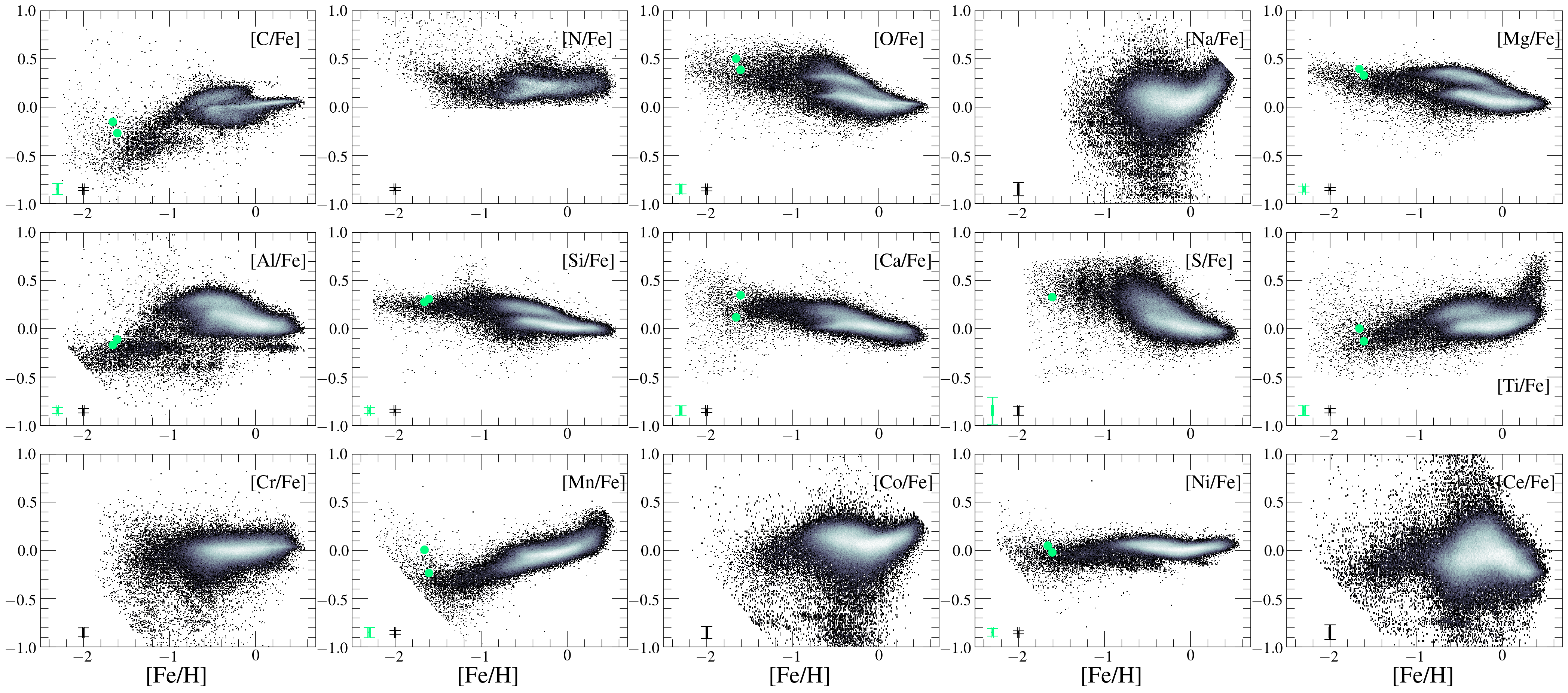}
\caption{Pontus stars (light green points) in every chemical composition plane studied in this work, from carbon to cerium. We note that for the C and N plots, we restrict our sample to stars with 1 $<$ log$g$ $<$ 2, to account for dredge-up effects in the red giant branch, and for all abundances, we also make the following cuts: \texttt{X$_{-}$FE$_{-}$FLAG} = 0 and \texttt{X$_{-}$FE$_{-}$ERR} $<$ 0.15. For the case of N, Na, Cr, Co, and Ce, the \texttt{X$_{-}$FE$_{-}$ERR} $<$ 0.15 restriction removes the two Pontus stars, and for S it removes one.}
\label{pontus_abundances_all}
\end{figure*}

\section{Chemical and kinematic properties of halo substructures}

\begin{table*}
\setlength{\tabcolsep}{8pt}
\centering
\caption{Mean orbital parameter values for each substructure studied in this work. We note that the the angular momentum, actions, orbital energy, eccentricity, maximum height above the plane, perigalacticon, and apogalacticon are computed using the \citet{McMillan2017} potential.}
\begin{tabular}{ |p{2.2cm}|p{4.cm}|p{4.2cm}|p{4.3cm}}
\hline
 Name & (v$_{\mathrm{R}}$,v$_{\phi}$,v$_{\mathrm{z}}$) &(L$_{\mathrm{x}}$,L$_{\mathrm{y}}$,L$_{z}$) &  (J$_{\mathrm{r}}$,J$_{\phi}$,J$_{\mathrm{z}}$)\\
  & (kms$^{-1}$) & (kpc kms$^{-1}$) & (kpc kms$^{-1}$)\\
\hline
\hline
Heracles & (1.98\pm106.31, 11.66\pm68.19, 2.35\pm90.40)& (--1.63\pm148.57, --3.76\pm190.74, 26.47\pm117.23)&(104.85\pm77.97, 29.42\pm116.92, 160.81\pm115.25)\\
\hline
\textit{Gaia}-Enceladus/Sausage & (--3.02\pm179.09, 6.87\pm46.65, 5.69\pm123.71)& (16.37\pm469.05, 69.90\pm866.59, 15.38\pm260.42)&  (1043.02\pm449.16, 29.81\pm260.10, 735.78\pm650.55)\\
\hline
Sagittarius & (184.78\pm66.70, 105.56\pm124.02, 124.94\pm65.12)& (460.23\pm345.39, 3982.96\pm1388.07, 735.78\pm860.83)&  (1164.12\pm575.14, 713.83\pm871.68, 3030.59\pm1236.89)\\
\hline
Helmi stream &(--34.50\pm131.06, 150.67\pm60.45, --63.37\pm164.74) &(--449.24\pm1210.39, 106.21\pm1666.54, 1165.73\pm274.68)&(793.40\pm1352.24, 1177.70\pm280.38, 1353.60\pm569.47)\\
\hline
Sequoia (\textit{M19}) &(33.02\pm165.33, --172.78\pm103.35, 17.01\pm106.54)&(546.64\pm855.13, 124.07\pm889.85, --1483.30\pm778.86)&(1140.20\pm682.65, --1460.83\pm774.12, 518.01\pm406.26)\\
\hline
Sequoia (\textit{K19}) &(35.72\pm166.22, --224.72\pm58.49, 54.78\pm122.73)& (824.20\pm986.23, --182.94\pm844.54, --1886.80\pm399.91)& (830.49\pm427.55, --1816.35\pm396.41, 662.92\pm441.86)\\
\hline
Sequoia (\textit{N20}) & (35.77\pm131.25, --172.51\pm90.16, 8.25\pm134.44)& (652.20\pm736.62, 236.21\pm1025.36, --1446.04\pm783.43)&(770.00\pm759.03, --1423.37\pm780.40, 699.91\pm609.53)\\
\hline
Thamnos &(13.05\pm87.02, --118.38\pm40.33, --12.71\pm71.43)& (92.86\pm294.91, 133.09\pm461.47, --849.80\pm192.88)& (266.58\pm91.32, --829.93\pm189.28, 187.87\pm148.79)\\
\hline
Aleph & (--4.04\pm38.55, 212.50\pm23.29, --14.04\pm33.63)&(--345.82\pm656.52, --124.49\pm369.04, 2040.06\pm422.91)&(44.06\pm35.72, 2055.17\pm427.09, 188.46\pm10.46)\\
\hline
LMS-1 & (5.61\pm93.13, 125.65\pm34.12, 16.73\pm94.98)&(--201.77\pm698.43, 68.87\pm901.47, 824.52\pm131.36)&(254.87\pm110.66, 836.90\pm132.36, 660.16\pm470.59)\\
\hline
Arjuna & (17.19\pm172.19, --176.91\pm88.75, 45.49\pm131.23)&(605.69\pm1156.87, --207.79\pm948.02, --1417.31\pm665.88) &(1281.13\pm1734.03, --1396.31\pm659.71, 738.17\pm625.25)\\
\hline
I'itoi & (19.61\pm99.41, --156.35\pm63.47, 59.13\pm133.86)&(562.58\pm1064.28, --375.26\pm1092.75, --1360.07\pm422.32)&(455.74\pm548.87, --1338.53\pm419.21, 897.37\pm615.83)\\
\hline
Nyx & (133.85\pm21.27, 158.01\pm26.21, 5.73\pm48.88)&(--57.72\pm164.31, 17.24\pm365.47, 1242.37\pm276.75)&(291.19\pm104.29, 1258.34\pm278.25, 55.49\pm54.31)\\
\hline
\hline
\end{tabular}
\label{taborbits}
\end{table*}

\vspace{0.5cm}

\begin{table*}
\setlength{\tabcolsep}{8pt}
\centering
\caption{Table~\ref{taborbits} continued.}
\begin{tabular}{ |p{2.2cm}|p{3.4cm}|p{3.3cm}}
\hline
 Name & E &(ecc,z$_{\mathrm{max}}$,$R_{\mathrm{peri}}$,$R_{\mathrm{apo}}$)\\
  & (km$^{2}$s$^{-2}$) & (--,kpc,kpc,kpc)\\
\hline
\hline
Heracles &--224860.22\pm15322.44&(0.81\pm0.12 ,1.67\pm0.85, 0.33\pm0.24, 3.01\pm1.02) \\
\hline
\textit{Gaia}-Enceladus/Sausage &--142394.72\pm12519.88&(0.93\pm0.06, 9.84\pm6.14, 0.61\pm1.03, 17.15\pm5.22)\\
\hline
Sagittarius &--107654.44\pm17153.96&(0.69\pm0.20, 35.74\pm17.68, 8.81\pm7.34, 38.12\pm17.86)\\
\hline
Helmi stream &--127688.22\pm18244.58&(0.65\pm0.15, 16.37\pm9.37, 3.97\pm1.36, 24.05\pm20.67)\\
\hline
Sequoia (\textit{M19}) &--126706.32\pm17892.71&(0.73\pm0.10, 11.11\pm9.01, 4.13\pm2.49, 26.55\pm14.52)\\
\hline
Sequoia (\textit{K19}) &--120838.38\pm7843.52& (0.62\pm0.11, 10.31\pm4.98, 5.75\pm1.72, 25.34\pm6.45)\\
\hline
Sequoia (\textit{N20}) &--133482.69\pm19836.49& (0.65\pm0.12, 10.06\pm8.72, 4.25\pm2.31, 21.94\pm14.94)\\
\hline
Thamnos & --169972.44\pm5539.53& (0.57\pm0.10, 3.10\pm1.60, 2.47\pm0.70, 9.05\pm1.08)\\
\hline
Aleph &--146048.56\pm9553.77&(0.18\pm0.07, 4.06\pm0.56, 8.46\pm2.00, 12.06\pm2.73)\\
\hline
LMS-1 &--156766.37\pm10493.48&(0.61\pm0.06, 7.04\pm2.75, 2.59\pm0.54, 10.89\pm1.95)\\
\hline
Arjuna &--125782.70\pm21774.65&(0.72\pm0.13, 14.80\pm20.52, 4.08\pm2.41, 29.62\pm31.21)\\
\hline
I'itoi &--136512.56\pm18198.76&(0.55\pm0.16, 10.87\pm6.53, 4.84\pm1.98, 18.09\pm10.51)\\
\hline
Nyx &--161398.10\pm8604.29&(0.50\pm0.08, 1.58\pm0.85, 3.78\pm0.95, 11.20\pm1.98)\\
\hline
\hline
\end{tabular}
\end{table*}

\begin{table*}
\setlength{\tabcolsep}{8pt}
\centering
\caption{Mean $\langle$[X/Fe]$\rangle$ abundance values for every halo substructure identified in this study. As we only identified 2(1) stars belonging to Pontus(Icarus), we exclude these halo substructures from this table.}
\begin{tabular}{ p{1.1cm}|p{1.45cm}|p{1.45cm}|p{1.45cm}|p{1.45cm}|p{1.45cm}|p{1.45cm}}
\hline
$\langle$[X/Fe]$\rangle$ & Heracles & \textit{Gaia}-Enceladus/ Sausage &  Sagittarius & Helmi stream & Sequoia (\textit{M19}) & Sequoia (\textit{K19}) \\
\hline
\hline
$\langle$[C/Fe]$\rangle$ &--0.30\pm0.18 &--0.30\pm0.29 & --0.38\pm0.14& --0.35\pm0.36& --0.31\pm0.32& --0.34\pm0.31\\
\hline
$\langle$[N/Fe]$\rangle$ & 0.26\pm0.18&0.19\pm0.37 & 0.04\pm0.14& 0.22\pm0.33& 0.13\pm0.34& 0.10\pm0.36\\
\hline
$\langle$[O/Fe]$\rangle$ & 0.32\pm0.09&0.24\pm0.19 &--0.01\pm0.08 & 0.27\pm0.16& 0.26\pm0.17& 0.28\pm0.17\\
\hline
$\langle$[Na/Fe]$\rangle$ & --0.01\pm0.42& 0.02\pm0.51& --0.34\pm0.27& 0.21\pm0.48& 0.14\pm0.50& 0.27\pm0.57\\
\hline
$\langle$[Mg/Fe]$\rangle$ & 0.28\pm0.06&0.16\pm0.14 &--0.03\pm0.08 &0.19\pm0.14& 0.17\pm0.13& 0.17\pm0.12\\
\hline
$\langle$[Al/Fe]$\rangle$ & --0.13\pm0.09& --0.21\pm0.23&--0.44\pm0.08 &--0.25\pm0.21& --0.29\pm0.17& --0.31\pm0.15\\
\hline
$\langle$[Si/Fe]$\rangle$ & 0.25\pm0.06& 0.14\pm0.13& --0.05\pm0.10& 0.16\pm0.12& 0.15\pm0.10& 0.14\pm0.09\\
\hline
$\langle$[Ca/Fe]$\rangle$ & 0.19\pm0.12&0.15\pm0.18 & 0.01\pm0.07& 0.14\pm0.16 & 0.16\pm0.17& 0.16\pm0.23\\
\hline
$\langle$[Ti/Fe]$\rangle$ & --0.04\pm0.13& --0.04\pm0.20&--0.15\pm0.09 & --0.11\pm0.20&--0.07\pm0.21 & --0.09\pm0.18\\
\hline
$\langle$[Cr/Fe]$\rangle$ & --0.13\pm0.33&--0.10\pm0.40 & --0.07\pm0.19& --0.04\pm0.36& --0.04\pm0.41& --0.02\pm0.42\\
\hline
$\langle$[Mn/Fe]$\rangle$ & --0.32\pm0.14& --0.28\pm0.22&--0.20\pm0.10 & --0.23\pm0.20& --0.33\pm0.18& --0.27\pm0.20\\
\hline
$\langle$[Co/Fe]$\rangle$ & --0.19\pm0.36&--0.09\pm0.44 & --0.16\pm0.15& --0.06\pm0.37& --0.08\pm0.44& --0.00\pm0.50\\
\hline
$\langle$[Ni/Fe]$\rangle$ & --0.03\pm0.08&--0.07\pm0.15 &--0.14\pm0.06 & --0.06\pm0.10& --0.11\pm0.10& --0.13\pm0.11\\
\hline
$\langle$[Ce/Fe]$\rangle$ & --0.15\pm0.24& --0.11\pm0.36&--0.00\pm0.26 & --0.12\pm0.34& --0.12\pm0.42& --0.17\pm0.39\\
\hline
$\langle$[Fe/H]$\rangle$ & --1.30\pm0.21& --1.18\pm0.42&--0.72\pm0.34 &--1.39\pm0.55 & --1.41\pm0.36& --1.53\pm0.45\\
\hline
\hline
\end{tabular}
\label{tababundances}
\end{table*}

\begin{table*}
\setlength{\tabcolsep}{12pt}
\centering
\caption{Table~\ref{tababundances} continued.}
\begin{tabular}{p{1.1cm}|p{1.45cm}|p{1.45cm}|p{1.45cm}|p{1.45cm}|p{1.45cm}|p{1.45cm}|p{1.45cm}}
\hline
$\langle$[X/Fe]$\rangle$ & Sequoia (\textit{N20}) & Thamnos & Aleph & LMS-1 & Arjuna & I'itoi & Nyx \\
\hline
\hline
$\langle$[C/Fe]$\rangle$ & --0.29\pm0.43& --0.17\pm0.28&0.01\pm0.07& --0.46\pm0.23& --0.36\pm0.24& --0.38\pm0.56&0.07\pm0.09\\
\hline
$\langle$[N/Fe]$\rangle$ & 0.23\pm0.37& 0.12\pm0.36& 0.17\pm0.08& 0.29\pm0.39& 0.11\pm0.26& 0.27\pm0.53& 0.08\pm0.16\\
\hline
$\langle$[O/Fe]$\rangle$ & 0.30\pm0.17& 0.43\pm0.13& 0.19\pm0.09& 0.38\pm0.11& 0.25\pm0.17& 0.41\pm0.11& 0.33\pm0.11\\
\hline
$\langle$[Na/Fe]$\rangle$ & 0.43\pm0.53& 0.24\pm0.57& 0.02\pm0.26 &0.38\pm0.49 & 0.04\pm0.48& 0.71\pm0.56&0.04\pm0.32\\
\hline
$\langle$[Mg/Fe]$\rangle$ & 0.22\pm0.12& 0.33\pm0.08& 0.17\pm0.06& 0.28\pm0.09 & 0.16\pm0.11& 0.34\pm0.06&0.32\pm0.08\\
\hline
$\langle$[Al/Fe]$\rangle$ &--0.33\pm0.18& --0.06\pm0.23& 0.12\pm0.07& --0.30\pm0.10 & --0.28\pm0.16& --0.32\pm0.08&0.24\pm0.12\\
\hline
$\langle$[Si/Fe]$\rangle$ & 0.17\pm0.10& 0.30\pm0.10& 0.10\pm0.05&0.22\pm0.07 & 0.16\pm0.10& 0.23\pm0.07&0.21\pm0.06\\
\hline
$\langle$[Ca/Fe]$\rangle$ & 0.16\pm0.24& 0.26\pm0.18& 0.06\pm0.04& 0.10\pm0.27& 0.16\pm0.11& 0.21\pm0.47& 0.17\pm0.08\\
\hline
$\langle$[Ti/Fe]$\rangle$ & --0.09\pm0.29& 0.01\pm0.17& 0.07\pm0.10&--0.11\pm0.20& --0.10\pm0.16& 0.06\pm0.34& 0.15\pm0.09\\
\hline
$\langle$[Cr/Fe]$\rangle$ & 0.08\pm0.47& --0.05\pm0.40& --0.04\pm0.11 & --0.12\pm0.40&--0.18\pm0.36 & 0.23\pm0.39&--0.05\pm0.16\\
\hline
$\langle$[Mn/Fe]$\rangle$ & --0.31\pm0.22& --0.37\pm0.18&  --0.08\pm0.05&--0.30\pm0.22& --0.35\pm0.16& 0.05\pm0.15& --0.15\pm0.09\\
\hline
$\langle$[Co/Fe]$\rangle$ & 0.05\pm0.46& 0.02\pm0.41& 0.02\pm0.22& --0.02\pm0.52&--0.15\pm0.42 &0.17\pm0.45 &0.06\pm0.26\\
\hline
$\langle$[Ni/Fe]$\rangle$ &  --0.10\pm0.14& --0.04\pm0.12&  0.03\pm0.02& --0.05\pm0.12& --0.09\pm0.09& --0.01\pm0.14&0.06\pm0.04\\
\hline
$\langle$[Ce/Fe]$\rangle$ & --0.13\pm0.41& --0.13\pm0.31& --0.12\pm0.29& --0.17\pm0.25&--0.18\pm0.40& 0.02\pm0.35&--0.16\pm0.29\\
\hline
$\langle$[Fe/H]$\rangle$ & --1.80\pm0.10&--1.53\pm0.36& --0.53\pm0.17&--1.83\pm0.25& --1.33\pm0.13 & --2.16\pm0.12&--0.55\pm0.29 \\
\hline
\hline
\end{tabular}
\end{table*}

\clearpage

\section{$\alpha$-elements}
\label{app_alphas}

\begin{figure*}
\includegraphics[width=\textwidth]{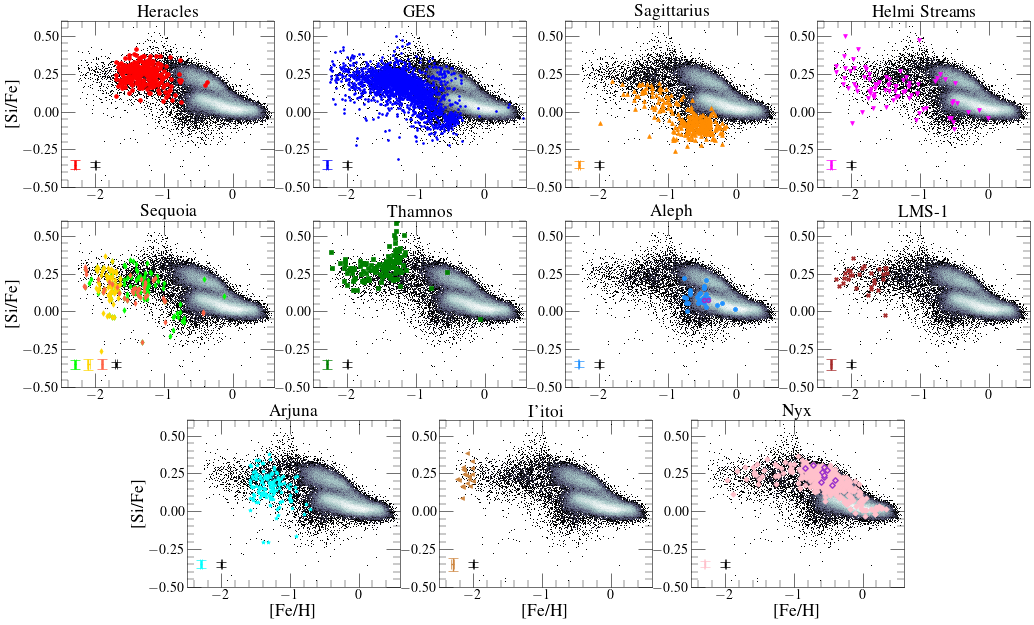}
\caption{The same illustration as Fig.~\ref{mgfes} in the Si-Fe plane. }
    \label{sife}
\end{figure*}

\begin{figure*}
\includegraphics[width=\textwidth]{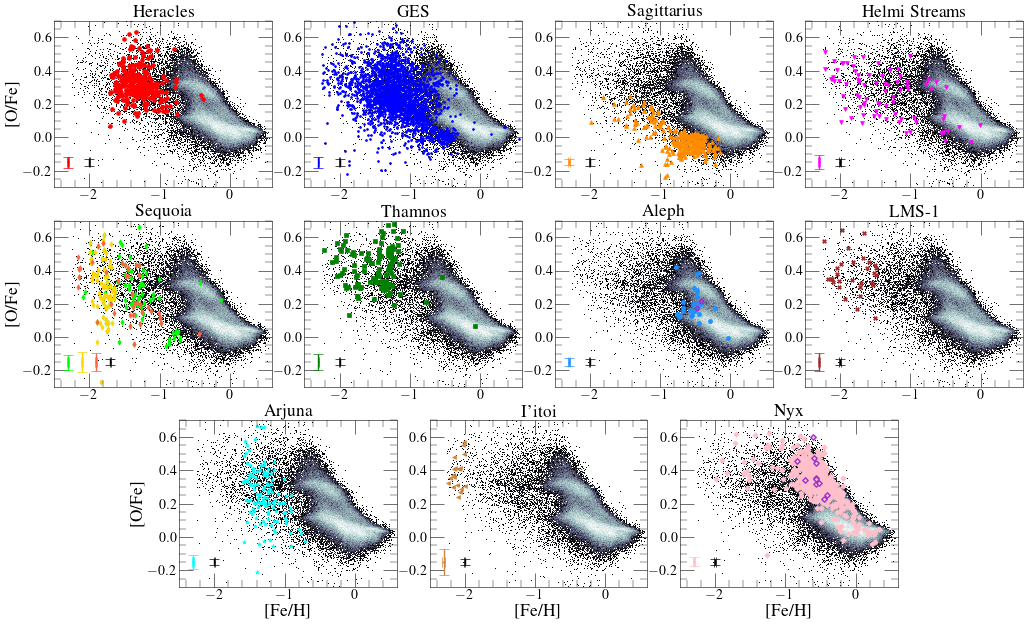}
\caption{The same illustration as Fig.~\ref{mgfes} in the O-Fe plane. }
    \label{ofe}
\end{figure*}

\begin{figure*}
\includegraphics[width=\textwidth]{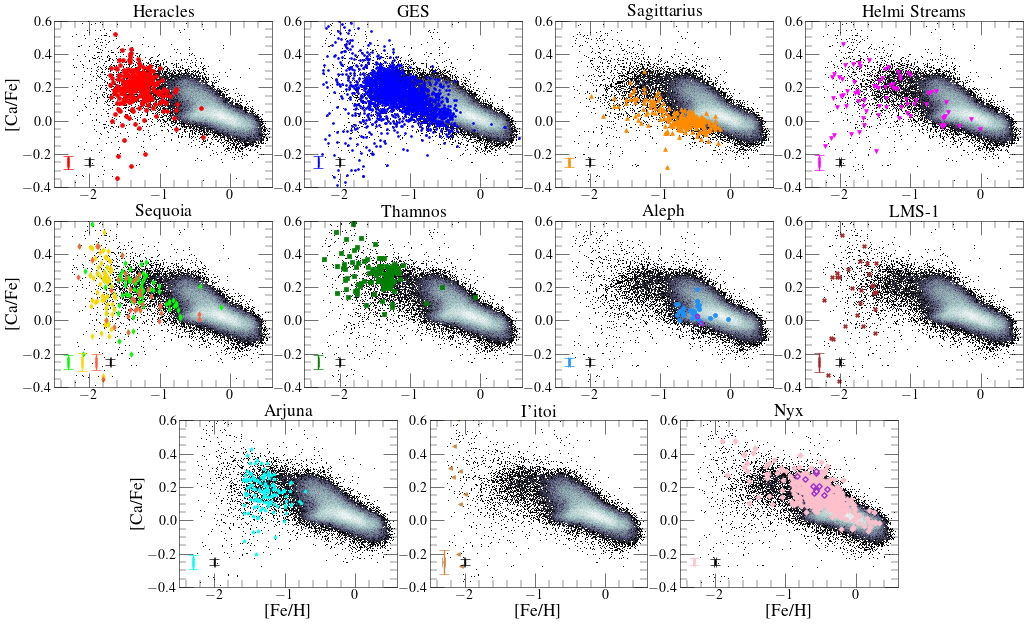}
\caption{The same illustration as Fig.~\ref{mgfes} in the Ca-Fe plane. }
    \label{cafe}
\end{figure*}

\begin{figure*}
\includegraphics[width=\textwidth]{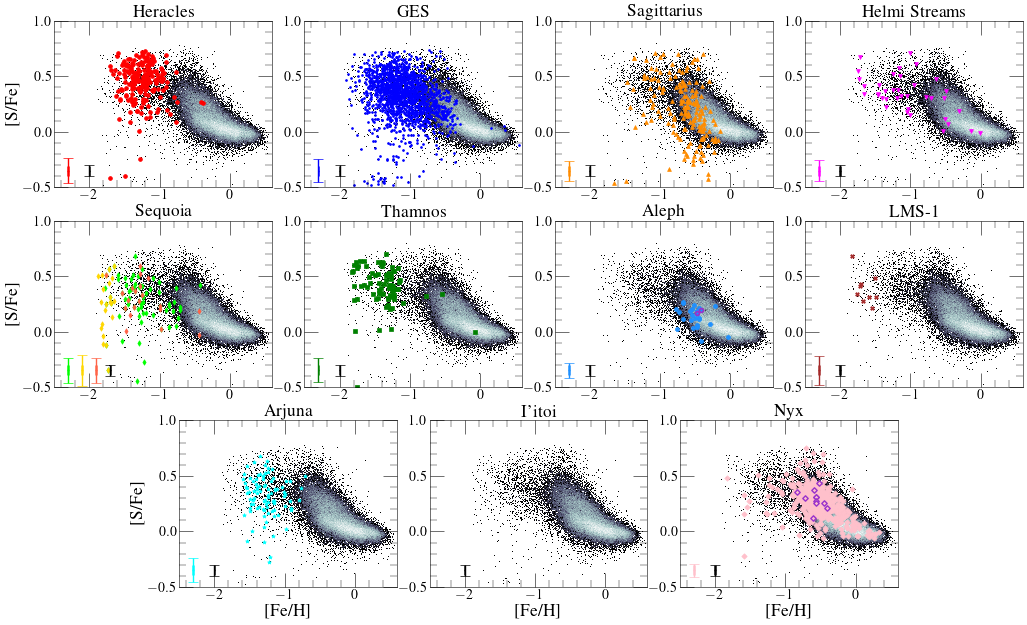}
\caption{The same illustration as Fig.~\ref{mgfes} in the S-Fe plane. We note that the grid limit appears clearly in this plane at the lowest [Fe/H] and highest [S/Fe] values. }
    \label{sfe}
\end{figure*}

\begin{figure*}
\includegraphics[width=\textwidth]{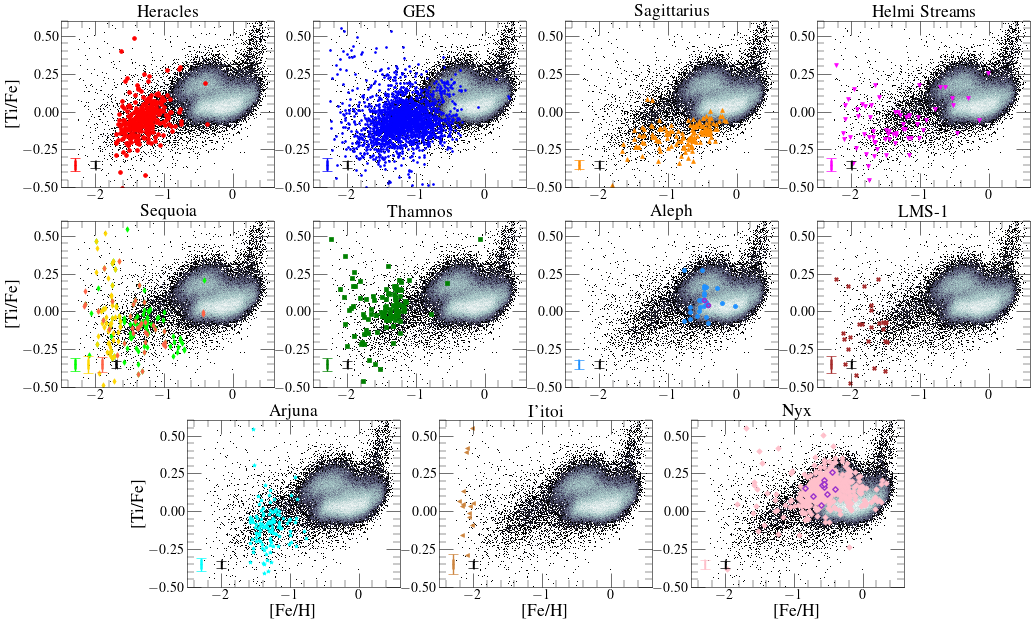}
\caption{The same illustration as Fig.~\ref{mgfes} in the Ti-Fe plane. }
    \label{tife}
\end{figure*}

\section{(C+N)/Fe}
\label{app_carbon_nitrogen}

\begin{figure*}
\includegraphics[width=\textwidth]{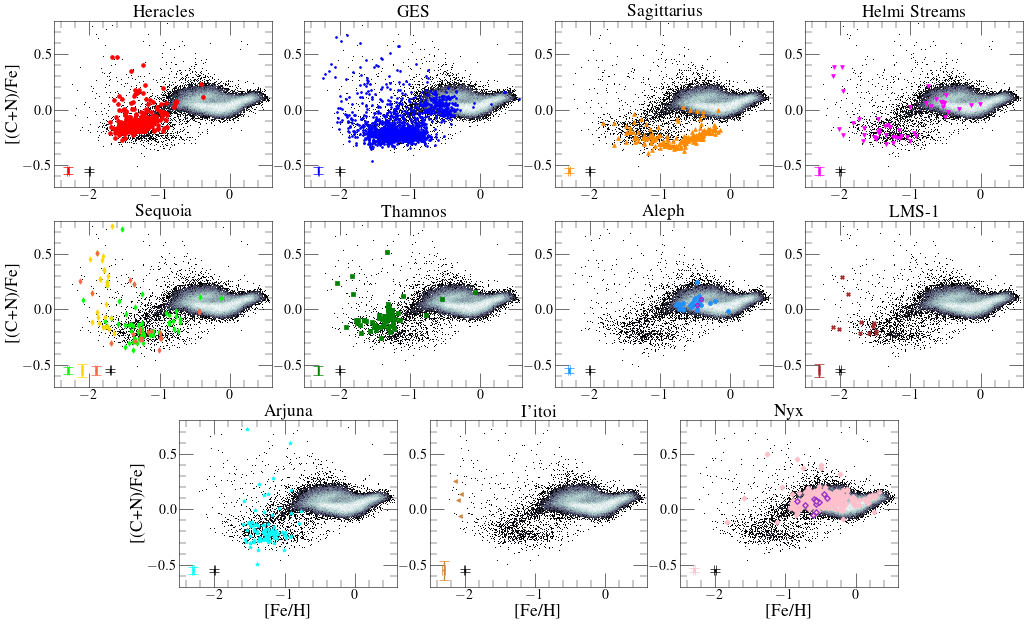}
\caption{The same illustration as Fig.~\ref{mgfes} in the (C+N)-Fe plane.}
    \label{cnfe}
    
\end{figure*}

\section{Iron-peak elements}
\label{app_ironpeak}

\begin{figure*}
\includegraphics[width=\textwidth]{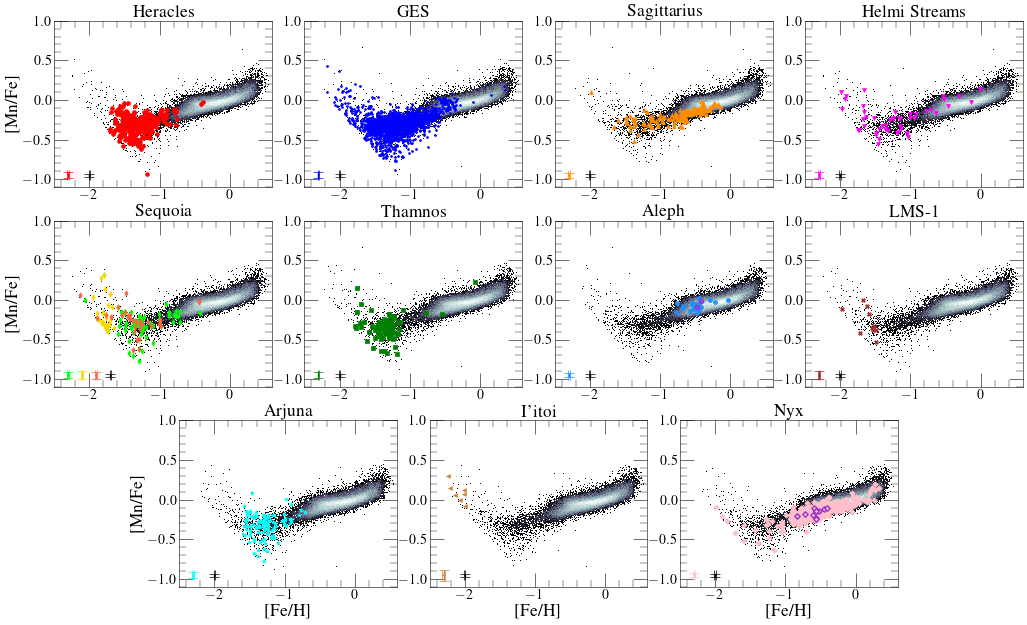}
\caption{The same illustration as Fig.~\ref{mgfes} in the Mn-Fe plane. We note that the grid limit appears clearly in this plane at the lowest [Fe/H] values.}
    \label{mnfe}
\end{figure*}

\begin{figure*}
\includegraphics[width=\textwidth]{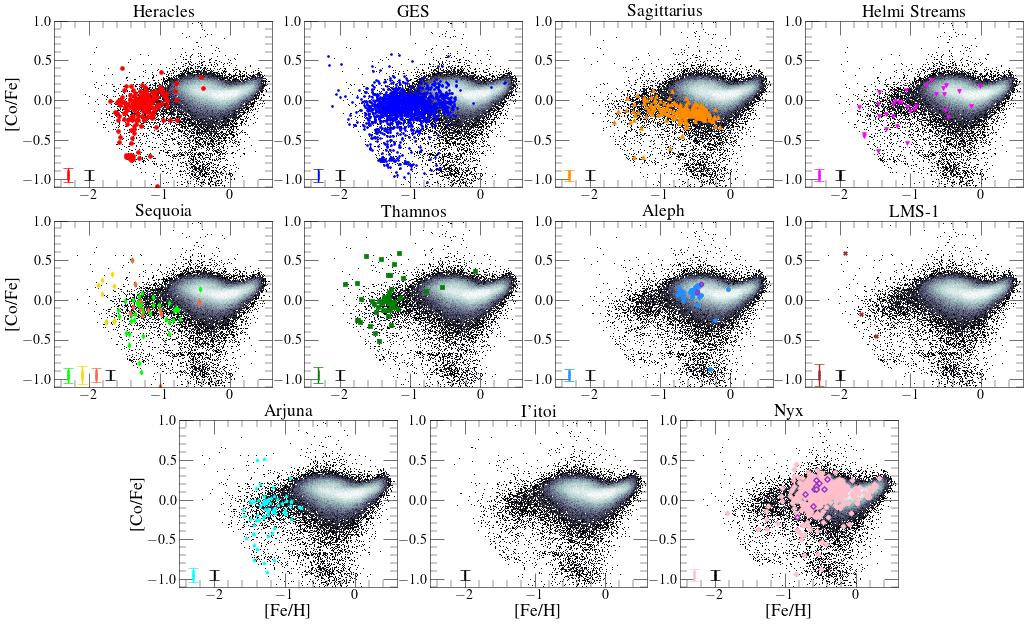}
\caption{The same illustration as Fig.~\ref{mgfes} in the Co-Fe plane. We note that the grid limit appears clearly in this plane at the lowest [Fe/H] values.}
    \label{cofe}
\end{figure*}

\begin{figure*}
\includegraphics[width=\textwidth]{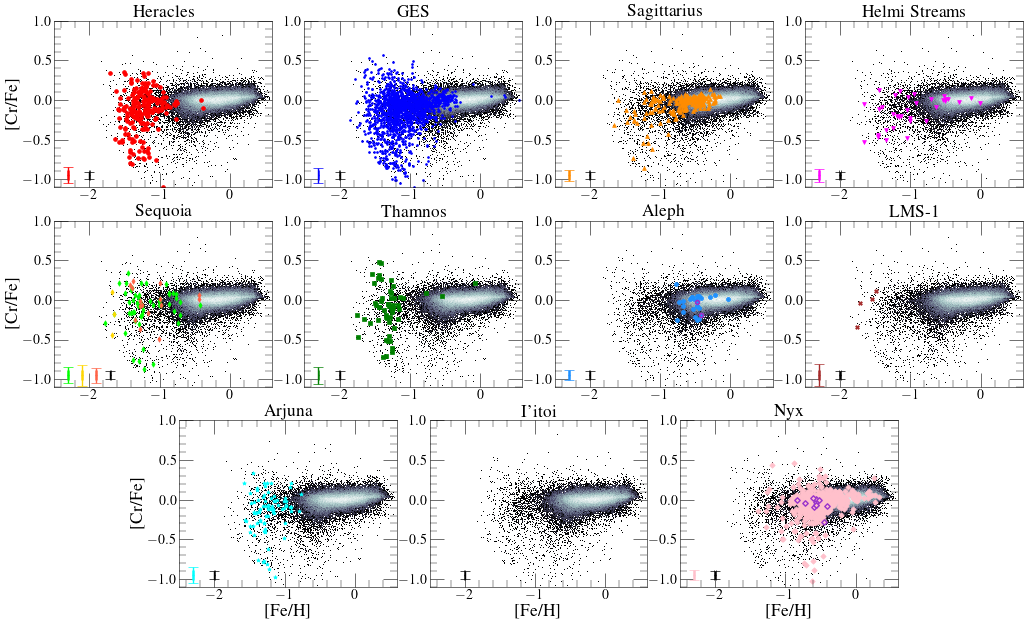}
\caption{The same illustration as Fig.~\ref{mgfes} in the Cr-Fe plane. We note that the grid limit appears clearly in this plane at the lowest [Fe/H] values.}
    \label{crfe}
\end{figure*}

\section{Odd-Z elements}
\label{app_oddZ}

\begin{figure*}
\includegraphics[width=\textwidth]{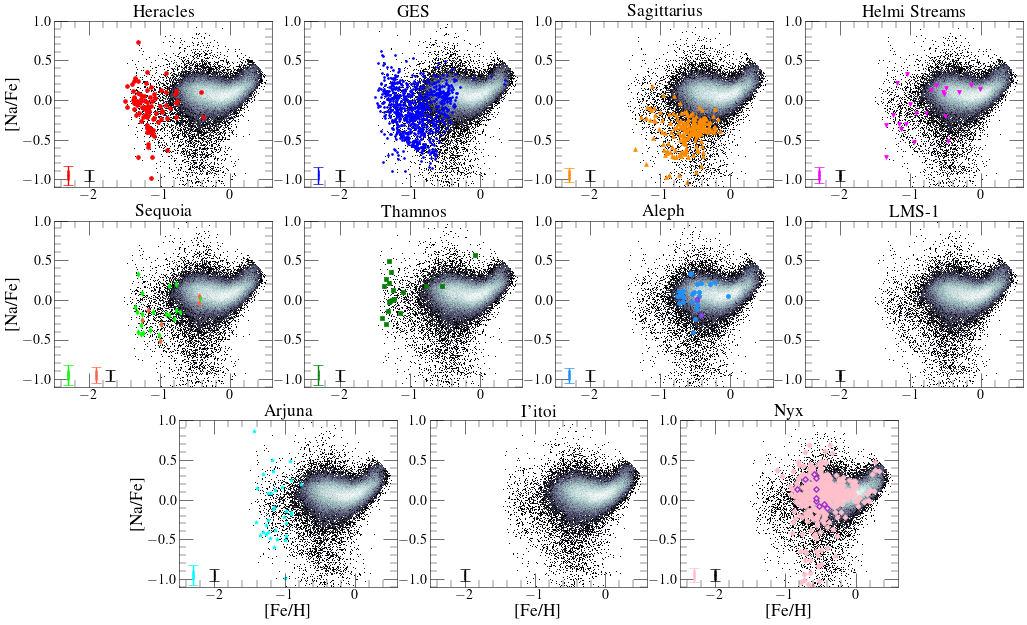}
\caption{The same illustration as Fig.~\ref{mgfes} in the Na-Fe plane. We note that the grid limit appears clearly in this plane at the lowest [Fe/H] values.}
    \label{nafe}
\end{figure*}

\begin{figure*}
\includegraphics[width=\textwidth]{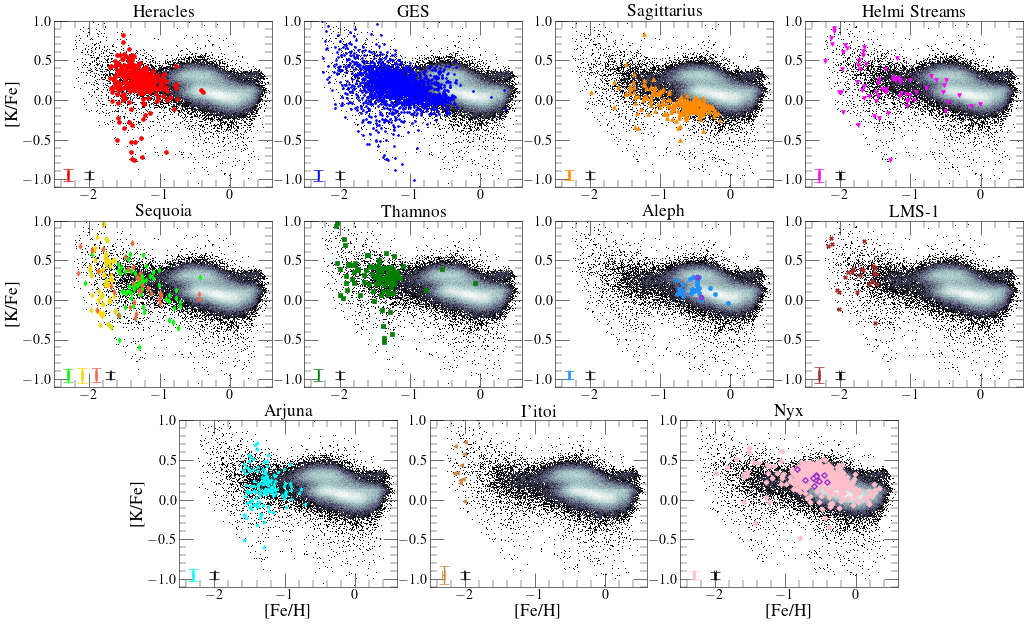}
\caption{The same illustration as Fig.~\ref{mgfes} in the K-Fe plane. We note that the grid limit appears clearly in this plane at the lowest [Fe/H] values.}
    \label{kfe}
\end{figure*}

\section{[Fe/H] used to compare substructures}
\label{appen_confusion}
\begin{figure*}
\includegraphics[width=0.75\textwidth]{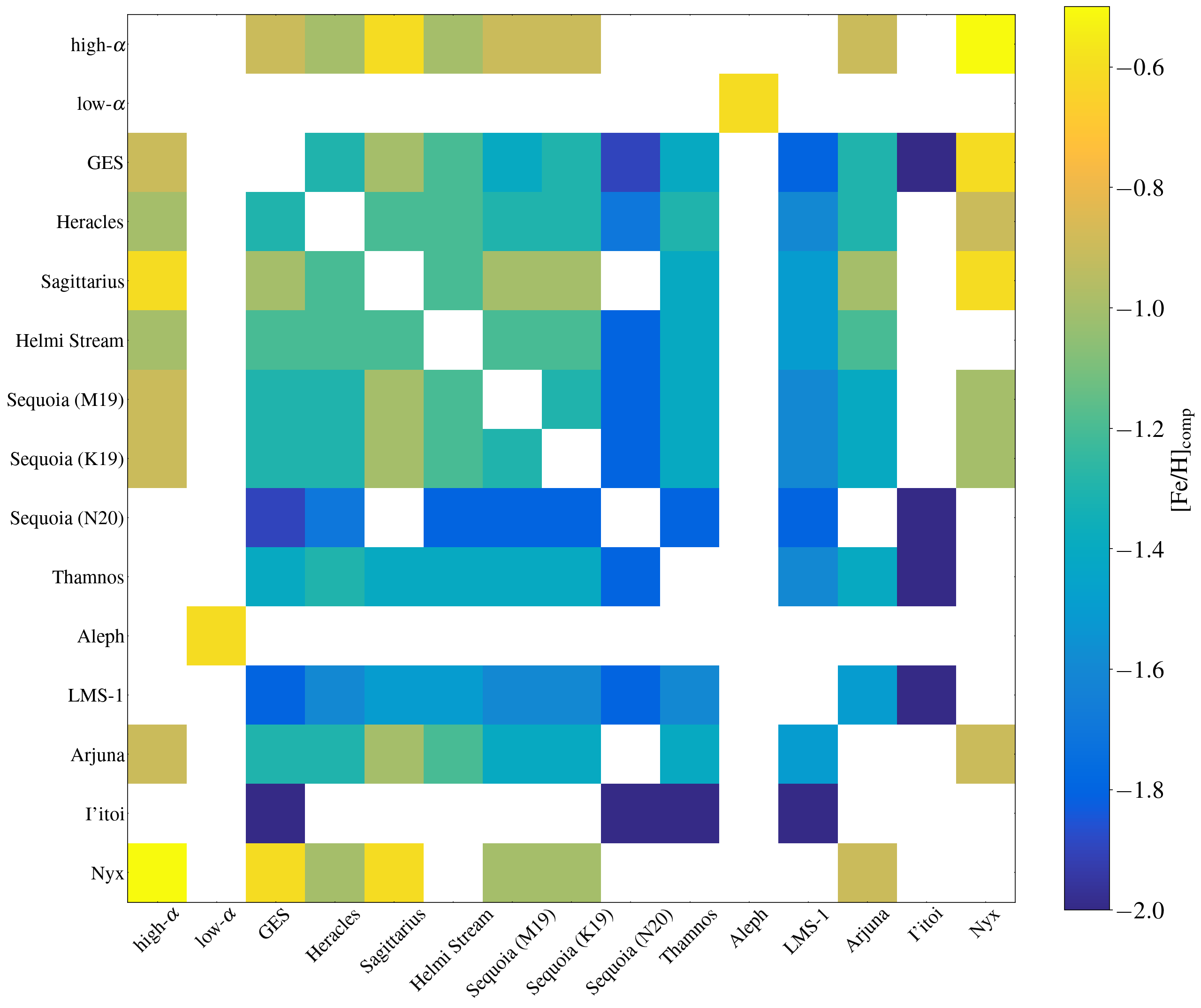}
\caption{[Fe/H]$_{\mathrm{comp}}$ used to obtain the results from Fig~\ref{confusion_matrix} when comparing every halo substructure with all the other substructures and with a high-/low-$\alpha$ disc sample.}
    \label{confused_matrix_feh}
\end{figure*}

\bsp	
\label{lastpage}
\end{document}